\documentclass[%
 reprint,
 amsmath,amssymb,
 aps,
]{revtex4-2}

\usepackage{graphicx}                                             
\usepackage{dcolumn}                                              
\usepackage{bm}                                                   
\usepackage[colorlinks=true, allcolors=blue]{hyperref}            
\usepackage[nameinlink,capitalize]{cleveref}                      

\usepackage{xr}

\makeatletter
\newcommand*{\addFileDependency}[1]{
  \typeout{(#1)}
  \@addtofilelist{#1}
  \IfFileExists{#1}{}{\typeout{No file #1.}}
}
\makeatother

\crefname{section}{Sec.}{Sec.}

\begin{document}

\preprint{APS/123-QED}

\title{Mechanism of Local Lattice Distortion Effects on Vacancy Migration Barriers in FCC Alloys}

\author{Zhucong Xi}
\altaffiliation[]{Department of Materials Science and Engineering, University of Michigan, Ann Arbor, Michigan, 48109, USA}
\author{Mingfei Zhang}
\altaffiliation[]{Department of Materials Science and Engineering, University of Michigan, Ann Arbor, Michigan, 48109, USA}
\author{Louis G. Hector Jr.}
\altaffiliation[]{GM Global Technical Center, General Motors Company, Warren, Michigan, 48092, USA}
\author{Amit Misra}
\altaffiliation[]{Department of Materials Science and Engineering, University of Michigan, Ann Arbor, Michigan, 48109, USA}
\author{Liang Qi}
\altaffiliation[]{Department of Materials Science and Engineering, University of Michigan, Ann Arbor, Michigan, 48109, USA}
\email{qiliang@umich.edu}

\date{\today}

\begin{abstract}
Accurate prediction of vacancy migration energy barriers, $\Delta E_{\text{a}}$, in multi-component alloys is extremely challenging yet critical for development of diffusional transformation kinetics needed to model alloy behavior in many technological applications. In this paper, results from $\Delta E_{\text{a}}$ and the energy driving force $\Delta E$ of many ($>$1000) vacancy migration events calculated using density functional theory and nudged elastic band method show large changes ($\sim$ 1 eV) of $\Delta E_{\text{a}}$ in different local chemical environments of the model face-centered cubic (FCC) Al-Mg-Zn alloys. Due to local lattice distortion effects induced by solute atoms (such as Mg) with different sizes than the matrix element (Al), the changes of $\Delta E_{\text{a}}$ for one type of migrating atoms originate primarily from fluctuations of $\Delta e_{\text{a}}\equiv \Delta E_{\text{a}} - \frac{1}{2}\Delta E$ (instead of $\frac{1}{2}\Delta E$ according to the widely used Kinetic Ising model). To understand these fluctuations, a quartic function of the reaction coordinate is shown to accurately describe the energy landscape of the minimum energy path (MEP) for each vacancy migration event studied in this paper. Analyses of the quartic function show that $\Delta e_{\text{a}}$ can be approximated with $\Delta e_{\text{a}} \approx \alpha k_fD^2$, where $\alpha \sim 0.022$ is a constant value of all types of migrating atoms in Al lattice. Here $D$ is the distance of a migrating atom between two adjacent equilibrium positions and $k_f$ is the average vibration spring constant of this atom at these two equilibrium positions. $k_f$ and $D$ quantitatively describe the lattice distortion effects on the curvatures and locations of the MEP at its initial and final states in different local chemical environments. We also used the local lattice occupations as inputs to train surrogate models to predict the coefficients of the quartic function, which accurately and efficiently output both $\Delta E_{\text{a}}$ and $\Delta E$ as the necessary inputs for the mesoscale studies of diffusional transformation in Al-Mg-Zn alloys.
\end{abstract}

\maketitle


\section{Introduction}
\label{sec:Intro}
Diffusion kinetics in metallic alloys and associated material mechanisms (e.g., aging), which control properties such as strength and ductility, are critically dependent upon vacancy-mediated migration of matrix atoms and substitutional solutes\cite{borgenstam2000dictra, pogatscher2014diffusion}. A migrating species in an alloy encounters complex and varying local chemical environments, especially in multicomponent alloys, which in turn change the energy barrier $\Delta E_{\text{a}}$ of a vacancy migration event between two adjacent lattice sites\cite{van2005first,mantina2009first,Osetsky16,Zhao16,Thomas20}. Accurate descriptions of such local chemical effects on $\Delta E_{\text{a}}$ are necessary to construct the kinetic master equations in mesoscale methods, such as kinetic Monte Carlo (kMC)\cite{clouet2004nucleation,sha2005kinetic,soisson2010atomistic,miyoshi2019temperature}, phase-field crystal (PFC)\cite{Elder02,Fallah16}, and diffusive molecular dynamics (DMD) simulations\cite{Li11DMD}, to study diffusion and precipitation. However, potentially large variations of local chemical environments present significant challenges which have yet to be overcome. 

\begin{figure}[ht]
    \centering
    \includegraphics[width=0.48\textwidth]{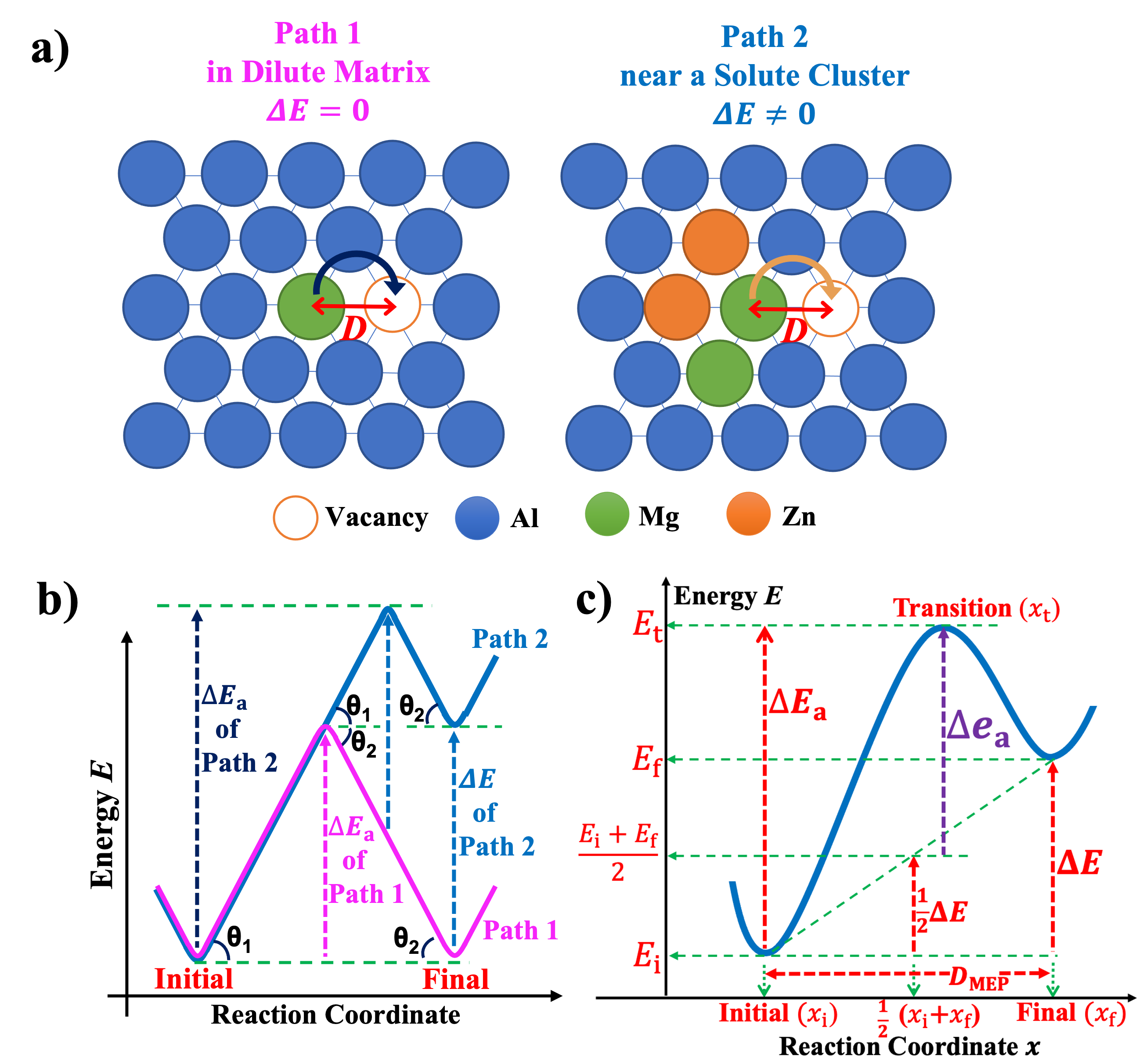}
    \caption{Models of energy barrier $\Delta E_{\text{a}}$ and driving force $\Delta E$ of vacancy migrations in Al alloys. (a): Two vacancy (open circle) migration events between adjacent lattice sites in Al alloys. Colored solid circles indicate the chemical elements on occupied sites. (b): Simplified energy landscape plots of the minimum energy paths (MEPs) for two events in (a) based on the two assumptions described in the text. (c): An energy landscape plot of the MEP for Path 2 in (b) without the two assumptions.}
    \label{fig:Ea_MEP}
\end{figure}

A typical strategy to predict $\Delta E_{\text{a}}$ in different local chemical environments is the Kinetic Ising Model detailed by two vacancy migration events with the same migrating atom in \cref{fig:Ea_MEP} (a) and (b)\cite{soisson2010atomistic} (Path 1 and 2). Path 1 occurs in a dilute local environment with zero energetic driving force $\Delta E$ and its $\Delta E_{\text{a}}$ is easily obtainable using first-principles calculations\cite{Messina14,wu2016high}. Path 2 represents a general vacancy migration case with non-zero $\Delta E$. \cref{fig:Ea_MEP} (b) describes the energy landscape of the minimum energy path (MEP) for each of these two events based on two assumptions. First, the MEP curves are approximated as linear functions of the reaction coordinate with almost the same slope ($\theta_1=\theta_2$) away from both the initial and final states; second, the only changes from Path 1 to 2 are that the MEP curves near the final state shift rigidly along the energy coordinate (Y-axis) by $\Delta E$ of Path 2. Therefore, it is easy to demonstrate that $\Delta E_{\text{a}}$ of Path 2 is equal to one-half of its $\Delta E$ plus $\Delta E_{\text{a}}$ of Path 1. In practice, $\Delta E$ of Path 2 can be predicted by the bond counting model\cite{rautiainen1999influence} or cluster expansion (CE) methods\cite{sanchez1984generalized,zhang2016cluster}, which use the local lattice occupations as inputs with parameters fitted based on first-principles calculations. This strategy to predict $\Delta E_{\text{a}}$ as a linear function of $\Delta E$ for a general vacancy migration event was used to model many metallic alloys\cite{rautiainen1999influence,sha2005kinetic,vincent2008precipitation,soisson2010atomistic,pareige2011kinetic,miyoshi2019temperature}. 

The above two assumptions on MEPs can be incorrect. \cref{fig:Ea_MEP} (c) illustrates the detailed MEP plot of Path 2 without the two assumptions: the distance between the initial and final states along the reaction coordinate, defined as $D_{\text{MEP}}$ in \cref{fig:Ea_MEP} (c), can vary due to the lattice distortion induced by changes of local chemical compositions; in addition, the MEP curves near the initial states can have different shapes (such as local curvatures) compared with those at the final states. These variations change the position of the transition state along the reaction coordinate and its energy. Thus, a robust model of $\Delta E_{\text{a}}$ should provide accurate descriptions of MEPs and the corresponding transition states\cite{soisson2000monte,clouet2004nucleation,soisson2007cu,daniels2020hybrid}. One strategy is to investigate $\Delta e_{\text{a}}$ defined as the transition-state energy ($E_{\text{t}}$) relative to the average of the initial-state ($E_{\text{i}}$) and final-state ($E_{\text{f}}$) energies: 
\begin{equation}
    \label{eq:BEP}
    \Delta e_{\text{a}} \equiv E_{\text{t}} - \frac{1}{2}(E_{\text{i}}+E_{\text{f}})= \Delta E_{\text{a}} - \frac{1}{2} \Delta E
\end{equation}
$\Delta e_{\text{a}}$ is a variable and a function of local lattice occupations. This function of $\Delta e_{\text{a}}$ can be fit using a local cluster expansion method\cite{van2001first,zhang2016cluster,goiri2019role}. $\Delta E_{\text{a}}$ is then obtained by the summation of $\Delta e_{\text{a}}$ and $\frac{1}{2}\Delta E$. Note that the Kinetic Ising Model is recovered if $\Delta e_{\text{a}}$ is a fixed value as $\Delta E_{\text{a}}$ in Path 1 of \cref{fig:Ea_MEP} (a). This method requires sufficient samples of transition states to construct the training data set for fitting $\Delta e_{\text{a}}$. The quantitative understanding of the mechanisms that determine $\Delta e_{\text{a}}$ and $\Delta E_{\text{a}}$ can benefit the selections of the representative vacancy migration cases for fitting and verifying the functions of $\Delta e_{\text{a}}$ and $\Delta E_{\text{a}}$ in different local chemical environments, which are critical for the investigations of diffusion kinetics in multiple precipitation stages of advanced alloys.

To clarify the mechanisms that determine $\Delta E_{\text{a}}$ and $\Delta e_{\text{a}}$, we applied high-throughput first-principles calculations to study vacancy migrations in model Al-Mg-Zn systems. As the 7XXX series of aerospace grade Al alloys, they achieve high strengths ($\sim$700 MPa) after appropriate heat treatments\cite{Liddicoat10}, but applications outside of aerospace are  limited since  solute clustering during natural aging limits formability\cite{sha2004early,liu2015effect,Huo16AAForm,Chatterjee22}. This issue exists in several types of Al alloys, and it can be mitigated if vacancy-mediated diffusion can be understood and manipulated \cite{WOLVERTON20075867,ZUROB2009141,pogatscher2014diffusion,werinos2016design} since this controls solute clustering.

We performed density functional theory (DFT) calculations for many ($>$ 1000) model Al-Mg-Zn alloys. MEPs and $\Delta E_{\text{a}}$ of vacancy migrations were computed via DFT plus the climbing image nudged elastic band (CI-NEB) method\cite{henkelman2000climbing,henkelman2000improved}. Details of DFT+CI-NEB methods are described in \cref{sec:DFT_NEB}. Our results in \cref{sec:dE_Ea} show that large fluctuations ($\sim1$ eV) of $\Delta E_{\text{a}}$ under different local chemical environments originate primarily from changes in $\Delta e_{\text{a}}$ rather than the commonly assumed variations of $\Delta E$, which are typically small (mostly $\pm\sim0.2$ eV). A quartic function of the reaction coordinate ($x$ in \cref{fig:Ea_MEP} (c)) is proposed in \cref{sec:Quartic} to accurately describe and analyze the MEPs of all investigated vacancy migration events. Analyses in \cref{sec:Distortion} reveal a new result which is that $\Delta e_{\text{a}}$ is linearly correlated to $k_f D^2$: $D$ is the Cartesian distance of a migrating atom between two adjacent equilibrium positions illustrated by the double-headed arrows in \cref{fig:Ea_MEP} (a), and $k_f$ is the average vibration spring constant of this atom at these two equilibrium positions. $D$ and $k_f$ are parameters that quantify the local lattice distortion effects, on, respectively, the locations and shapes of the MEP at local energy minimum states. Specifically, $D$ is correlated with $D_{\text{MEP}}$ in \cref{fig:Ea_MEP} (c) and $k_f$ is related to the second derivatives of the MEP curves at the local energy minimum states in \cref{fig:Ea_MEP} (c). Both $D$ and $k_f$ can be calculated relatively easily without accurate descriptions of MEPs obtained from the DFT + CI-NEB method. Details of the calculation methods for $D$ and $k_f$ are described in \cref{sec:D_kf}.

In \cref{sec:Surrogate}, based on our DFT+CI-NEB calculations, surrogate models using local lattice occupations as inputs are proposed to predict the coefficients of the quartic function of the vacancy migration MEP in Al-Mg-Zn alloys. This leads to a new approach to accurately and efficiently predict the MEPs and the corresponding $\Delta E_{\text{a}}$ and $\Delta E$ as functions of local chemical compositions. With this new method to estimate $\Delta E_{\text{a}}$ and $\Delta E$, more accurate mesoscale studies, such as kMC, can be conducted. Finally, discussion of the major developments in the paper and conclusions are provided in \cref{sec:Con}.

\section{Methods}
\label{sec:Methods}

\subsection{Transition-State Calculations}
\label{sec:DFT_NEB}
\begin{figure}[ht]
    \centering
    \includegraphics[width=0.48\textwidth]{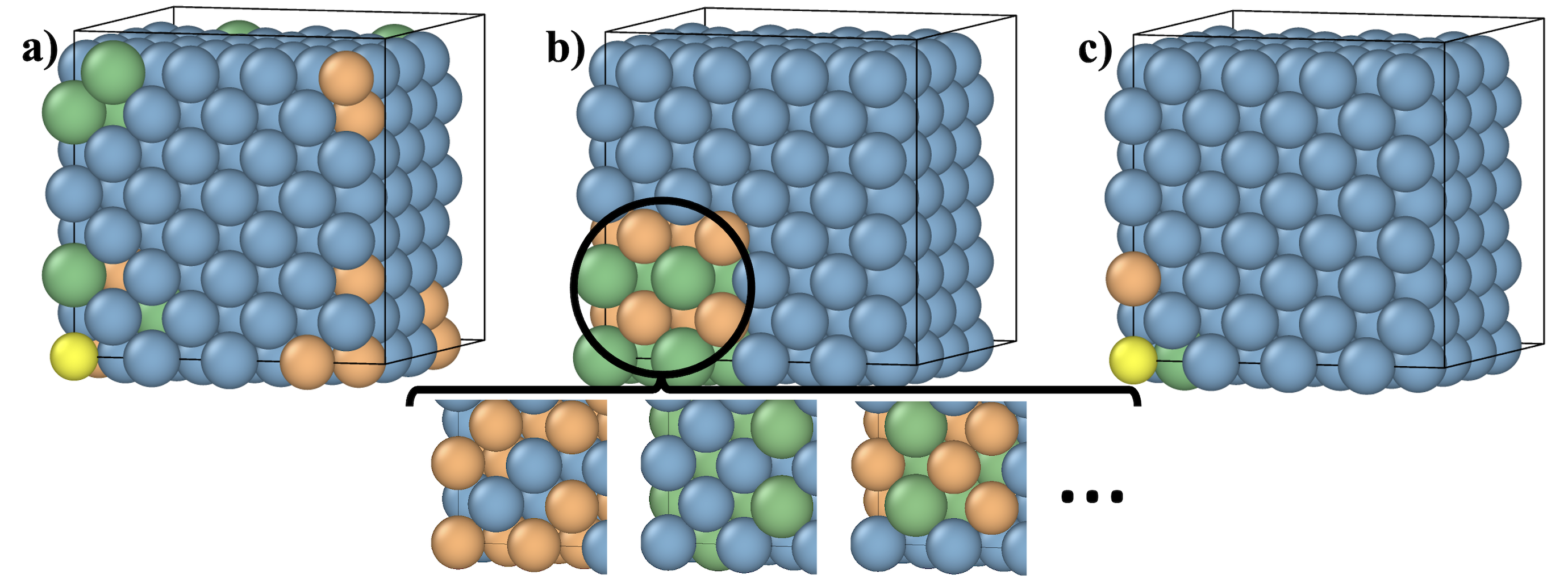}
    \caption{Schematic diagrams of the model Al alloys supercells used to calculate vacancy migration barriers. Blue, orange, and green spheres represent Al, Zn, and Mg atoms, respectively. Yellow spheres represent vacancy sites. (a): A typical $4\times4\times4$ FCC supercell with random distributions of solute (Mg and Zn) atoms. (b): A typical $4\times4\times4$ FCC supercell with a $2\times2\times2$ ordered cluster structure embedded on Al matrix. (c): A typical $4\times4\times4$ FCC supercell with a single solute atom (Zn) embedded on a neighboring site around the vacancy and the migrating atom (Mg).}
    \label{fig:supercells}
\end{figure}

To compute migration energy barriers and the minimum energy paths (MEP) of vacancy migrations, we performed high-throughput density functional theory (DFT) calculations for model Al-Mg binary alloys, Al-Zn binary alloys, and Al-Mg-Zn ternary alloys. The energies of the atomic configurations at the initial and the final states ($E_\text{i}$ and $E_\text{f}$) were first calculated with the Vienna Ab-Initio Simulation Package (VASP) \cite{kresse1996efficiency,kresse1996efficient}, with all-electron projector-augmented wave potentials (PAW) method with the Perdew-Burke-Ernzerhof (PBE) exchange-correlation functional\cite{BLOCHL94PAW,Perdew96PBE}. All calculations used $4 \times 4 \times 4$ supercells, constructed from the FCC Al unit cell, with 255 atoms and 1 vacancy. All supercells used to calculate the vacancy migration barriers can be divided into three categories. The configurations in the first category, as shown in \cref{fig:supercells} (a), are randomly generated solid solution structures with different local concentrations of solute atoms (Mg and Zn) around the vacancy site and the migrating atom (Al, Mg, or Zn). These structures simulate vacancy diffusion in the solid-solution state. For the configurations in the second category, as shown in \cref{fig:supercells} (b), either $2\times2\times2$ or $2\times 2 \times 4$ ordered cluster structures are embedded in the $4\times4\times4$ pure Al matrix. The data from these configurations are designed to describe the vacancy moving inside the precipitates or along the boundary between the ordered precipitates and the solid-solution Al matrix. These ordered structures were chosen from proposed Guinier–Preston (GP) zone precipitates \cite{berg2001gp} and ordered (L1$_\text{0}$, L1$_\text{2}$, L1$_\text{0}^*$, W2, CH, and Z1) intermetallic structures on an FCC lattice\cite{zhuravlev2017phase}. The third category, as shown in \cref{fig:supercells} (c), consists of supercells with a single solute atom (Mg or Zn) embedded in the lattice of neighboring sites (including $1^{\text{st}}$, $2^{\text{nd}}$, and $3^{\text{rd}}$ nearest neighbors) of the vacancy and the migrating atom in the $4\times4\times4$ pure Al matrix. These configurations address the effect of a single solute atom on the vacancy migration barrier. 

For all DFT calculations, the total energies for supercells of the initial and final states were converged to $10^{-6}$ eV/cell for the ionic relaxation loop and $10^{-7}$ eV for the electronic self-consistency loop using a plane-wave cutoff energy of 450.0 eV  and Methfessel–Paxton smearing of 0.4 eV. A $2 \times 2 \times 2$ k-point grid was applied for all supercells. Each grid was generated using the Monkhorst-Pack scheme \cite{monkhorst1976special}. The K-Points convergence tests are summarized in Supplementary Note 1. The supercell sizes in all vacancy migration investigations were always fixed as four times that of the FCC lattice constant of the model 7XXX series Al alloy. Hence. a supercell of $4 \times 4 \times 4$ conventional FCC cells with 256 atoms (based upon the FCC unit cell) was used to calculate the lattice constant. The supercell consisted of 244 Al atoms, 7 Mg atoms, and 5 Zn atoms, which were within the range of compositions of 7075 Al alloys. Lattice occupations inside this supercell were optimized by the special quasi-random structures (SQS) method using the Alloy Theoretic Automated Toolkit (ATAT)\cite{zunger1990special}. The lattice constant of this SQS-optimized supercell was 4.046 {\r{A}} after DFT relaxation of nuclear coordinates and the cell volume. This value is close to the lattice constant of a pure Al crystal at 0K (4.041 {\r{A}} from DFT calculations with the same setups described in this section). The effects of lattice constant variations on vacancy migration barriers are described in Supplementary Note 2. 

For each vacancy migration event, the energy of the transition state ($E_\text{t}$) and the energy barrier ($\Delta E_{\text{a}} \equiv E_\text{t} - E_\text{i}$) were gathered by utilizing the climbing image nudged elastic band (CI-NEB) method after evaluating the energy difference ($\Delta E$) by using $E_\text{f}$ minus $E_\text{i}$. This was accomplished with VASP and the Transition States Tools (VTST) package \cite{henkelman2000climbing,henkelman2000improved}. Five images between the relaxed initial and final images were set. The artificial spring constant was set to $5$ eV/{\r{A}}$^\text{2}$. The electronic self consistency-loop breaking criteria was set to $10^{-4}$ eV and the force convergence criteria for all models was set to be less than $0.05$ eV/{\r{A}}. The force-based quick-min optimizer provided by VTST was used for the CI-NEB calculations\cite{sheppard2008optimization}. Verifications of the transition state via phonon calculations is described in Supplementary Note 3. Finally, 2500 $\Delta E$ and $\Delta E_{\text{a}}$ pairs were obtained from 1250 CI-NEB calculations by considering both forward and backward vacancy migrations.

\subsection{Calculations of Migration Distances and Vibration Spring Constants}
\label{sec:D_kf}
As mentioned in \cref{sec:Intro} and discussed in \cref{sec:Distortion}, the migration distances $D$ and vibration spring constants $k_f$ of the migrating atoms are critical to describe the local lattice distortion effects on the vacancy migration MEPs and their $\Delta E_{\text{a}}$. In this subsection, the detailed methods to calculate them and the related parameters are presented. First, the relative distance between the initial and final states along the MEP, $D_{\text{MEP}}$, as indicated in \cref{fig:Ea_MEP} (c), can be obtained from the outputs of the CI-NEB calculations. Here we set $N$ as the number of intermediate images inserted between the initial and final states, and $I_j$ represents the configuration of the $j^{\text{th}}$ intermediate image in the CI-NEB calculations. Specifically, $I_0=I_{\text{i}}$ and $I_{N+1}=I_{\text{f}}$ denote the initial and final configurations, respectively. $D_{\text{RHD}}(I_{a}, I_{b})$ is a function that returns the magnitude of the relative high-dimensional distance between $I_{a}$ and $I_{b}$\cite{henkelman2000climbing,henkelman2000improved}:
\begin{equation}
\label{eq:D_RHD}
    D_{\text{RHD}}(I_{a}, I_{b}) = \sqrt{\sum\limits_{k=1}^{N_{\text{atom}}} \left(\left(\pmb x_{b, k} - \pmb x_{a, k}\right)^\text{T} \left(\pmb x_{b, k} - \pmb x_{a, k}\right) \right)}
\end{equation}
Here, $\pmb x_{j, k}$ is a three-dimensional vector representing the Cartesian positions of the $k^{\text{th}}$ atom in the $j^{\text{th}}$ image, and $N_{\text{atom}}$ is the total number of atoms in each configuration. Since only 5 intermediate images were chosen between the relaxed initial and final images for all CI-NEB calculations in this study, $D_{\text{MEP}}$ reduces to:
\begin{equation}
\label{eq:D_MEP}
    D_{\text{MEP}} = \sum\limits_{i=0}^{N=5}D_{\text{RHD}}(I_{j}, I_{j+1})
\end{equation}
Alternatively, the migration distance $D$ of a migrating atom between two adjacent equilibrium positions (its Cartesian positions in initial and final states) can be directly calculated as:
\begin{equation}
\label{eq:D}
    D = \sqrt{ \left(\pmb x_{\text{f}} - \pmb x_{\text{i}}\right)^\text{T} \left(\pmb x_{\text{f}} - \pmb x_{\text{i}}\right)}
\end{equation}
Here $\pmb x_{\text{i}}$ and $\pmb x_{\text{f}}$ denote the Cartesian position of the migrating atom in the equilibrium initial and final states, respectively. Because most atoms are almost stationary during the vacancy migration process, there are strong correlations between $D_{\text{MEP}}$ and $D$, so the value of $D$ is utilized to quantify the lattice distortion effects on the MEP and the corresponding $\Delta e_{\text{a}}$/$\Delta E_{\text{a}}$ for each vacancy migration case in \cref{sec:Distortion}. Additional discussion of the correlations between $D_{\text{MEP}}$ and $D$ is in Supplementary Note 4.

The vibration spring constants $k_f$ of migrating atoms are calculated based on the Hessian matrix $\pmb H$, which is the matrix of the second derivatives of the energy with respect to the atomic positions, obtained using the finite difference method implemented in VASP. $\pmb H$ should be a $3N_{\text{atom}}$ dimensional matrix if all $N_{\text{atom}}$ atoms can be displaced in the supercell. In principle, $k_f$ at the initial and final states can be acquired by finding the eigenvalues of $\pmb H$, of which the corresponding eigenvectors describe the motions of atoms along the MEP of the vacancy migration. Using the Harmonic approximation, the energy landscape, $V$, of the MEP at the initial and final states can be expressed as $V =\frac{1}{2}k_f x^2$. Here $x$ is the displacement along the MEP. 

However, it is expensive to calculate $\pmb H$ for all the investigated cases in this study if all 255 atoms in a supercell are displaced. Since most atoms are nearly stationary during the vacancy migration process, we can approximate the value of $k_f$ by fixing the positions of atoms far away from the vacancy during the calculation of $\pmb H$. In this study, only the migrating atom is displaced during the calculation of $\pmb H$ for the initial and final states, but all other atoms are fixed. The calculated vibration spring constant values under this fixed-atom condition were obtained for both the initial and final states, and the average value was used in \cref{sec:Distortion} to estimate the lattice distortion effect on the MEP and the corresponding $\Delta e_{\text{a}}$ and $\Delta E_{\text{a}}$ for each vacancy migration case. More accurate $k_f$ can be obtained if more atoms in the supercells were displaced during the calculation of $\pmb H$. Detailed discussions of $k_f$ calculation is summarized in Supplementary Note 5.

\section{Results}
\subsection{DFT Calculations of $\Delta E_{\text{a}}$ and $\Delta E$}
\label{sec:dE_Ea}

\begin{figure}[ht]
    \centering
    \includegraphics[width=0.48\textwidth]{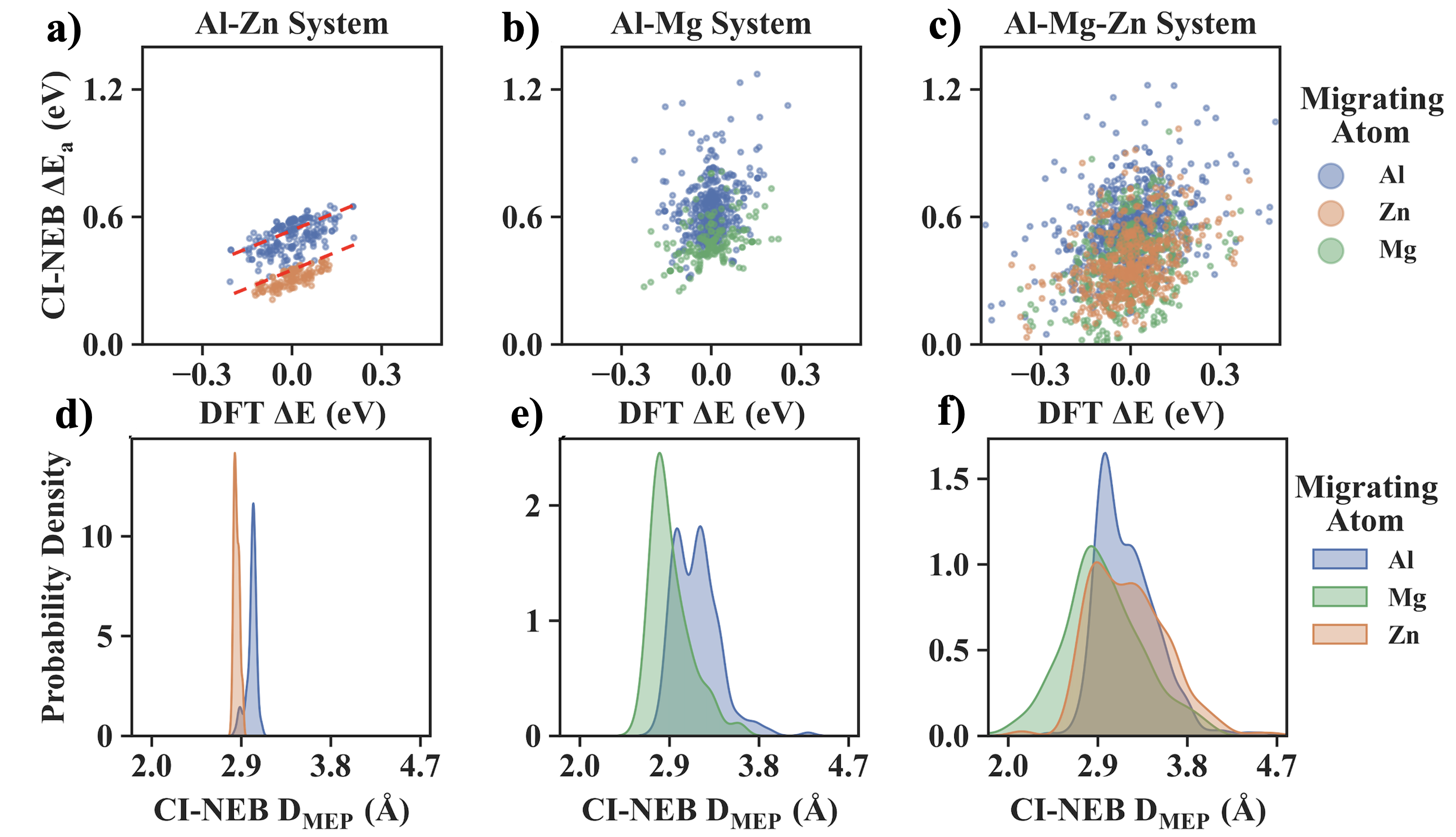}
    \caption{Correlations between $\Delta E_{\text{a}}$ and $\Delta E$ for vacancy migration events in Al alloys. (a)-(c): Correlations between $\Delta E_{\text{a}}$ and $\Delta E$ for vacancy migration events from DFT + CI-NEB calculations. Migrating atoms are Al (blue dots), Zn (orange dots), and Mg (green dots) in binary Al-Zn (d), Al-Mg (e), and ternary Al-Mg-Zn (f) systems, respectively. The scattering distributions in (e) and (f) indicate the variations of $\Delta E_{\text{a}}$ do not only depend on the variations of $\Delta E$ as suggested in \cref{fig:Ea_MEP} (b). (g)-(i): Kernel density estimations \cite{parzen1962estimation} of $D_{\text{MEP}}$, the distance between the initial and final states defined in \cref{fig:Ea_MEP} (c), are plotted for migrating Al (blue), Zn (orange), and Mg (green) atoms in Al-Zn (g), Al-Mg (h), and Al-Mg-Zn (i) systems, respectively. Large variations of $D_{\text{MEP}}$ in (h) and (i) suggest strong lattice distortion effects in different local chemical environments. The color coding applies to subsequent figures.}
    \label{fig:database}
\end{figure}

Correlations of $\Delta E_a$ (Y-axis) and $\Delta E$ (X-axis) from our computational results are plotted in \cref{fig:database} (a) to (c) for different types of migrating atoms. In the Al-Zn binary system (\cref{fig:database} (a)), $\Delta E_a$ and $\Delta E$ data are scattered; however, the data still follows (approximately) linear relationships for both the migrating Al and Zn atoms, respectively. Simple linear regressions suggest that the slope of each fitted straight line is close to $\frac{1}{2}$, so $\Delta e_{\text{a}}$ is approximately a constant according to \cref{eq:BEP} and the Kinetic Ising Model is still approximately valid for vacancy migrations in the Al-Zn binary system. However, as seen in \cref{fig:database} (b) and (c), the $\Delta E_a$ and $\Delta E$ data become significantly scattered, and a linear relationship does not apply when Mg is added as a solute element for all types of migrating atoms (Al, Zn and Mg) in both binary Al-Mg and ternary Ag-Mg-Zn systems. In these cases, $\Delta E$ values are still distributed in a similar range as those in the Al-Zn system in \cref{fig:database} (a), mostly from $\sim -0.2$ eV to $\sim 0.2$ eV. However, $E_a$ values are scattered in much wider ranges from almost 0 eV up to $\sim 1$ eV. This indicates that the fluctuations of $\Delta E_{\text{a}}$ in different local chemical environments are mostly dependent upon changes in $\Delta e_{\text{a}}$ rather than the small variations of $\Delta E$ according to \cref{eq:BEP}. These deviations demonstrate that even simple vacancy migrations in alloys with close-packed lattices are complex; hence the assumptions behind \cref{fig:Ea_MEP} (b) are incorrect and the Kinetic Ising Model is not broadly applicable. 

One problem in \cref{fig:Ea_MEP} (b) is the neglect of changes in MEP along the reaction coordinate axis. Here, we define the distance between the initial and final states along the reaction coordinate as $D_{\text{MEP}}$ as indicated in \cref{fig:Ea_MEP} (c). The reaction coordinate $x$ and $D_{\text{MEP}}$ for all investigated cases result from DFT + CI-NEB calculations\cite{henkelman2000climbing,henkelman2000improved}. \cref{fig:database} (d), (e), and (f) show the kernel density estimations \cite{parzen1962estimation}, which indicate smoothed probability distributions, of $D_{\text{MEP}}$ for different types of migrating atoms (Al, Mg and Zn) in all migration events. In the Al-Zn binary system (\cref{fig:database} (d)), $D_{\text{MEP}}$ values are centered-distributed with negligible standard deviation $\sigma_{D_{\text{MEP}}}$ for both migrating Al atoms ($\sigma_{D_{\text{MEP}}}=$0.035 \r{A}) and migrating Zn atoms ($\sigma_{D_{\text{MEP}}}=$0.018 \r{A}). These distributions indicate occupations of Zn atoms near vacancy sites induce small lattice distortions. However, in both Al-Mg systems (\cref{fig:database} (e)) and Al-Mg-Zn systems (\cref{fig:database} (f)), $\sigma_{D_{\text{MEP}}}$ is much larger for all migrating Al, Mg, and Zn atoms ($\sigma_{D_{\text{MEP}}}=\sim$ 0.2 \r{A} in all cases). These distributions indicate occupations of Mg atoms near vacancy sites induce relatively large lattice distortions. These lattice distortions are understandable because of the atomic size differences and large fluctuations of local Mg/Zn concentrations for all investigated supercells. The size of Mg atoms is much larger than those of Zn and Al atoms (the radii of Mg and Zn and Al atoms are 1.50, 1.35, and 1.25 $\text{\AA}$\cite{Slater64}, respectively), so the lattice distortion effects due to Mg atoms in the Al matrix are much stronger than those due to Zn atoms. The large fluctuations of local Mg/Zn concentrations originate from the multiple types of supercells used in our calculations as shown in \cref{fig:supercells}, which correspond to different precipitation stages of Al alloys. A key question is how to construct accurate MEP models illustrated in \cref{fig:Ea_MEP} (c) to accommodate the lattice distortion effects if we want to understand the physical mechanisms behind $\Delta e_{\text{a}}$ and $\Delta E_a$ variations.

\subsection{Quartic Functions of the MEP}
\label{sec:Quartic}

\begin{figure}[ht]
    \centering
    \includegraphics[width=0.48\textwidth]{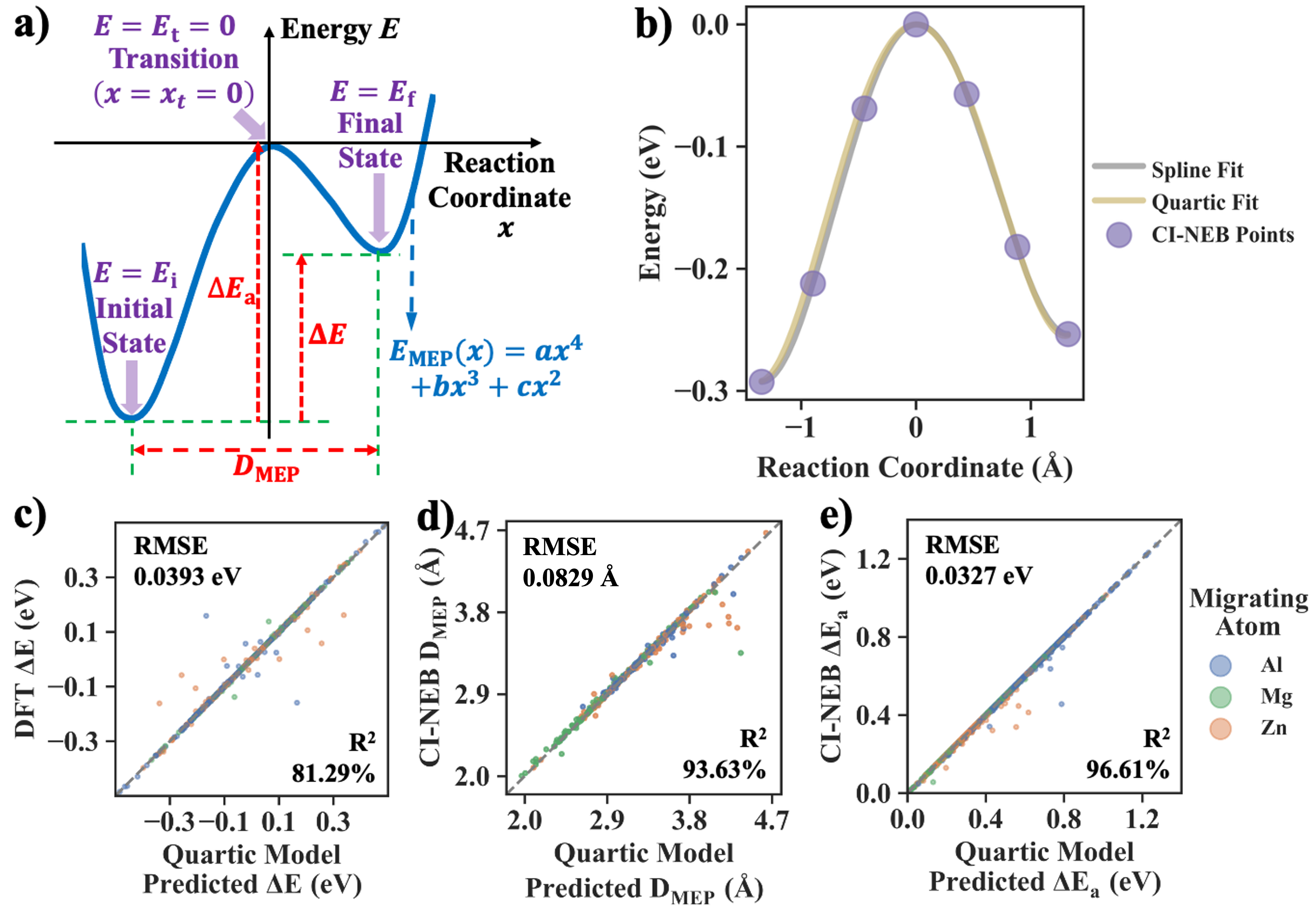}
    \caption{The quartic function in \cref{eq:quartic} is used to fit vacancy migration MEPs from DFT + CI-NEB calculations. (a) Schematic plot of $E_{\text{MEP}}(x)$ of \cref{eq:quartic} showing $\Delta E$, $D_{\text{MEP}}$, and $\Delta E_{\text{a}}$. (b) A specific example where \cref{eq:quartic} is used to fit an MEP curve from a DFT + CI-NEB calculation. (c)-(e): Comparisons of $\Delta E$ (c), $D_{\text{MEP}}$ (d), and $\Delta E_{\text{a}}$ (e) from DFT + CI-NEB calculations and those predicted from the fitted \cref{eq:quartic} for all Al (blue), Zn (orange), and Mg (green) migrating atoms in all investigated supercells. The root-mean-square error (RMSE) is denoted at the upper left, and the number at the bottom-right corner shows the coefficient of determination R$^{2}$ (close to 100$\%$ means high accuracy). Small RMSE and large R$^2$ values in (c)-(e) demonstrate that \cref{eq:quartic} is accurate to describe vacancy migration MEPs. The same RMSE and R$^2$ symbols are used in \cref{fig:correlations} and \cref{fig:predictions}.}
    \label{fig:quartic}
\end{figure}

An accurate and quantitative model to describe the MEP in \cref{fig:Ea_MEP} (c) has to satisfy several physical conditions, including zero first derivative at initial ($x_{\text{i}}$), transition ($x_{\text{t}}$) and final ($x_{\text{f}}$) states. Thus, we propose that the energy landscape of a general vacancy migration MEP, as a function of the reaction coordinate $x$ with a single local energy maximum, is described by a simple quartic function, $E_{\text{MEP}}(x)$:
\begin{equation}
    \label{eq:quartic}
    E_{\text{MEP}}(x) = ax^4+bx^3+cx^2
\end{equation}
Here, the coefficients ($a$, $b$, and $c$) are assumed to depend on the local lattice occupations near a vacancy/adjacent migrating atom pair. Values of $x_{\text{i}}$, $x_{\text{t}}$, and $x_{\text{f}}$ are determined by the zero-first-derivative requirements mentioned above. The first derivative of \cref{eq:quartic} is $E_{\text{MEP}}^{\prime}(x) = 4ax^3+3bx^2+2cx$, which roots $x_0=0$, $x_1=\frac{-3b-\sqrt{9b^2-32ac}}{8a}$, and $x_2=\frac{-3b+\sqrt{9b^2-32ac}}{8a}$. When $a>0$ and $c<0$, \cref{eq:quartic} has two local minima and one local maximum, which corresponds to the shape of the energy landscape along the MEP in \cref{fig:Ea_MEP} (c). As plotted in \cref{fig:quartic} (a), we can shift the energy landscape to make the transition state at the origin point by denoting the position of the transition state at $x_{\text{t}} = 0$ and denoting its energy on the MEP $E_{\text{MEP}}(x_{\text{t}}=0)=0$. Then we make the positions of the initial state and final state at two local minima as $x_{\text{i}} = x_1$ and  $x_{\text{f}} = x_2$. If \cref{eq:quartic} is accurate enough to describe the MEP for each migration event and its coefficients ($a$, $b$, and $c$) can be predicated, $\Delta E_{\text{a}}$, $\Delta E$, and $D_{\text{MEP}}$ of the corresponding migration event can be predicted as suggested in \cref{fig:quartic} (a) (then its $\Delta e_{\text{a}}$ is from \cref{eq:BEP}).

Thus, by assuming \cref{eq:quartic} is accurate to describe all the MEPs from our DFT+CI-NEB calculations, we applied a least-squares fitting method with a weight matrix to fit the coefficients $a$, $b$, and $c$ for each migration event. Since CI-NEB methods use a series of images along the reaction path to calculate MEPs, we can not only collect the energetics of the initial, final, and transition states, but also those of other intermediate images, which are at certain coordinates along  MEPs. The following conditions are included in the quartic equation fitting: the energies of initial and final states predicted by the quartic function equal those from DFT calculations, $E_{\text{MEP}}(x_{\text{i}})=E_{\text{i}}$ and $E_{\text{MEP}}(x_{\text{f}})=E_{\text{f}}$; the energies of the other intermediate images equal to those from the DFT+CI-NEB calculations, $E_{\text{MEP}}(x_j)=E_{j}$; the first derivatives at initial and final states zero, $E_{\text{MEP}}^{\prime}(x_{\text{i}})=0$ and $E_{\text{MEP}}^{\prime}(x_{\text{f}})=0$. Thus, the following equation can be obtained:
\begin{equation}
\label{eq:fit}
\begin{pmatrix}
x_{\text{i}}^4 & x_{\text{i}}^3 & x_{\text{i}}^2 \\
\vdots & \vdots & \vdots \\
x_j^4 & x_j^3 & x_j^2 \\
\vdots & \vdots & \vdots \\
x_{\text{f}}^4 & x_{\text{f}}^3 & x_{\text{f}}^2 \\
4x_{\text{i}}^3 & 3x_{\text{i}}^2 & 2x_{\text{i}} \\
4x_{\text{f}}^3 & 3x_{\text{f}}^2 & 2x_{\text{f}} 
\end{pmatrix}
\begin{pmatrix}
a\\ 
b\\
c
\end{pmatrix} = 
\begin{pmatrix}
E_{\text{i}}\\ 
\vdots\\
E_j\\
\vdots\\
E_{\text{f}}\\
0\\0
\end{pmatrix}
\end{equation}
Here $x_j$ is the location of the $j^{\text{th}}$ intermediate image along the reaction coordinate and $E_j$ is its energy relative to the transition state (since $x_{\text{t}} = 0$ and $E_{\text{MEP}}(x_{\text{t}})=0$ according to \cref{eq:quartic}). All values of $x_j$ and $E_j$ are directly from DFT + CI-NEB calculations. We denote the left matrix as $\pmb X$, the quartic coefficients vector as $\pmb \beta$, and the right vector as $\pmb y$ for \cref{eq:fit}, which can be re-written as $\pmb X \pmb \beta = \pmb y$. To find the best description of each MEP, the weighted linear regression is applied, which is a generalization of ordinary least squares: 
\begin{equation}
\label{eq:mep_matrix}
\left(\pmb X^\text{T} \pmb W \pmb X \right) \pmb{\hat{\beta}}  = \pmb X^\text{T} \pmb W \pmb y
\end{equation}
Here, $\pmb W$ is a diagonal matrix, with each of its elements representing a weighting coefficient used for each data point. The estimated quartic coefficients vector is $\pmb{\hat{\beta}} = \left(\pmb X^\text{T} \pmb W \pmb X \right)^{-1} \pmb X^\text{T} \pmb W \pmb y$. To emphasize the accuracy of the computed energy terms $\Delta E$ and $\Delta E_{\text{a}}$, we increase the weight elements of the first condition mentioned above to large finite numbers and retain other weight elements equal to $1$. Each MEP curve of all DFT-CI-NEB calculations was fitted by \cref{eq:mep_matrix}.

\cref{fig:quartic} (b) shows an MEP curve from the DFT + CI-NEB calculation is accurately described by both the standard spline fitting and our quartic fitting curve based on \cref{eq:quartic}. Overall, \cref{fig:quartic} (c), (d), and (e) depict close matches between $\Delta E$, $D_{\text{MEP}}$, and $\Delta E_{\text{a}}$ from direct DFT + CI-NEB calculations (Y-axis) and those from the quartic function $E_{\text{MEP}}(x)$ with fitted coefficients (X-axis), respectively. Low values of the root-mean-square error (RMSE) (close to 0) and high values of the coefficient of determination $R^2$ (close to 100$\%$) confirm that \cref{eq:quartic} is accurate and robust enough to describe the MEP of vacancy migrations in Al-Mg-Zn alloys while incorporating the requisite physics associated with the vacancy migration MEPs.

\begin{figure}[ht]
     \centering
     \includegraphics[width=0.48\textwidth]{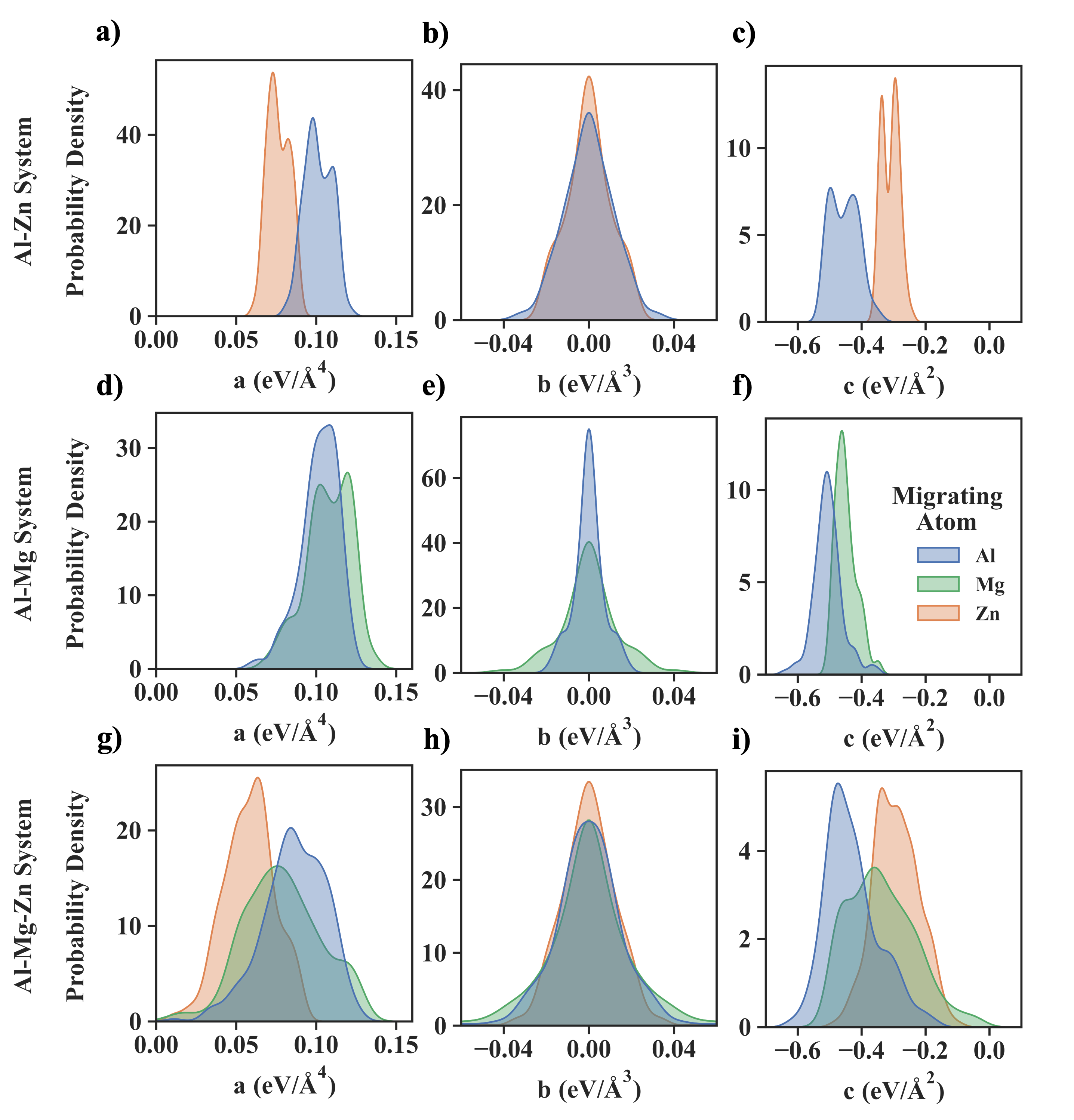}
     \caption{Kernel density estimations of fitted coefficients ($a$, $b$, and $c$) of \cref{eq:quartic} for all Al (blue), Zn (orange), and Mg (green) migrating atoms in all investigated supercells. (a)-(c): Probability densities of $a$, $b$, and $c$ in Al-Zn systems. (d)-(f): Probability densities of $a$, $b$, and $c$ in Al-Mg systems.  (g)-(i): Probability densities of $a$, $b$, and $c$ in Al-Mg-Zn systems. The narrow probability densities variations of $b$ in all investigated supercells is consistent with the small variations of $\Delta E$ in \cref{fig:database} (a)-(c).}
     \label{fig:coefficients}
\end{figure}

\cref{fig:coefficients} shows the kernel density estimations of the fitted coefficients of \cref{eq:quartic} for all vacancy migration cases in different alloy systems (\cref{fig:coefficients} (a)-(c) in Al-Zn binary systems, \cref{fig:coefficients} (d)-(f) in Al-Mg binary systems, and \cref{fig:coefficients} (g)-(i) in Al-Mg-Zn ternary systems). The results show the distributions of $a$ and $c$ vary significantly for different types of migrating atoms in three alloy systems. The wide ranges in $a$ and $c$ indicate the shapes of the MEPs in \cref{fig:Ea_MEP} (c) and \cref{fig:quartic} (a) can change significantly because both $a$ and $c$ determine the coordinates and curvatures of MEPs at local energy minimum states (as discussed in \cref{sec:Distortion}, the ratio of $a$ to $c$ is also important to determine the MEPs and the values of $\Delta E$/$D_{\text{MEP}}$/$\Delta e_{\text{a}}$, so these distributions of $a$ and $c$ can not be used to explain the differences between Al-Zn alloys and Al-Mg/Al-Mg-Zn alloys in \cref{fig:database}). Alternatively, the distributions of $b$ for all types of migrating atoms in all alloy systems are always in narrow ranges close to zero, which is consistent with the small variations of $\Delta E$ in \cref{fig:Ea_MEP} (d) ($\Delta E = 0$ if $b=0$). This special feature of $b$ provides us a relatively easy and accurate way to predict $\Delta e_{\text{a}}$ based on lattice distortion effects as follows.

\subsection{Estimations of $\Delta e_{\text{a}}$ based on Lattice Distortion Effects}
\label{sec:Distortion}

\begin{figure}[ht]
     \centering
     \includegraphics[width=0.48\textwidth]{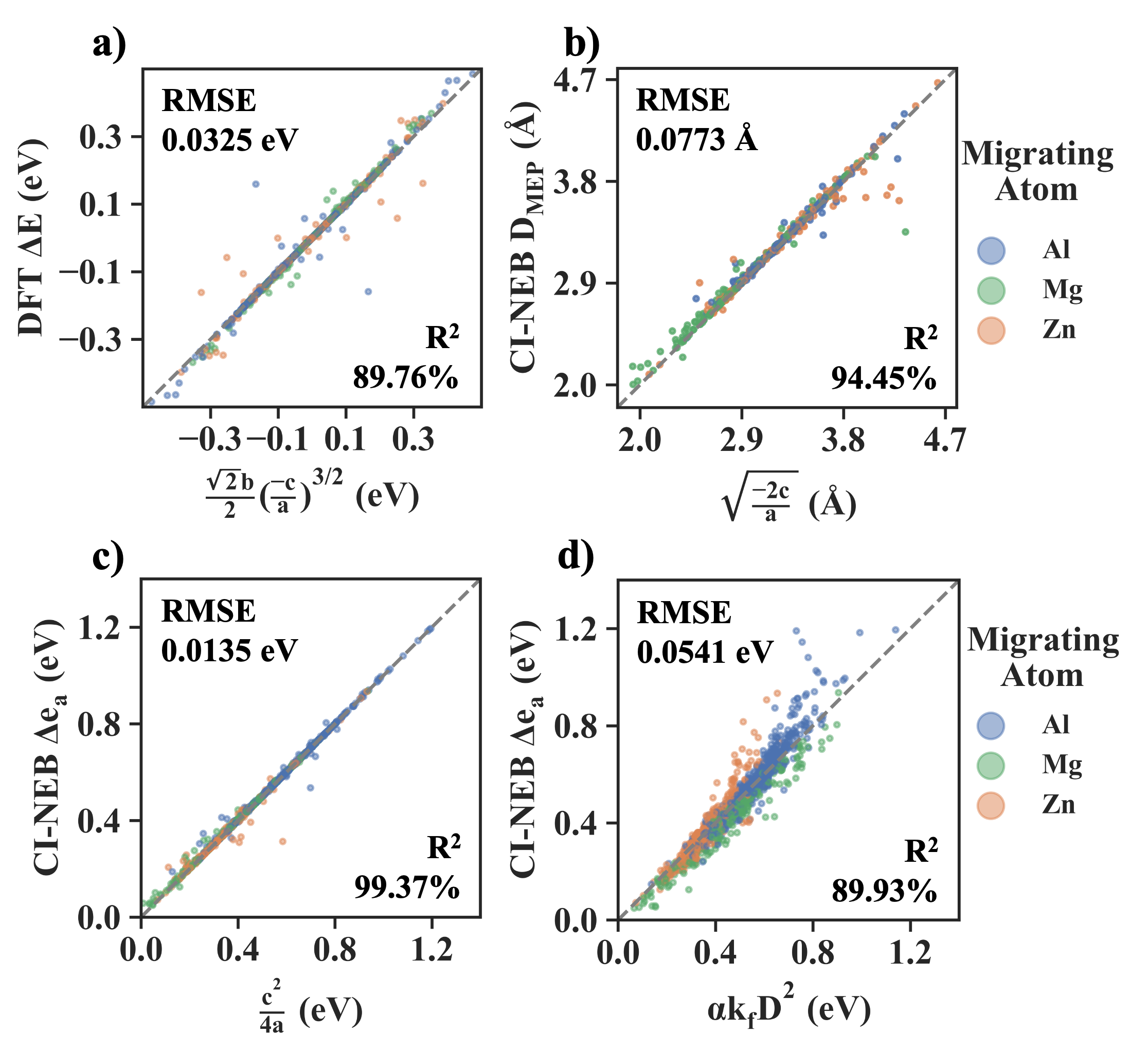}
     \caption{Methods to estimate $\Delta E$ (a), $D_{\text{MEP}}$ (b), and $\Delta e_{\text{a}}$  (c) of vacancy migration MEPs. (a)-(c) Comparisons between DFT+CI-NEB calculated $\Delta E$ (a), $D_{\text{MEP}}$ (b), and $\Delta e_{\text{a}}$ (c) and those estimated based on \cref{eq:quartic} and \cref{eq:approx_quartic}. (d) Correlations between DFT+CI-NEB calculated $\Delta e_{\text{a}}$ and $k_fD^2$ to verify \cref{eq:e0}.}
     \label{fig:correlations}
\end{figure}

As indicated by \cref{fig:coefficients}, we can assume $b\approx 0$ giving
\begin{equation}
    \label{eq:approx_quartic}
    E_{\text{MEP}}(x) \approx ax^4+cx^2
\end{equation}
This is the same free energy formalism of second-order phase transitions in Landau theory\cite{Landau}. Thus, $x_{\text{i}} \approx -\sqrt{\frac{-c}{2a}}$, $x_{\text{t}} = 0$, and $x_{\text{f}} \approx \sqrt{\frac{-c}{2a}}$, respectively. Accordingly, $D_{\text{MEP}} = x_{\text{f}} - x_{\text{i}} \approx \sqrt{\frac{-2c}{a}}$ and $\Delta e_{\text{a}} \approx -ax_{\text{i}}^4-cx_{\text{i}}^2 \approx \frac{c^2}{4a}$. We can also estimate $\Delta E \approx 2bx_{\text{f}}^3 \approx \frac{\sqrt{2}}{2}(\frac{-c}{a})^{\frac{3}{2}}b$ based on \cref{eq:quartic}. These approximate relations are confirmed in \cref{fig:correlations} (a)-(c) by the linear correlations between the DFT + CI-NEB results ($\Delta E$, $D_{\text{MEP}}$, and $\Delta e_{\text{a}}$ on the Y-axis) and their estimations based on \cref{eq:quartic} and \cref{eq:approx_quartic} (the X-axis), respectively.

In addition, the second derivative of the MEP at the local-minimum states $E_{\text{MEP}}''(x_{\text{i}})\approx E_{\text{MEP}}''(x_{\text{f}})\approx -4c $ according to \cref{eq:approx_quartic}. Thus, we can get an approximate relation as $\Delta e_{\text{a}} \approx \frac{E_{\text{MEP}}''(x_{\text{i}})D_{\text{MEP}}^2}{32}$. Accurate $E_{\text{MEP}}''$ and $D_{\text{MEP}}$ are obtained from the MEP curves produced by the DFT + CI-NEB calculations. However, because the migrating atom moves a distance ($>$ $\sim$2 $\text{\AA}$ in the Al lattice) much larger than the other atoms for a general migration event, the motion of the migrating atom is the most important factor for the reaction coordinate $x$. Therefore, we assume $E_{\text{MEP}}''$ is proportional to the average vibration spring constant $k_f$ of the migrating atom at the initial and final states, and we also assume $D_{\text{MEP}}$ is proportional to the distance $D$ of the same atom at these two states as illustrated in \cref{fig:Ea_MEP} (a). These assumptions give:
\begin{equation}
    \label{eq:e0}
    \Delta e_{\text{a}} \approx \alpha k_fD^2
\end{equation}
where $\alpha$ is a unitless constant. Because the migrating atom moves a distance much larger than all other atoms, the variations of $k_f$ and $D$ can be used to approximate the local lattice distortion effects on the shape and locations of local-minimum states along a MEP. These two parameters are obtained from DFT calculations of $\pmb H$ and the coordinate of the migrating atom in fully relaxed structures as described in \cref{sec:D_kf}. The validity of \cref{eq:e0} is confirmed by comparing $\Delta e_{\text{a}}$ from DFT + CI-NEB calculations (Y-axis) and $\alpha k_fD^2$ (X-axis) for all migrating atoms in all investigated supercells in \cref{fig:correlations} (d). This shows that \cref{eq:e0} with the same $\alpha$ value ($\approx 0.022$ fitted by \cref{fig:correlations} (d)) works for all Al, Mg, Zn migrating atoms in these Al alloys. \cref{eq:e0} therefore provides an efficient way to estimate $\Delta e_{\text{a}}$ and $\Delta E_{\text{a}}$ using standard DFT calculations without the CI-NEB method.

\begin{table*}[ht]
    \centering 
    \caption{Results of vacancy migrations in a dilute Al matrix (at most one solute atom in a supercell) are listed as the migration barrier $\Delta E_{\text{a}} = \Delta e_{\text{a}}$, the average vibration spring constant $k_f$ of the migrating atom at the initial and final states calculated under this fixed-atom condition described in \cref{sec:D_kf}, the high dimensional distance along the minimum energy path between the initial and final states $D_{\text{MEP}}$ defined in \cref{eq:D_MEP}, the Cartesian distance of the migrating atom between initial and final states $D$ defined in \cref{eq:D}, the values of $k_f D^2$, and the coefficient $\alpha \equiv \frac{\Delta e_{\text{a}}}{k_fD^2}$ of selected migrating atoms. As a reference, the value of $\alpha$ in \cref{eq:e0} fitted from the whole database of vacancy migrations is $0.0220$ as shown in \cref{fig:correlations} (d).}
    \label{tbl:dea}
    \vspace{5mm} 
    \begin{tabular}{|c|c|c|c|c|c|c|}
\hline
Migrating atom & $\Delta E_{\text{a}} = \Delta e_{\text{a}}$ (eV) & $k_f$ (eV/\r{A}$^\text{2}$) & $D_{\text{MEP}}$ (\r{A}) & $D$ (\r{A}) & $k_f D^2$ (eV) & $\alpha\equiv\frac{\Delta e_{\text{a}}}{k_fD^2}$ \\
\hline
Al & 0.58 & 3.60 & 3.00 & 2.75 & 27.28 & 0.0213 \\
\hline
Mg & 0.47 & 3.39 & 2.78 & 2.58 & 22.62 & 0.0207 \\
\hline
Zn & 0.34 & 2.02 & 2.80 & 2.69 & 14.66 & 0.0233 \\
\hline
    \end{tabular}
\end{table*}

To further verify the generality and accuracy of \cref{eq:e0}, we compute $k_f$ and $D$ of specific examples of vacancy migration with different types of migrating atoms in a dilute Al alloy in \cref{tbl:dea}. In these cases, except for the migrating atom, there is no solute atom in the supercell so that the initial and final states are equivalent since the MEP is symmetric on two sides of the transition state. Thus, the migration energetic driving force $\Delta E = 0$ and $\Delta e_{\text{a}} \equiv \Delta E_{\text{a}} - \frac{1}{2}\Delta E = \Delta E_{\text{a}}$ according to \cref{eq:BEP}.
We also present these results of $\Delta E_{\text{a}} = \Delta e_{\text{a}}$ from DFT+CI-NEB calculations in \cref{tbl:dea}. These results show that, for all types of migrating atoms (Al, Mg, and Zn) in a dilute Al matrix, the ratio of $\Delta e_{\text{a}}$ to $k_f D^2$ is almost a constant value close to the $\alpha$ value (0.022) fitted from all migration cases with different values of $\Delta E$, thereby further supporting the generality and validity of \cref{eq:e0}.

\subsection{Surrogate Models to Predict the MEP}
\label{sec:Surrogate}

\begin{figure}[ht]
     \centering
     \includegraphics[width=0.25\textwidth]{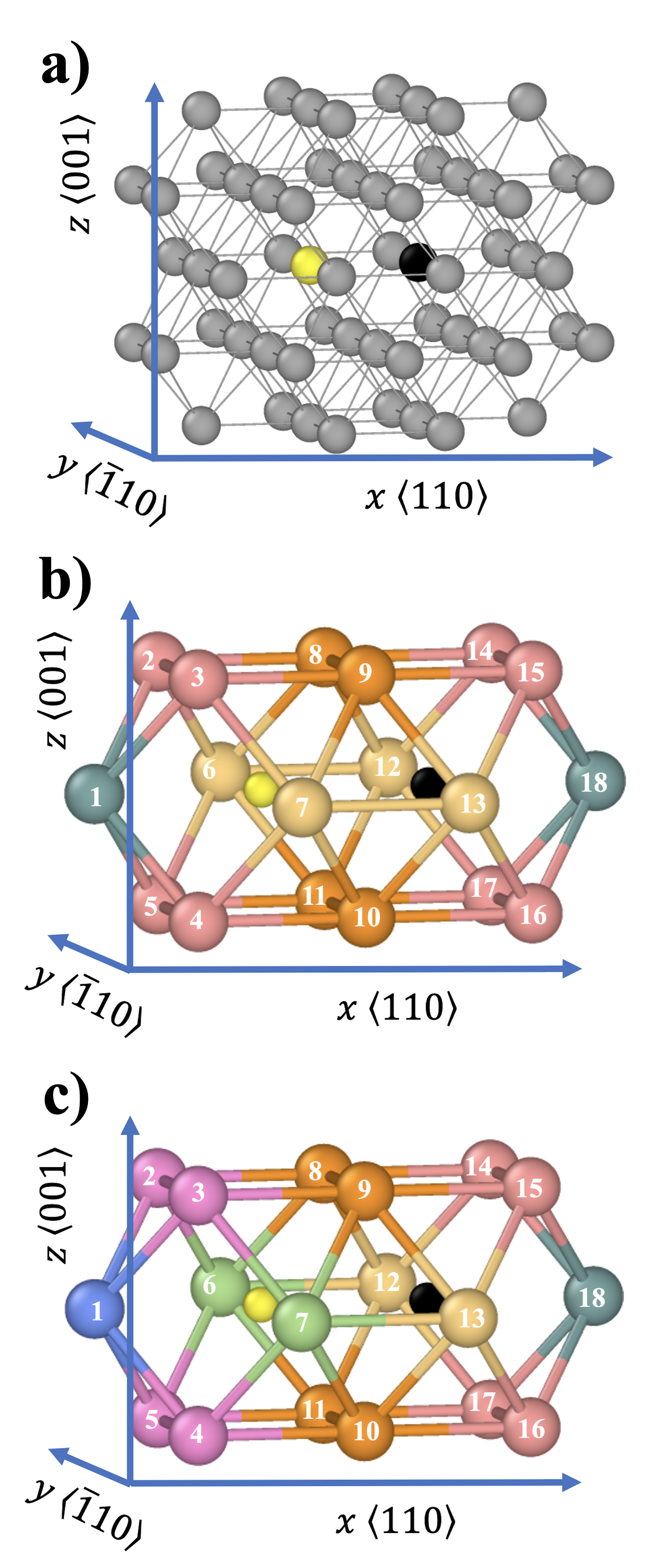}
     \caption{Illustrations of local lattice sites related to the vacancy migration and their symmetry properties considered in the surrogate models of coefficients of \cref{eq:quartic}. (a): The plot of the $1^{\text{st}}$, $2^{\text{nd}}$, and $3^{\text{rd}}$ nearest neighboring sites of the vacancy (black) and the migrating atom (yellow) aligned along the $\langle110\rangle$ direction (x-axis). The vacancy and the migrating atom are plotted in the same way in (b) and (c). (b) and (c): Effects of $mmm$ and $mm2$ point group symmetry applied on the $1^{\text{st}}$ neighboring lattice sites, respectively. Atoms with the same color are at the symmetrically equivalent lattice sites, so their contributions to the inputs of the surrogate models are averaged together. Here the $mmm$ point group shown in (b) has a mirror symmetry plane perpendicular to the $\langle\bar{1}10\rangle$ (y-axis), a mirror symmetry plane perpendicular to $\langle001\rangle$ (z-axis), and a mirror symmetry plane perpendicular to $\langle110\rangle$ direction (x-axis). The $mm2$ point group shown in (c) has a mirror symmetry plane perpendicular to the $\langle\bar{1}10\rangle$ (y-axis), a mirror symmetry plane perpendicular to $\langle001\rangle$ (z-axis), and 2-fold symmetry along $\langle110\rangle$ direction (x-axis).}
     \label{fig:encoding}
\end{figure}

Although \cref{eq:e0} can be helpful to estimate $\Delta e_{\text{a}}$ and $\Delta E_{\text{a}}$ without using the computationally expensive CI-NEB method for Al-Mg-Zn and potentially other multicomponent FCC alloys, it still requires DFT calculations that need considerable computational resources. To study diffusion and precipitation in mesoscale methods such as kMC simulations\cite{clouet2004nucleation,sha2005kinetic}, we still need to accurately and efficiently predict $\Delta E_{\text{a}}$ and $\Delta E$ in different local chemical environments. A practical approach is to construct surrogate models that can predict the coefficients ($a$, $b$ and $c$) of \cref{eq:quartic} with respect to the local lattice occupations, then the properties of the MEP ($\Delta E_{\text{a}}$, $\Delta E$, and $D_{\text{MEP}}$) can be automatically obtained based on \cref{eq:quartic} and \cref{fig:quartic} (a). The general strategy to construct these surrogate models and train them based on DFT+CI-NEB results is described as follows.

To train the surrogate models for coefficients of \cref{eq:quartic}, 2000 training data points of $\Delta E$ and $\Delta E_{\text{a}}$ pairs (plus the corresponding supercell configurations of the initial and final states) were chosen randomly from the total 2500 data points generated from the DFT+CI-NEB calculations. The remaining 500 data points were utilized as testing data to evaluate the predictive accuracy of the surrogate models. All of the data were divided into three different groups based on the chemical type of the migrating atom (Al, Mg or Zn)\cite{lindsey1994microstructural, rautiainen1998modelling}. The input information was chosen to be the type of the migrating atom and the type of all atoms on the $1^{\text{st}}$, $2^{\text{nd}}$, and $3^{\text{rd}}$ nearest-neighbor lattice sites relative to the vacancy site before and after the migration event, as shown \cref{fig:encoding} (a). This difference between the even-order-term coefficients of \cref{eq:quartic} ($a$ and $c$) and the odd-order-term coefficient ($b$) suggests we should use different symmetry constraints to construct the inputs of the surrogate models for coefficients. Thus, the input information (only $1^{\text{st}}$ nearest-neighbor lattice sites are shown) for surrogate models of $a$ and $c$ is constructed based on the symmetry operations of the $mmm$ point group shown in \cref{fig:encoding} (b), and input for the surrogate model of $b$ is constructed based on the symmetry operations of the $mm2$ point group shown in \cref{fig:encoding} (c). 

In both \cref{fig:encoding} (b) and (c), the vacancy (black color) and the migration atom (yellow color) are aligned along the $\langle110\rangle$ direction (x-axis). For the $mmm$ point group illustrated in \cref{fig:encoding} (b), there is a mirror symmetry plane perpendicular to the $\langle\bar{1}10\rangle$ (y-axis), a mirror symmetry plane perpendicular to the $\langle001\rangle$ (z-axis), and a mirror symmetry plane perpendicular to the $\langle110\rangle$ direction (x-axis). For the $mm2$ point group illustrated in \cref{fig:encoding} (c), there is a mirror symmetry plane perpendicular to the $\langle\bar{1}10\rangle$ (y-axis), a mirror plane perpendicular to the $\langle001\rangle$ (z-axis), a 2-fold rotation axis along the $\langle110\rangle$ direction (x-axis). Thus, the neighboring sites can be divided into different sets based on their symmetry relative to the vacancy and the migrating atom. As shown in \cref{fig:encoding} (b) and (c), the $mmm$ point group sorts the $1^{\text{st}}$ nearest neighbor sites into 4 sets; the $mm2$ point group sorts the $1^{\text{st}}$ nearest neighbor sites into 7 sets; each set of lattice sites is plotted in the same color. The same strategy is applied to $2^{\text{nd}}$, and $3^{\text{rd}}$ nearest neighbors and 2-atom clusters (atoms at two lattice sites not apart than $3^{\text{rd}}$-nearest-neighbor distance (4.955 {\r{A}})) as well. Atoms and clusters that are symmetrically equivalent should have the same contribution to the inputs of the surrogate models for coefficients of \cref{eq:quartic}.

Besides the symmetry effect, the encoding strategy of the lattice occupations has a significant impact on the surrogate model. In this work, we applied the one-hot encoding method \cite{xie2018crystal,chen2021direct} to construct feature vectors to describe types of single atoms and 2-atom clusters. The advantage of using the one-hot encoding for categorical data is that since it represents each type of the variable by a unique digit, there is no quantitative relationship between the values of variables. Hence, one-hot encoding without introducing any fictional ordinal relationship can be more accurate. The symmetry properties related to vacancy migrations in the FCC lattice illustrated in \cref{fig:encoding} are applied to construct these feature vectors. Because $a$ and $c$ of \cref{eq:quartic} are the coefficients of fourth-order and second-order terms, respectively, each should have the same values for the forward and backward migration processes in a vacancy migration case. However, $b$ of \cref{eq:quartic} is the coefficient of a third-order term, so it should have the opposite values in forward and backward migration processes. Consequently, the feature vectors for the surrogate models of $a$ and $c$ are constructed based on the symmetry operations of the $mmm$ point group as illustrated in \cref{fig:encoding} (b), and the feature vectors for the surrogate model of $b$ are constructed based on the symmetry operations of the $mm2$ point group as illustrated in \cref{fig:encoding} (c). In both \cref{fig:encoding} (b) and (c), the symmetrically equivalent lattice sites are of the same color, so the contributions of chemical elements on these symmetrically equivalent sites to the feature vectors should be averaged. 

Examples of feature vectors and the symmetry constraints on feature vectors are described as follows. We used a feature vector $\pmb v \in \mathbb{R}^3$ to represent the chemical type of a single atom: 
\begin{equation}
\begin{split}
\pmb v_{\text{Al}} &= \begin{pmatrix} 1,&0,&0\end{pmatrix} \\ 
\pmb v_{\text{Mg}} &= \begin{pmatrix} 0,&1,&0\end{pmatrix} \\
\pmb v_{\text{Zn}} &= \begin{pmatrix} 0,&0,&1\end{pmatrix} 
\end{split}
\end{equation}
For 2-atom clusters, if both of two lattice sites are from the same symmetry-equivalent sets (two sites with the same color in \cref{fig:encoding} (b) or (c)), such as the cluster of atom 8 and atom 9 shown in \cref{fig:encoding} (b) or (c), then their orientation and order relative to the vacancy site and the migrating atom can be neglected. Therefore, there are 6 combinations in total to put different types of chemical elements in these two sites. We used a feature vector $\pmb v \in \mathbb{R}^6$ to represent each type:
\begin{equation}
\begin{split}
 \pmb v_{\text{Al-Al}} &= \begin{pmatrix} 1,&0,&0,&0,&0,&0\end{pmatrix} \\
 \pmb v_{\text{Al-Mg}} &= \begin{pmatrix} 0,&1,&0,&0,&0,&0\end{pmatrix} \\
 \pmb v_{\text{Al-Zn}} &= \begin{pmatrix} 0,&0,&1,&0,&0,&0\end{pmatrix} \\
 \pmb v_{\text{Mg-Mg}} &= \begin{pmatrix} 0,&0,&0,&1,&0,&0\end{pmatrix} \\
 \pmb v_{\text{Mg-Zn}} &= \begin{pmatrix} 0,&0,&0,&0,&1,&0\end{pmatrix} \\
 \pmb v_{\text{Zn-Zn}} &= \begin{pmatrix} 0,&0,&0,&0,&0,&1\end{pmatrix} 
\end{split}
\end{equation}
However, if two lattice sites are from different symmetry sets, for instance, the cluster of atom 3 and atom 9 shown in \cref{fig:encoding} (b) or (c), then their orientations and order can affect the vacancy migration energetics. Therefore, there are 9 combinations to put different types of chemical elements in these two sites. This required use of a feature vector $\pmb v \in \mathbb{R}^9$ to represent each type:
\begin{equation}
\begin{split}
\pmb v_{\text{Al-Al}} &= \begin{pmatrix} 1,&0,&0,&0,&0,&0,&0,&0,&0\end{pmatrix} \\
\pmb v_{\text{Al-Mg}} &= \begin{pmatrix} 0,&1,&0,&0,&0,&0,&0,&0,&0\end{pmatrix} \\
\pmb v_{\text{Al-Zn}} &= \begin{pmatrix} 0,&0,&1,&0,&0,&0,&0,&0,&0\end{pmatrix} \\
\pmb v_{\text{Mg-Al}} &= \begin{pmatrix} 0,&0,&0,&1,&0,&0,&0,&0,&0\end{pmatrix} \\
\pmb v_{\text{Mg-Mg}} &= \begin{pmatrix} 0,&0,&0,&0,&1,&0,&0,&0,&0\end{pmatrix} \\
\pmb v_{\text{Mg-Zn}} &= \begin{pmatrix} 0,&0,&0,&0,&0,&1,&0,&0,&0\end{pmatrix} \\
\pmb v_{\text{Zn-Al}} &= \begin{pmatrix} 0,&0,&0,&0,&0,&0,&1,&0,&0\end{pmatrix} \\
\pmb v_{\text{Zn-Mg}} &= \begin{pmatrix} 0,&0,&0,&0,&0,&0,&0,&1,&0\end{pmatrix} \\
\pmb v_{\text{Zn-Zn}} &= \begin{pmatrix} 0,&0,&0,&0,&0,&0,&0,&0,&1\end{pmatrix}
\end{split}
\end{equation}

After using feature vectors to label single atoms and 2-atom clusters on the local lattice occupations near the vacancy and the migrating atom, we can average the one-hot feature encoding vectors from the clusters that share the same symmetry. A feature vector that represents the averaged information can be obtained. For example, if the 18 first-nearest-neighboring sites shown in \cref{fig:encoding} (b) have the following lattice occupations ($\sigma_i$, where $i$ is the site index plotted in \cref{fig:encoding} (b)): $\sigma_1=$ Al, $\sigma_2= $ Mg, $\sigma_3=$ Al, $\sigma_4=$ Al, $\sigma_5=$ Zn, $\sigma_6=$ Mg, $\sigma_7=$ Al, $\sigma_8=$ Al, $\sigma_9=$ Zn, $\sigma_{10}=$ Mg, $\sigma_{11}=$ Zn, $\sigma_{12}=$ Al, $\sigma_{13}=$ Al, $\sigma_{14}=$ Al, $\sigma_{15}=$ Al, $\sigma_{16}=$ Mg, $\sigma_{17}=$ Zn and $\sigma_{18}=$ Al, respectively, then four feature vectors can be obtained for the single-atom occupations in 4 sets of $1^{\text{st}}$ nearest neighbor sites by considering the $mmm$ point group:
\begin{equation}
\begin{split}
\hat{\pmb v}_1 &= \frac{1}{2} \pmb{v}_{Al} + \frac{1}{2} \pmb v_{Al} \\
               &= \begin{pmatrix} 1, & 0, & 0\end{pmatrix} \\
\hat{\pmb v}_2 &= \frac{1}{8} \pmb v_{Mg} + \frac{1}{8} \pmb v_{Al} + \frac{1}{8} \pmb v_{Al} + \frac{1}{8} \pmb v_{Zn} \\
               &+ \frac{1}{8} \pmb v_{Al} + \frac{1}{8} \pmb v_{Al} + \frac{1}{8} \pmb v_{Mg} + \frac{1}{8} \pmb v_{Zn} \\
               &= \begin{pmatrix} 0.5, & 0.25, & 0.25\end{pmatrix}\\
\hat{\pmb v}_3 &= \frac{1}{4} \pmb{v}_{Mg} + \frac{1}{4} \pmb{v}_{Al} + \frac{1}{4} \pmb v_{Al} + \frac{1}{4} \pmb v_{Al} \\
               &=\begin{pmatrix} 0.75, & 0.25, & 0\end{pmatrix}\\
\hat{\pmb v}_4 &= \frac{1}{4} \pmb{v}_{Al} + \frac{1}{4} \pmb{v}_{Zn} + \frac{1}{4} \pmb v_{Mg} + \frac{1}{4} \pmb v_{Zn} \\
               &=\begin{pmatrix} 0.25, & 0.25, & 0.5\end{pmatrix}
\end{split}
\end{equation}
Here, each $\mathbb{R}^3$ feature vector of a single atom is multiplied by a weighting factor $\frac{1}{n_{\text{s}}}$, where $n_{\text{s}}$ is the number of symmetry-equivalent sites in each of these 4 sets. Concatenating these feature vectors together, we can obtain a combined feature vector $\hat{\pmb v} = \begin{pmatrix} \hat{\pmb v}_1, & \hat{\pmb v}_1, & \hat{\pmb v}_1, & \hat{\pmb v}_4\end{pmatrix} \in \mathbb{R}^{12}$. When we extended this method to 2-atoms clusters within $3^{\text{rd}}$ nearest neighboring distance among all lattice sites shown in \cref{fig:encoding} (a), we obtained the combined feature vectors that describe the local environment of a vacancy migration event. 

The dimensionalities of the combined feature vectors of lattice occupations in lattice sites of \cref{fig:encoding} (a) were 1401 based on the $mm2$ point group symmetry operations and 711 based on the $mmm$ point group symmetry operations. These large dimensionalities were at the same scale as the size of our three training datasets (for three different elements of migrating atoms), which reflect a typical downside of one-hot encoding: it tends to create multicollinearity among individual variables because it creates multiple new variables. However, we can apply principal component analysis (PCA) to reduce the dimensionality of the feature vectors. Overall, using one-hot encoding and the PCA method together, we can eliminate potential quantitative relationships and multicollinearity between the individual variables at the same time, which significantly increases the accuracy and robustness of the surrogate model. More details regarding the dimensionality and PCA methods are described in Supplementary Note 6 and Supplementary Note 7.

After the dimensionality reduction, the ridge regression (linear least squares with the $L_2$ regularization) was applied to the training data. It can be described in the form of least squares as:
\begin{equation}
\label{eq:ridge}
\pmb {\hat{X}}  \pmb{\hat{\beta}}_\text{ridge}  = \pmb y
\end{equation}
where, the estimated parameters $\pmb{\hat{\beta}}_\text{ridge}$ minimizes the objective function:
\begin{equation}
\label{eq:ridge_obj}
\min\limits_{\pmb{\beta}} \left\{ |\ \pmb y - \pmb {\hat{X}}  \pmb{\beta}\|_2^2 + \lambda |\pmb{\beta}\|_2^2 \right\}
\end{equation}
Here, $\pmb {\hat{X}}$ is the dimension-reduced feature space. Each row in $\pmb {\hat{X}}$ represents a dimension-reduced feature vector, and it has $m$ rows in total, where $m$ is the size of the training dataset. $\pmb y$ is a vector that contains the results of the targeted coefficients $a$, $b$ or $c$. Since there are two constraints ($a>0$ and $c<0$) to make sure that \cref{eq:quartic} represents the MEP in \cref{fig:Ea_MEP} (c), elements in $\pmb y$ can be $\log(a)$, $b$ or $\log(-c)$ for each data point. The scalar $\lambda$ is a user-defined regularization parameter, which was set to $1$ in our calculations. Based on \cref{eq:ridge_obj}, the estimated parameters vector is $\pmb{\hat{\beta}}_\text{ridge} =\left(\pmb X^\text{T} \pmb X + \lambda \pmb I \right)^{-1} \pmb X^\text{T} \pmb y$, where $\pmb I$ is an identity matrix.

\begin{figure}[ht]
     \centering
     \includegraphics[width=0.48\textwidth]{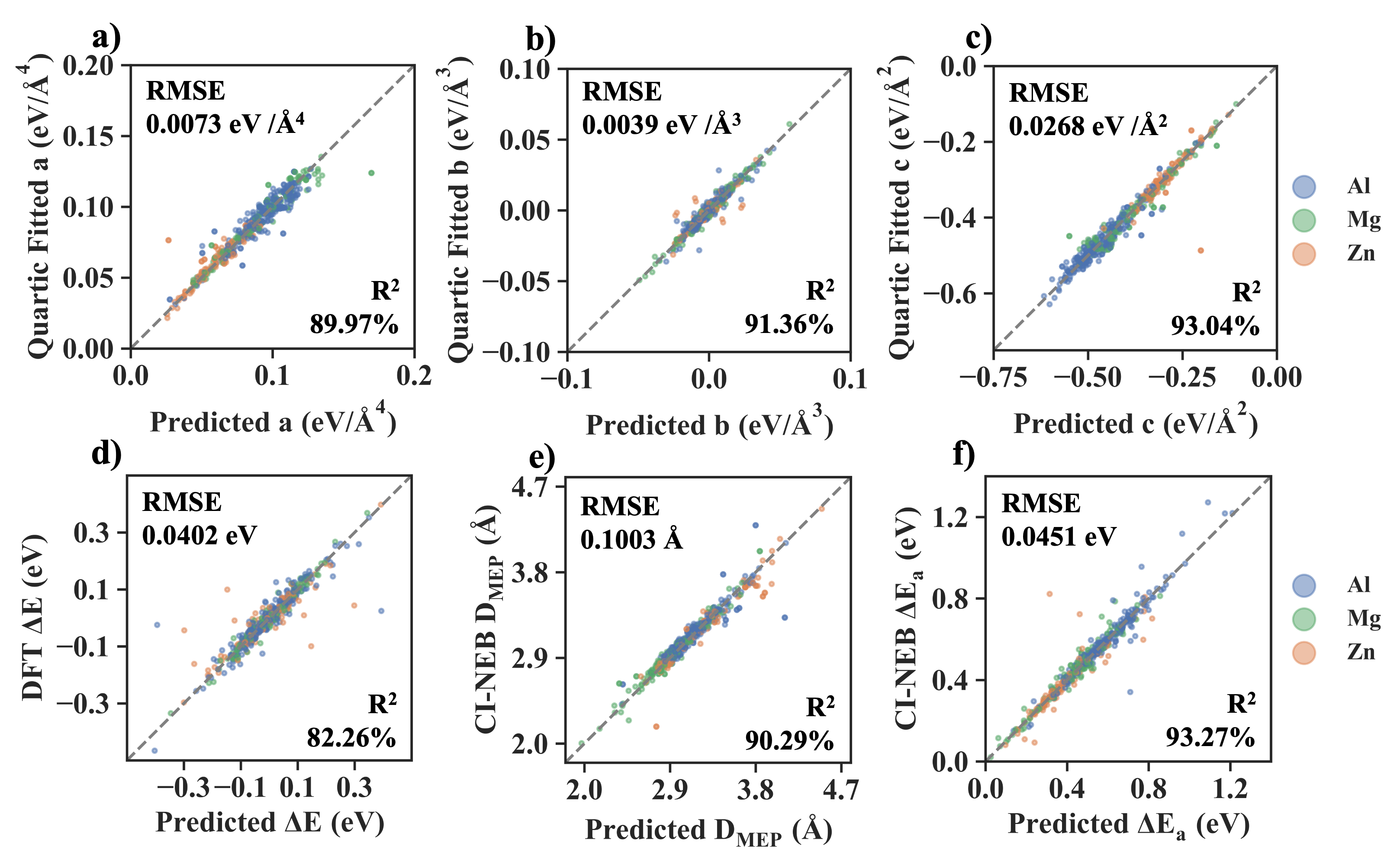}
     \caption{Performances of surrogate models to predict vacancy migration energetics based on \cref{eq:quartic}. (a)-(c): Comparisons between directly fitted results and predictions from our surrogate models for $E_{\text{MEP}}(x)$ coefficients ($a$, $b$, and $c$ in \cref{eq:quartic}). (d)-(f): Comparisons between DFT+CI-NEB calculated results and predictions based on our surrogate-model-predicted $E_{\text{MEP}}(x)$ for $\Delta E$ (d), $D_{\text{MEP}}$ (e), and $\Delta E_{\text{a}}$ (f).}
     \label{fig:predictions}
\end{figure}

After training the surrogate models to output the coefficients $a$, $b$ and $c$ based on the local lattice occupations, we can calculate the properties of the MEP ($\Delta E_{\text{a}}$, $\Delta E$, and $D_{\text{MEP}}$) from \cref{eq:quartic} as follows: $\Delta E = E_{\text{MEP}}(x_{\text{f}})-E_{\text{MEP}}(x_{\text{i}}) = b^2(x_{\text{f}}^3-x_{\text{i}}^3) = \frac{b \left(9 b^2-32 a c\right)^{3/2}}{256 a^3}$, $D_{\text{MEP}}=x_{\text{f}}-x_{\text{i}}=\frac{\sqrt{9 b^2-32 a c}}{4 a}$, and $\Delta E_{\text{a}} =-E_{\text{MEP}}(x_{\text{i}})$. Details of the training performance of surrogate models are described in Supplementary Note 8.

\cref{fig:predictions} (a)-(c) show how predictions of the coefficients of \cref{eq:quartic} from our surrogate models (X-axis) match with the coefficients of \cref{eq:quartic} directly fitted based on DFT+CI-NEB results (Y-axis) for 500 test cases chosen randomly from the total 2500 DFT+CI-NEB calculations. With the predicted coefficients, the values of $\Delta E$, $D_{\text{MEP}}$, and $\Delta E_{\text{a}}$ can then be calculated based on \cref{eq:quartic} and \cref{fig:quartic} (a). \cref{fig:predictions} (d)-(f) compare these predicated values from surrogate models (X-axis) with $\Delta E$, $D_{\text{MEP}}$, and $\Delta E_{\text{a}}$ directly from DFT+CI-NEB calculations (Y-axis). All plots in \cref{fig:predictions} indicate accurate matches between the surrogate model predictions and DFT+CI-NEB calculations, with low RMSE and high R$^2$ values (close or larger than 90$\%$). Particularly, the RMSE values of both $\Delta E$ and $\Delta E_{\text{a}}$ values are less than 0.04 eV, indicating our surrogate models can give precise descriptions of both the energetic driving force and energy barrier of vacancy migrations in complex local chemical environments. Using the one-hot encoding methods to describe the local lattice occupations as the inputs, these surrogate models can be easily implemented into kMC simulations for studies of early-stage precipitation kinetics in Al-Mg-Zn alloys.

\section{Discussion and Conclusions}
\label{sec:Con}

Several previous studies support the generality of our studies of lattice distortion effects on vacancy migration barriers. For example, \cref{eq:e0} is similar to the general linear correlation between $\Delta E_{\text{a}}$ and $a_l^3B_0$ ($a_l$ is the lattice constant and $B_0$ is the bulk modulus) for many pure metals with stable (such as Al, Ni, Cu, and Pt) or metastable (such as Fe and Ti) FCC structures\cite{Flynn68,Angsten14}. Both $k_f$ and $B_0$ are related to second derivatives of the energy landscape at local-minimum states. As another example, strong correlations between site distortions and Li-ion migration barriers and correlations between Li-ion vibrational frequencies and Li-ion migration barriers were found separately in superionic conductors with antiperovskite structures (related to FCC lattice)\cite{Chen21LiIon}. Yet another example is that an equation similar to \cref{eq:e0} was proposed to estimate the local free energy barriers in glass materials\cite{Hall87}. These results suggest \cref{eq:quartic} and \cref{eq:e0} can be applied to atomic migrations in many other materials with FCC and similar crystal structures if each migration MEP only has one local energy maximum as plotted in \cref{fig:Ea_MEP} (c). Thus, not only are these equations (\cref{eq:quartic} and \cref{eq:e0}) and the related surrogate models suitable for describing the energetics of vacancy migrations in multicomponent Al alloys, but they can also be applied in other multicomponent alloys such as high entropy alloys (HEAs) and the related concept of complex concentrated alloys (CCAs), where there can be strong lattice distortion effects on diffusion kinetics due to fluctuations in local chemical compositions\cite{tsai2013sluggish,Zhao16,Osetsky16,osetsky2020tunable,Thomas20}.

The surrogate models to predict coefficients of \cref{eq:quartic} can be further improved from different aspects. First, only the feature vectors related to 2-atom clusters have been considered. We have confirmed that the accuracy of the surrogate models can increase if the feature vectors related to 3-atom clusters are considered (the R$^2$ values of the predictions of $\Delta E$ can be more than 90$\%$ in these cases). Second, high-order methods other than the linear ridge regression can be applied to train the surrogate models. However, since these surrogate models will be implemented into kMC simulations, these improvement strategies may increase the  computational cost significantly and impede the ability of the kMC simulations to study the relatively long-time and large-scale diffusion and precipitation kinetics. Thus, the trade-off between accuracy and efficiency should be carefully considered for the construction of these surrogate models. These decisions can be made if kMC simulations are performed and compared with experimental validations, which will be the subject of future research.

In addition, physical mechanisms (including the symmetry properties discussed in \cref{sec:Distortion} and \cref{sec:Surrogate}) will be applied to discover more efficient approaches to construct the DFT+CI-NEB data set to train the surrogate models. For example, \cref{eq:e0} provides a criterion to select the representative data with appropriate distributions of $\Delta e_{\text{a}}$ and $\Delta E_{\text{a}}$ as the training data set. Last but not least, the generality of our surrogate models based on \cref{eq:quartic} for different alloy compositions should also be verified. We have performed the DFT+CI-NEB calculations and analyses of quaternary Al-Mg-Zn-X alloy systems, where X is the alloying element possibly affecting the vacancy migration kinetics. Our preliminary results show that surrogate models based on \cref{eq:quartic} can also accurately describe the MEPs and the related $\Delta E_{\text{a}}$/$\Delta E$ in these quaternary alloy systems, which will be discussed in our future work.

In summary, the major conclusions of this study are 
\begin{enumerate}
\item DFT+CI-NEB calculations provide energy barriers $\Delta E_{\text{a}}$ and driving forces $\Delta E$ of many ($>$ 1000) vacancy migration events in different local chemical environments within the face-centered cubic (FCC) lattices of Al-Mg-Zn alloys.

\item The widely applied Kinetic Ising model \cite{soisson2010atomistic}, which states $\Delta E_{\text{a}} = \Delta e_{\text{a}} + \frac{1}{2} \Delta E$ and $\Delta e_{\text{a}}$ is a constant value for one type of migrating atom in different local chemical environments inside a lattice, is not broadly applicable to FCC alloys, such as multicomponent Al alloys (Al-Mg system and Al-Mg-Zn system). This is because of the local lattice distortion effects resulting from changes in the chemical environment experienced by a migrating atom. Only Zn atoms near vacancy cites in Al lattices induce small lattice distortions due to the relatively small size difference between Al and Zn atoms\cite{Slater64}. Alternatively, large fluctuations ($\sim$ 1 eV) of $\Delta E_{\text{a}}$ in Al-Mg and Al-Mg-Zn alloys originate primarily from changes in $\Delta e_{\text{a}} = \Delta E_{\text{a}} - \frac{1}{2} \Delta E$ due to local lattice distortion effects because of the relatively large size of Mg atoms compared with Al and Zn atoms\cite{Slater64}. Here $\Delta e_{\text{a}}$ can be regarded as the transition-state energy ($E_{\text{t}}$) relative to the average of the initial-state ($E_{\text{i}}$) and final-state ($E_{\text{f}}$) energies \cite{van2001first}.
 
\item Based upon comparisons with DFT+CI-NEB results, a quartic function of the reaction coordinate $x$, $E_{\text{MEP}}(x) = ax^4+bx^3+cx^2$, accurately describes the energy landscape of the minimum energy path (MEP) for each vacancy migration event in the FCC lattice, where $E_{\text{MEP}}(x)$ of a vacancy migration event only has a single maximum at the transition state.

\item The quartic functions of the MEPs in Al-Mg-Zn alloys suggest that $\Delta e_{\text{a}}$ of all types of migrating atoms in the FCC lattice of Al can be approximated with $\Delta e_{\text{a}} \approx \alpha k_fD^2$, where $\alpha \sim 0.022$ is a constant value. Here $D$ is the distance of a migrating atom between two adjacent equilibrium positions and $k_f$ is the average vibration spring constant of this atom at these two equilibrium positions. This relation provide a a significant speedup in estimating $\Delta E_{\text{a}}$ without computational costly CI-NEB calculations since $k_f$ is calculated rapidly by displacing only the migrating atom from its equilibrium positions.

\item Surrogate models using local lattice occupations as inputs were trained to predict the coefficients of the quartic function. The quartic function can then predict both $\Delta E_{\text{a}}$ and $\Delta E$ with the ab-initio accuracy but without additional DFT or CI-NEB calculations. The efficient and accurate predictions of $\Delta E_{\text{a}}$ and $\Delta E$ using these surrogate models will facilitate mesoscale studies, such as kinetic Monte Carlo simulations, of diffusional transformations that are critical for the processing and applications of Al-Mg-Zn-based and other FCC alloys, such as the solute clustering and early-stage precipitations during the natural aging of 7XXX series of Al alloys\cite{sha2004early,liu2015effect,Huo16AAForm,Chatterjee22}.

\end{enumerate}

\section*{Acknowledgement}
This research is support by NSF-DMR-GOALI, Award Number: 1905421. The calculations were performed by using the Extreme Science and Engineering Discovery Environment (XSEDE) Stampede2 at the TACC through allocation TG-DMR190035.

\bibliography{main}

\begin{thebibliography}{58}%
\makeatletter
\providecommand \@ifxundefined [1]{%
 \@ifx{#1\undefined}
}%
\providecommand \@ifnum [1]{%
 \ifnum #1\expandafter \@firstoftwo
 \else \expandafter \@secondoftwo
 \fi
}%
\providecommand \@ifx [1]{%
 \ifx #1\expandafter \@firstoftwo
 \else \expandafter \@secondoftwo
 \fi
}%
\providecommand \natexlab [1]{#1}%
\providecommand \enquote  [1]{``#1''}%
\providecommand \bibnamefont  [1]{#1}%
\providecommand \bibfnamefont [1]{#1}%
\providecommand \citenamefont [1]{#1}%
\providecommand \href@noop [0]{\@secondoftwo}%
\providecommand \href [0]{\begingroup \@sanitize@url \@href}%
\providecommand \@href[1]{\@@startlink{#1}\@@href}%
\providecommand \@@href[1]{\endgroup#1\@@endlink}%
\providecommand \@sanitize@url [0]{\catcode `\\12\catcode `\$12\catcode
  `\&12\catcode `\#12\catcode `\^12\catcode `\_12\catcode `\%12\relax}%
\providecommand \@@startlink[1]{}%
\providecommand \@@endlink[0]{}%
\providecommand \url  [0]{\begingroup\@sanitize@url \@url }%
\providecommand \@url [1]{\endgroup\@href {#1}{\urlprefix }}%
\providecommand \urlprefix  [0]{URL }%
\providecommand \Eprint [0]{\href }%
\providecommand \doibase [0]{https://doi.org/}%
\providecommand \selectlanguage [0]{\@gobble}%
\providecommand \bibinfo  [0]{\@secondoftwo}%
\providecommand \bibfield  [0]{\@secondoftwo}%
\providecommand \translation [1]{[#1]}%
\providecommand \BibitemOpen [0]{}%
\providecommand \bibitemStop [0]{}%
\providecommand \bibitemNoStop [0]{.\EOS\space}%
\providecommand \EOS [0]{\spacefactor3000\relax}%
\providecommand \BibitemShut  [1]{\csname bibitem#1\endcsname}%
\let\auto@bib@innerbib\@empty
\bibitem [{\citenamefont {Borgenstam}\ \emph {et~al.}(2000)\citenamefont
  {Borgenstam}, \citenamefont {H{\"o}glund}, \citenamefont {{\AA}gren},\ and\
  \citenamefont {Engstr{\"o}m}}]{borgenstam2000dictra}%
  \BibitemOpen
  \bibfield  {author} {\bibinfo {author} {\bibfnamefont {A.}~\bibnamefont
  {Borgenstam}}, \bibinfo {author} {\bibfnamefont {L.}~\bibnamefont
  {H{\"o}glund}}, \bibinfo {author} {\bibfnamefont {J.}~\bibnamefont
  {{\AA}gren}},\ and\ \bibinfo {author} {\bibfnamefont {A.}~\bibnamefont
  {Engstr{\"o}m}},\ }\bibfield  {title} {\bibinfo {title} {Dictra, a tool for
  simulation of diffusional transformations in alloys},\ }\href@noop {}
  {\bibfield  {journal} {\bibinfo  {journal} {Journal of phase equilibria}\
  }\textbf {\bibinfo {volume} {21}},\ \bibinfo {pages} {269} (\bibinfo {year}
  {2000})}\BibitemShut {NoStop}%
\bibitem [{\citenamefont {Pogatscher}\ \emph {et~al.}(2014)\citenamefont
  {Pogatscher}, \citenamefont {Antrekowitsch}, \citenamefont {Werinos},
  \citenamefont {Moszner}, \citenamefont {Gerstl}, \citenamefont {Francis},
  \citenamefont {Curtin}, \citenamefont {L{\"o}ffler},\ and\ \citenamefont
  {Uggowitzer}}]{pogatscher2014diffusion}%
  \BibitemOpen
  \bibfield  {author} {\bibinfo {author} {\bibfnamefont {S.}~\bibnamefont
  {Pogatscher}}, \bibinfo {author} {\bibfnamefont {H.}~\bibnamefont
  {Antrekowitsch}}, \bibinfo {author} {\bibfnamefont {M.}~\bibnamefont
  {Werinos}}, \bibinfo {author} {\bibfnamefont {F.}~\bibnamefont {Moszner}},
  \bibinfo {author} {\bibfnamefont {S.~S.}\ \bibnamefont {Gerstl}}, \bibinfo
  {author} {\bibfnamefont {M.}~\bibnamefont {Francis}}, \bibinfo {author}
  {\bibfnamefont {W.}~\bibnamefont {Curtin}}, \bibinfo {author} {\bibfnamefont
  {J.~F.}\ \bibnamefont {L{\"o}ffler}},\ and\ \bibinfo {author} {\bibfnamefont
  {P.~J.}\ \bibnamefont {Uggowitzer}},\ }\bibfield  {title} {\bibinfo {title}
  {Diffusion on demand to control precipitation aging: application to al-mg-si
  alloys},\ }\href@noop {} {\bibfield  {journal} {\bibinfo  {journal} {Physical
  Review Letters}\ }\textbf {\bibinfo {volume} {112}},\ \bibinfo {pages}
  {225701} (\bibinfo {year} {2014})}\BibitemShut {NoStop}%
\bibitem [{\citenamefont {Van~der Ven}\ and\ \citenamefont
  {Ceder}(2005)}]{van2005first}%
  \BibitemOpen
  \bibfield  {author} {\bibinfo {author} {\bibfnamefont {A.}~\bibnamefont
  {Van~der Ven}}\ and\ \bibinfo {author} {\bibfnamefont {G.}~\bibnamefont
  {Ceder}},\ }\bibfield  {title} {\bibinfo {title} {First principles
  calculation of the interdiffusion coefficient in binary alloys},\ }\href@noop
  {} {\bibfield  {journal} {\bibinfo  {journal} {Physical review letters}\
  }\textbf {\bibinfo {volume} {94}},\ \bibinfo {pages} {045901} (\bibinfo
  {year} {2005})}\BibitemShut {NoStop}%
\bibitem [{\citenamefont {Mantina}\ \emph {et~al.}(2009)\citenamefont
  {Mantina}, \citenamefont {Wang}, \citenamefont {Chen}, \citenamefont {Liu},\
  and\ \citenamefont {Wolverton}}]{mantina2009first}%
  \BibitemOpen
  \bibfield  {author} {\bibinfo {author} {\bibfnamefont {M.}~\bibnamefont
  {Mantina}}, \bibinfo {author} {\bibfnamefont {Y.}~\bibnamefont {Wang}},
  \bibinfo {author} {\bibfnamefont {L.}~\bibnamefont {Chen}}, \bibinfo {author}
  {\bibfnamefont {Z.}~\bibnamefont {Liu}},\ and\ \bibinfo {author}
  {\bibfnamefont {C.}~\bibnamefont {Wolverton}},\ }\bibfield  {title} {\bibinfo
  {title} {First principles impurity diffusion coefficients},\ }\href@noop {}
  {\bibfield  {journal} {\bibinfo  {journal} {Acta Materialia}\ }\textbf
  {\bibinfo {volume} {57}},\ \bibinfo {pages} {4102} (\bibinfo {year}
  {2009})}\BibitemShut {NoStop}%
\bibitem [{\citenamefont {Osetsky}\ \emph {et~al.}(2016)\citenamefont
  {Osetsky}, \citenamefont {Beland},\ and\ \citenamefont
  {Stoller}}]{Osetsky16}%
  \BibitemOpen
  \bibfield  {author} {\bibinfo {author} {\bibfnamefont {Y.~N.}\ \bibnamefont
  {Osetsky}}, \bibinfo {author} {\bibfnamefont {L.~K.}\ \bibnamefont
  {Beland}},\ and\ \bibinfo {author} {\bibfnamefont {R.~E.}\ \bibnamefont
  {Stoller}},\ }\bibfield  {title} {\bibinfo {title} {Specific features of
  defect and mass transport in concentrated fcc alloys},\ }\href
  {https://doi.org/10.1016/j.actamat.2016.06.018} {\bibfield  {journal}
  {\bibinfo  {journal} {ACTA MATERIALIA}\ }\textbf {\bibinfo {volume} {115}},\
  \bibinfo {pages} {364} (\bibinfo {year} {2016})}\BibitemShut {NoStop}%
\bibitem [{\citenamefont {Zhao}\ \emph {et~al.}(2016)\citenamefont {Zhao},
  \citenamefont {Stocks},\ and\ \citenamefont {Zhang}}]{Zhao16}%
  \BibitemOpen
  \bibfield  {author} {\bibinfo {author} {\bibfnamefont {S.}~\bibnamefont
  {Zhao}}, \bibinfo {author} {\bibfnamefont {G.~M.}\ \bibnamefont {Stocks}},\
  and\ \bibinfo {author} {\bibfnamefont {Y.}~\bibnamefont {Zhang}},\ }\bibfield
   {title} {\bibinfo {title} {Defect energetics of concentrated solid-solution
  alloys from ab initio calculations: Ni0.5co0.5, ni0.5fe0.5, ni0.8fe0.2 and
  ni0.8cr0.2},\ }\href {https://doi.org/10.1039/c6cp05161h} {\bibfield
  {journal} {\bibinfo  {journal} {PHYSICAL CHEMISTRY CHEMICAL PHYSICS}\
  }\textbf {\bibinfo {volume} {18}},\ \bibinfo {pages} {24043} (\bibinfo {year}
  {2016})}\BibitemShut {NoStop}%
\bibitem [{\citenamefont {Thomas}\ and\ \citenamefont
  {Patala}(2020)}]{Thomas20}%
  \BibitemOpen
  \bibfield  {author} {\bibinfo {author} {\bibfnamefont {S.~L.}\ \bibnamefont
  {Thomas}}\ and\ \bibinfo {author} {\bibfnamefont {S.}~\bibnamefont
  {Patala}},\ }\bibfield  {title} {\bibinfo {title} {Vacancy diffusion in
  multi-principal element alloys: The role of chemical disorder in the ordered
  lattice},\ }\href {https://doi.org/10.1016/j.actamat.2020.06.022} {\bibfield
  {journal} {\bibinfo  {journal} {ACTA MATERIALIA}\ }\textbf {\bibinfo {volume}
  {196}},\ \bibinfo {pages} {144} (\bibinfo {year} {2020})}\BibitemShut
  {NoStop}%
\bibitem [{\citenamefont {Clouet}\ \emph {et~al.}(2004)\citenamefont {Clouet},
  \citenamefont {Nastar},\ and\ \citenamefont {Sigli}}]{clouet2004nucleation}%
  \BibitemOpen
  \bibfield  {author} {\bibinfo {author} {\bibfnamefont {E.}~\bibnamefont
  {Clouet}}, \bibinfo {author} {\bibfnamefont {M.}~\bibnamefont {Nastar}},\
  and\ \bibinfo {author} {\bibfnamefont {C.}~\bibnamefont {Sigli}},\ }\bibfield
   {title} {\bibinfo {title} {Nucleation of al 3 zr and al 3 sc in aluminum
  alloys: From kinetic monte carlo simulations to classical theory},\
  }\href@noop {} {\bibfield  {journal} {\bibinfo  {journal} {Physical Review
  B}\ }\textbf {\bibinfo {volume} {69}},\ \bibinfo {pages} {064109} (\bibinfo
  {year} {2004})}\BibitemShut {NoStop}%
\bibitem [{\citenamefont {Sha}\ and\ \citenamefont
  {Cerezo}(2005)}]{sha2005kinetic}%
  \BibitemOpen
  \bibfield  {author} {\bibinfo {author} {\bibfnamefont {G.}~\bibnamefont
  {Sha}}\ and\ \bibinfo {author} {\bibfnamefont {A.}~\bibnamefont {Cerezo}},\
  }\bibfield  {title} {\bibinfo {title} {Kinetic monte carlo simulation of
  clustering in an al--zn--mg--cu alloy (7050)},\ }\href@noop {} {\bibfield
  {journal} {\bibinfo  {journal} {Acta materialia}\ }\textbf {\bibinfo {volume}
  {53}},\ \bibinfo {pages} {907} (\bibinfo {year} {2005})}\BibitemShut
  {NoStop}%
\bibitem [{\citenamefont {Soisson}\ \emph {et~al.}(2010)\citenamefont
  {Soisson}, \citenamefont {Becquart}, \citenamefont {Castin}, \citenamefont
  {Domain}, \citenamefont {Malerba},\ and\ \citenamefont
  {Vincent}}]{soisson2010atomistic}%
  \BibitemOpen
  \bibfield  {author} {\bibinfo {author} {\bibfnamefont {F.}~\bibnamefont
  {Soisson}}, \bibinfo {author} {\bibfnamefont {C.}~\bibnamefont {Becquart}},
  \bibinfo {author} {\bibfnamefont {N.}~\bibnamefont {Castin}}, \bibinfo
  {author} {\bibfnamefont {C.}~\bibnamefont {Domain}}, \bibinfo {author}
  {\bibfnamefont {L.}~\bibnamefont {Malerba}},\ and\ \bibinfo {author}
  {\bibfnamefont {E.}~\bibnamefont {Vincent}},\ }\bibfield  {title} {\bibinfo
  {title} {Atomistic kinetic monte carlo studies of microchemical evolutions
  driven by diffusion processes under irradiation},\ }\href@noop {} {\bibfield
  {journal} {\bibinfo  {journal} {Journal of Nuclear Materials}\ }\textbf
  {\bibinfo {volume} {406}},\ \bibinfo {pages} {55} (\bibinfo {year}
  {2010})}\BibitemShut {NoStop}%
\bibitem [{\citenamefont {Miyoshi}\ \emph {et~al.}(2019)\citenamefont
  {Miyoshi}, \citenamefont {Kimizuka}, \citenamefont {Ishii},\ and\
  \citenamefont {Ogata}}]{miyoshi2019temperature}%
  \BibitemOpen
  \bibfield  {author} {\bibinfo {author} {\bibfnamefont {H.}~\bibnamefont
  {Miyoshi}}, \bibinfo {author} {\bibfnamefont {H.}~\bibnamefont {Kimizuka}},
  \bibinfo {author} {\bibfnamefont {A.}~\bibnamefont {Ishii}},\ and\ \bibinfo
  {author} {\bibfnamefont {S.}~\bibnamefont {Ogata}},\ }\bibfield  {title}
  {\bibinfo {title} {Temperature-dependent nucleation kinetics of
  guinier-preston zones in al--cu alloys: An atomistic kinetic monte carlo and
  classical nucleation theory approach},\ }\href@noop {} {\bibfield  {journal}
  {\bibinfo  {journal} {Acta Materialia}\ }\textbf {\bibinfo {volume} {179}},\
  \bibinfo {pages} {262} (\bibinfo {year} {2019})}\BibitemShut {NoStop}%
\bibitem [{\citenamefont {Elder}\ \emph {et~al.}(2002)\citenamefont {Elder},
  \citenamefont {Katakowski}, \citenamefont {Haataja},\ and\ \citenamefont
  {Grant}}]{Elder02}%
  \BibitemOpen
  \bibfield  {author} {\bibinfo {author} {\bibfnamefont {K.}~\bibnamefont
  {Elder}}, \bibinfo {author} {\bibfnamefont {M.}~\bibnamefont {Katakowski}},
  \bibinfo {author} {\bibfnamefont {M.}~\bibnamefont {Haataja}},\ and\ \bibinfo
  {author} {\bibfnamefont {M.}~\bibnamefont {Grant}},\ }\bibfield  {title}
  {\bibinfo {title} {Modeling elasticity in crystal growth},\ }\bibfield
  {journal} {\bibinfo  {journal} {PHYSICAL REVIEW LETTERS}\ }\textbf {\bibinfo
  {volume} {88}},\ \href {https://doi.org/10.1103/PhysRevLett.88.245701}
  {10.1103/PhysRevLett.88.245701} (\bibinfo {year} {2002})\BibitemShut
  {NoStop}%
\bibitem [{\citenamefont {Fallah}\ \emph {et~al.}(2016)\citenamefont {Fallah},
  \citenamefont {Langelier}, \citenamefont {Ofori-Opoku}, \citenamefont
  {Raeisinia}, \citenamefont {Provatas},\ and\ \citenamefont
  {Esmaeili}}]{Fallah16}%
  \BibitemOpen
  \bibfield  {author} {\bibinfo {author} {\bibfnamefont {V.}~\bibnamefont
  {Fallah}}, \bibinfo {author} {\bibfnamefont {B.}~\bibnamefont {Langelier}},
  \bibinfo {author} {\bibfnamefont {N.}~\bibnamefont {Ofori-Opoku}}, \bibinfo
  {author} {\bibfnamefont {B.}~\bibnamefont {Raeisinia}}, \bibinfo {author}
  {\bibfnamefont {N.}~\bibnamefont {Provatas}},\ and\ \bibinfo {author}
  {\bibfnamefont {S.}~\bibnamefont {Esmaeili}},\ }\bibfield  {title} {\bibinfo
  {title} {Cluster evolution mechanisms during aging in al-mg-si alloys},\
  }\href {https://doi.org/10.1016/j.actamat.2015.09.027} {\bibfield  {journal}
  {\bibinfo  {journal} {ACTA MATERIALIA}\ }\textbf {\bibinfo {volume} {103}},\
  \bibinfo {pages} {290} (\bibinfo {year} {2016})}\BibitemShut {NoStop}%
\bibitem [{\citenamefont {Li}\ \emph {et~al.}(2011)\citenamefont {Li},
  \citenamefont {Sarkar}, \citenamefont {Cox}, \citenamefont {Lenosky},
  \citenamefont {Bitzek},\ and\ \citenamefont {Wang}}]{Li11DMD}%
  \BibitemOpen
  \bibfield  {author} {\bibinfo {author} {\bibfnamefont {J.}~\bibnamefont
  {Li}}, \bibinfo {author} {\bibfnamefont {S.}~\bibnamefont {Sarkar}}, \bibinfo
  {author} {\bibfnamefont {W.~T.}\ \bibnamefont {Cox}}, \bibinfo {author}
  {\bibfnamefont {T.~J.}\ \bibnamefont {Lenosky}}, \bibinfo {author}
  {\bibfnamefont {E.}~\bibnamefont {Bitzek}},\ and\ \bibinfo {author}
  {\bibfnamefont {Y.}~\bibnamefont {Wang}},\ }\bibfield  {title} {\bibinfo
  {title} {Diffusive molecular dynamics and its application to nanoindentation
  and sintering},\ }\bibfield  {journal} {\bibinfo  {journal} {PHYSICAL REVIEW
  B}\ }\textbf {\bibinfo {volume} {84}},\ \href
  {https://doi.org/10.1103/PhysRevB.84.054103} {10.1103/PhysRevB.84.054103}
  (\bibinfo {year} {2011})\BibitemShut {NoStop}%
\bibitem [{\citenamefont {Messina}\ \emph {et~al.}(2014)\citenamefont
  {Messina}, \citenamefont {Nastar}, \citenamefont {Garnier}, \citenamefont
  {Domain},\ and\ \citenamefont {Olsson}}]{Messina14}%
  \BibitemOpen
  \bibfield  {author} {\bibinfo {author} {\bibfnamefont {L.}~\bibnamefont
  {Messina}}, \bibinfo {author} {\bibfnamefont {M.}~\bibnamefont {Nastar}},
  \bibinfo {author} {\bibfnamefont {T.}~\bibnamefont {Garnier}}, \bibinfo
  {author} {\bibfnamefont {C.}~\bibnamefont {Domain}},\ and\ \bibinfo {author}
  {\bibfnamefont {P.}~\bibnamefont {Olsson}},\ }\bibfield  {title} {\bibinfo
  {title} {Exact ab initio transport coefficients in bcc
  $\mathrm{Fe}\ensuremath{-}x$ ($x=\mathrm{Cr}$, $\mathrm{Cu}$, $\mathrm{Mn}$,
  $\mathrm{Ni}$, $p$, $\mathrm{Si}$) dilute alloys},\ }\href
  {https://doi.org/10.1103/PhysRevB.90.104203} {\bibfield  {journal} {\bibinfo
  {journal} {Phys. Rev. B}\ }\textbf {\bibinfo {volume} {90}},\ \bibinfo
  {pages} {104203} (\bibinfo {year} {2014})}\BibitemShut {NoStop}%
\bibitem [{\citenamefont {Wu}\ \emph {et~al.}(2016)\citenamefont {Wu},
  \citenamefont {Mayeshiba},\ and\ \citenamefont {Morgan}}]{wu2016high}%
  \BibitemOpen
  \bibfield  {author} {\bibinfo {author} {\bibfnamefont {H.}~\bibnamefont
  {Wu}}, \bibinfo {author} {\bibfnamefont {T.}~\bibnamefont {Mayeshiba}},\ and\
  \bibinfo {author} {\bibfnamefont {D.}~\bibnamefont {Morgan}},\ }\bibfield
  {title} {\bibinfo {title} {High-throughput ab-initio dilute solute diffusion
  database},\ }\href@noop {} {\bibfield  {journal} {\bibinfo  {journal}
  {Scientific data}\ }\textbf {\bibinfo {volume} {3}},\ \bibinfo {pages} {1}
  (\bibinfo {year} {2016})}\BibitemShut {NoStop}%
\bibitem [{\citenamefont {Rautiainen}\ and\ \citenamefont
  {Sutton}(1999)}]{rautiainen1999influence}%
  \BibitemOpen
  \bibfield  {author} {\bibinfo {author} {\bibfnamefont {T.}~\bibnamefont
  {Rautiainen}}\ and\ \bibinfo {author} {\bibfnamefont {A.}~\bibnamefont
  {Sutton}},\ }\bibfield  {title} {\bibinfo {title} {Influence of the atomic
  diffusion mechanism on morphologies, kinetics, and the mechanisms of
  coarsening during phase separation},\ }\href@noop {} {\bibfield  {journal}
  {\bibinfo  {journal} {Physical Review B}\ }\textbf {\bibinfo {volume} {59}},\
  \bibinfo {pages} {13681} (\bibinfo {year} {1999})}\BibitemShut {NoStop}%
\bibitem [{\citenamefont {Sanchez}\ \emph {et~al.}(1984)\citenamefont
  {Sanchez}, \citenamefont {Ducastelle},\ and\ \citenamefont
  {Gratias}}]{sanchez1984generalized}%
  \BibitemOpen
  \bibfield  {author} {\bibinfo {author} {\bibfnamefont {J.~M.}\ \bibnamefont
  {Sanchez}}, \bibinfo {author} {\bibfnamefont {F.}~\bibnamefont
  {Ducastelle}},\ and\ \bibinfo {author} {\bibfnamefont {D.}~\bibnamefont
  {Gratias}},\ }\bibfield  {title} {\bibinfo {title} {Generalized cluster
  description of multicomponent systems},\ }\href@noop {} {\bibfield  {journal}
  {\bibinfo  {journal} {Physica A: Statistical Mechanics and its Applications}\
  }\textbf {\bibinfo {volume} {128}},\ \bibinfo {pages} {334} (\bibinfo {year}
  {1984})}\BibitemShut {NoStop}%
\bibitem [{\citenamefont {Zhang}\ and\ \citenamefont
  {Sluiter}(2016)}]{zhang2016cluster}%
  \BibitemOpen
  \bibfield  {author} {\bibinfo {author} {\bibfnamefont {X.}~\bibnamefont
  {Zhang}}\ and\ \bibinfo {author} {\bibfnamefont {M.~H.}\ \bibnamefont
  {Sluiter}},\ }\bibfield  {title} {\bibinfo {title} {Cluster expansions for
  thermodynamics and kinetics of multicomponent alloys},\ }\href@noop {}
  {\bibfield  {journal} {\bibinfo  {journal} {Journal of Phase Equilibria and
  Diffusion}\ }\textbf {\bibinfo {volume} {37}},\ \bibinfo {pages} {44}
  (\bibinfo {year} {2016})}\BibitemShut {NoStop}%
\bibitem [{\citenamefont {Vincent}\ \emph {et~al.}(2008)\citenamefont
  {Vincent}, \citenamefont {Becquart}, \citenamefont {Pareige}, \citenamefont
  {Pareige},\ and\ \citenamefont {Domain}}]{vincent2008precipitation}%
  \BibitemOpen
  \bibfield  {author} {\bibinfo {author} {\bibfnamefont {E.}~\bibnamefont
  {Vincent}}, \bibinfo {author} {\bibfnamefont {C.}~\bibnamefont {Becquart}},
  \bibinfo {author} {\bibfnamefont {C.}~\bibnamefont {Pareige}}, \bibinfo
  {author} {\bibfnamefont {P.}~\bibnamefont {Pareige}},\ and\ \bibinfo {author}
  {\bibfnamefont {C.}~\bibnamefont {Domain}},\ }\bibfield  {title} {\bibinfo
  {title} {Precipitation of the fecu system: A critical review of atomic
  kinetic monte carlo simulations},\ }\href@noop {} {\bibfield  {journal}
  {\bibinfo  {journal} {Journal of Nuclear Materials}\ }\textbf {\bibinfo
  {volume} {373}},\ \bibinfo {pages} {387} (\bibinfo {year}
  {2008})}\BibitemShut {NoStop}%
\bibitem [{\citenamefont {Pareige}\ \emph {et~al.}(2011)\citenamefont
  {Pareige}, \citenamefont {Roussel}, \citenamefont {Novy}, \citenamefont
  {Kuksenko}, \citenamefont {Olsson}, \citenamefont {Domain},\ and\
  \citenamefont {Pareige}}]{pareige2011kinetic}%
  \BibitemOpen
  \bibfield  {author} {\bibinfo {author} {\bibfnamefont {C.}~\bibnamefont
  {Pareige}}, \bibinfo {author} {\bibfnamefont {M.}~\bibnamefont {Roussel}},
  \bibinfo {author} {\bibfnamefont {S.}~\bibnamefont {Novy}}, \bibinfo {author}
  {\bibfnamefont {V.}~\bibnamefont {Kuksenko}}, \bibinfo {author}
  {\bibfnamefont {P.}~\bibnamefont {Olsson}}, \bibinfo {author} {\bibfnamefont
  {C.}~\bibnamefont {Domain}},\ and\ \bibinfo {author} {\bibfnamefont
  {P.}~\bibnamefont {Pareige}},\ }\bibfield  {title} {\bibinfo {title} {Kinetic
  study of phase transformation in a highly concentrated fe--cr alloy: Monte
  carlo simulation versus experiments},\ }\href@noop {} {\bibfield  {journal}
  {\bibinfo  {journal} {Acta Materialia}\ }\textbf {\bibinfo {volume} {59}},\
  \bibinfo {pages} {2404} (\bibinfo {year} {2011})}\BibitemShut {NoStop}%
\bibitem [{\citenamefont {Soisson}\ and\ \citenamefont
  {Martin}(2000)}]{soisson2000monte}%
  \BibitemOpen
  \bibfield  {author} {\bibinfo {author} {\bibfnamefont {F.}~\bibnamefont
  {Soisson}}\ and\ \bibinfo {author} {\bibfnamefont {G.}~\bibnamefont
  {Martin}},\ }\bibfield  {title} {\bibinfo {title} {Monte carlo simulations of
  the decomposition of metastable solid solutions: Transient and steady-state
  nucleation kinetics},\ }\href@noop {} {\bibfield  {journal} {\bibinfo
  {journal} {Physical Review B}\ }\textbf {\bibinfo {volume} {62}},\ \bibinfo
  {pages} {203} (\bibinfo {year} {2000})}\BibitemShut {NoStop}%
\bibitem [{\citenamefont {Soisson}\ and\ \citenamefont
  {Fu}(2007)}]{soisson2007cu}%
  \BibitemOpen
  \bibfield  {author} {\bibinfo {author} {\bibfnamefont {F.}~\bibnamefont
  {Soisson}}\ and\ \bibinfo {author} {\bibfnamefont {C.-C.}\ \bibnamefont
  {Fu}},\ }\bibfield  {title} {\bibinfo {title} {Cu-precipitation kinetics in
  $\alpha$- fe from atomistic simulations: Vacancy-trapping effects and
  cu-cluster mobility},\ }\href@noop {} {\bibfield  {journal} {\bibinfo
  {journal} {Physical Review B}\ }\textbf {\bibinfo {volume} {76}},\ \bibinfo
  {pages} {214102} (\bibinfo {year} {2007})}\BibitemShut {NoStop}%
\bibitem [{\citenamefont {Daniels}\ and\ \citenamefont
  {Bellon}(2020)}]{daniels2020hybrid}%
  \BibitemOpen
  \bibfield  {author} {\bibinfo {author} {\bibfnamefont {C.}~\bibnamefont
  {Daniels}}\ and\ \bibinfo {author} {\bibfnamefont {P.}~\bibnamefont
  {Bellon}},\ }\bibfield  {title} {\bibinfo {title} {Hybrid kinetic monte carlo
  algorithm for strongly trapping alloy systems},\ }\href@noop {} {\bibfield
  {journal} {\bibinfo  {journal} {Computational Materials Science}\ }\textbf
  {\bibinfo {volume} {173}},\ \bibinfo {pages} {109386} (\bibinfo {year}
  {2020})}\BibitemShut {NoStop}%
\bibitem [{\citenamefont {Van~der Ven}\ \emph {et~al.}(2001)\citenamefont
  {Van~der Ven}, \citenamefont {Ceder}, \citenamefont {Asta},\ and\
  \citenamefont {Tepesch}}]{van2001first}%
  \BibitemOpen
  \bibfield  {author} {\bibinfo {author} {\bibfnamefont {A.}~\bibnamefont
  {Van~der Ven}}, \bibinfo {author} {\bibfnamefont {G.}~\bibnamefont {Ceder}},
  \bibinfo {author} {\bibfnamefont {M.}~\bibnamefont {Asta}},\ and\ \bibinfo
  {author} {\bibfnamefont {P.}~\bibnamefont {Tepesch}},\ }\bibfield  {title}
  {\bibinfo {title} {First-principles theory of ionic diffusion with nondilute
  carriers},\ }\href@noop {} {\bibfield  {journal} {\bibinfo  {journal}
  {Physical Review B}\ }\textbf {\bibinfo {volume} {64}},\ \bibinfo {pages}
  {184307} (\bibinfo {year} {2001})}\BibitemShut {NoStop}%
\bibitem [{\citenamefont {Goiri}\ \emph {et~al.}(2019)\citenamefont {Goiri},
  \citenamefont {Kolli},\ and\ \citenamefont {Van~der Ven}}]{goiri2019role}%
  \BibitemOpen
  \bibfield  {author} {\bibinfo {author} {\bibfnamefont {J.~G.}\ \bibnamefont
  {Goiri}}, \bibinfo {author} {\bibfnamefont {S.~K.}\ \bibnamefont {Kolli}},\
  and\ \bibinfo {author} {\bibfnamefont {A.}~\bibnamefont {Van~der Ven}},\
  }\bibfield  {title} {\bibinfo {title} {Role of short-and long-range ordering
  on diffusion in ni-al alloys},\ }\href@noop {} {\bibfield  {journal}
  {\bibinfo  {journal} {Physical Review Materials}\ }\textbf {\bibinfo {volume}
  {3}},\ \bibinfo {pages} {093402} (\bibinfo {year} {2019})}\BibitemShut
  {NoStop}%
\bibitem [{\citenamefont {Liddicoat}\ \emph {et~al.}(2010)\citenamefont
  {Liddicoat}, \citenamefont {Liao}, \citenamefont {Zhao}, \citenamefont {Zhu},
  \citenamefont {Murashkin}, \citenamefont {Lavernia}, \citenamefont {Valiev},\
  and\ \citenamefont {Ringer}}]{Liddicoat10}%
  \BibitemOpen
  \bibfield  {author} {\bibinfo {author} {\bibfnamefont {P.~V.}\ \bibnamefont
  {Liddicoat}}, \bibinfo {author} {\bibfnamefont {X.-Z.}\ \bibnamefont {Liao}},
  \bibinfo {author} {\bibfnamefont {Y.}~\bibnamefont {Zhao}}, \bibinfo {author}
  {\bibfnamefont {Y.}~\bibnamefont {Zhu}}, \bibinfo {author} {\bibfnamefont
  {M.~Y.}\ \bibnamefont {Murashkin}}, \bibinfo {author} {\bibfnamefont {E.~J.}\
  \bibnamefont {Lavernia}}, \bibinfo {author} {\bibfnamefont {R.~Z.}\
  \bibnamefont {Valiev}},\ and\ \bibinfo {author} {\bibfnamefont {S.~P.}\
  \bibnamefont {Ringer}},\ }\bibfield  {title} {\bibinfo {title}
  {Nanostructural hierarchy increases the strength of aluminium alloys},\
  }\href {https://doi.org/10.1038/ncomms1062} {\bibfield  {journal} {\bibinfo
  {journal} {Nature Communications}\ }\textbf {\bibinfo {volume} {1}},\
  \bibinfo {pages} {63} (\bibinfo {year} {2010})}\BibitemShut {NoStop}%
\bibitem [{\citenamefont {Sha}\ and\ \citenamefont
  {Cerezo}(2004)}]{sha2004early}%
  \BibitemOpen
  \bibfield  {author} {\bibinfo {author} {\bibfnamefont {G.}~\bibnamefont
  {Sha}}\ and\ \bibinfo {author} {\bibfnamefont {A.}~\bibnamefont {Cerezo}},\
  }\bibfield  {title} {\bibinfo {title} {Early-stage precipitation in
  al--zn--mg--cu alloy (7050)},\ }\href@noop {} {\bibfield  {journal} {\bibinfo
   {journal} {Acta Materialia}\ }\textbf {\bibinfo {volume} {52}},\ \bibinfo
  {pages} {4503} (\bibinfo {year} {2004})}\BibitemShut {NoStop}%
\bibitem [{\citenamefont {Liu}\ \emph {et~al.}(2015)\citenamefont {Liu},
  \citenamefont {Li}, \citenamefont {Han}, \citenamefont {Deng},\ and\
  \citenamefont {Zhang}}]{liu2015effect}%
  \BibitemOpen
  \bibfield  {author} {\bibinfo {author} {\bibfnamefont {S.}~\bibnamefont
  {Liu}}, \bibinfo {author} {\bibfnamefont {C.}~\bibnamefont {Li}}, \bibinfo
  {author} {\bibfnamefont {S.}~\bibnamefont {Han}}, \bibinfo {author}
  {\bibfnamefont {Y.}~\bibnamefont {Deng}},\ and\ \bibinfo {author}
  {\bibfnamefont {X.}~\bibnamefont {Zhang}},\ }\bibfield  {title} {\bibinfo
  {title} {Effect of natural aging on quench-induced inhomogeneity of
  microstructure and hardness in high strength 7055 aluminum alloy},\
  }\href@noop {} {\bibfield  {journal} {\bibinfo  {journal} {Journal of Alloys
  and Compounds}\ }\textbf {\bibinfo {volume} {625}},\ \bibinfo {pages} {34}
  (\bibinfo {year} {2015})}\BibitemShut {NoStop}%
\bibitem [{\citenamefont {Huo}\ \emph {et~al.}(2016)\citenamefont {Huo},
  \citenamefont {Hou}, \citenamefont {Zhang},\ and\ \citenamefont
  {Zhang}}]{Huo16AAForm}%
  \BibitemOpen
  \bibfield  {author} {\bibinfo {author} {\bibfnamefont {W.}~\bibnamefont
  {Huo}}, \bibinfo {author} {\bibfnamefont {L.}~\bibnamefont {Hou}}, \bibinfo
  {author} {\bibfnamefont {Y.}~\bibnamefont {Zhang}},\ and\ \bibinfo {author}
  {\bibfnamefont {J.}~\bibnamefont {Zhang}},\ }\bibfield  {title} {\bibinfo
  {title} {Warm formability and post-forming microstructure/property of
  high-strength aa 7075-t6 al alloy},\ }\href
  {https://doi.org/https://doi.org/10.1016/j.msea.2016.08.054} {\bibfield
  {journal} {\bibinfo  {journal} {Materials Science and Engineering: A}\
  }\textbf {\bibinfo {volume} {675}},\ \bibinfo {pages} {44} (\bibinfo {year}
  {2016})}\BibitemShut {NoStop}%
\bibitem [{\citenamefont {Chatterjee}\ \emph {et~al.}(2022)\citenamefont
  {Chatterjee}, \citenamefont {Qi},\ and\ \citenamefont
  {Misra}}]{Chatterjee22}%
  \BibitemOpen
  \bibfield  {author} {\bibinfo {author} {\bibfnamefont {A.}~\bibnamefont
  {Chatterjee}}, \bibinfo {author} {\bibfnamefont {L.}~\bibnamefont {Qi}},\
  and\ \bibinfo {author} {\bibfnamefont {A.}~\bibnamefont {Misra}},\ }\bibfield
   {title} {\bibinfo {title} {In situ transmission electron microscopy
  investigation of nucleation of gp zones under natural aging in al-zn-mg
  alloy},\ }\href
  {https://doi.org/https://doi.org/10.1016/j.scriptamat.2021.114319} {\bibfield
   {journal} {\bibinfo  {journal} {Scripta Materialia}\ }\textbf {\bibinfo
  {volume} {207}},\ \bibinfo {pages} {114319} (\bibinfo {year}
  {2022})}\BibitemShut {NoStop}%
\bibitem [{\citenamefont {Wolverton}(2007)}]{WOLVERTON20075867}%
  \BibitemOpen
  \bibfield  {author} {\bibinfo {author} {\bibfnamefont {C.}~\bibnamefont
  {Wolverton}},\ }\bibfield  {title} {\bibinfo {title} {Solute–vacancy
  binding in aluminum},\ }\href
  {https://doi.org/https://doi.org/10.1016/j.actamat.2007.06.039} {\bibfield
  {journal} {\bibinfo  {journal} {Acta Materialia}\ }\textbf {\bibinfo {volume}
  {55}},\ \bibinfo {pages} {5867} (\bibinfo {year} {2007})}\BibitemShut
  {NoStop}%
\bibitem [{\citenamefont {Zurob}\ and\ \citenamefont
  {Seyedrezai}(2009)}]{ZUROB2009141}%
  \BibitemOpen
  \bibfield  {author} {\bibinfo {author} {\bibfnamefont {H.}~\bibnamefont
  {Zurob}}\ and\ \bibinfo {author} {\bibfnamefont {H.}~\bibnamefont
  {Seyedrezai}},\ }\bibfield  {title} {\bibinfo {title} {A model for the growth
  of solute clusters based on vacancy trapping},\ }\href
  {https://doi.org/https://doi.org/10.1016/j.scriptamat.2009.03.025} {\bibfield
   {journal} {\bibinfo  {journal} {Scripta Materialia}\ }\textbf {\bibinfo
  {volume} {61}},\ \bibinfo {pages} {141} (\bibinfo {year} {2009})}\BibitemShut
  {NoStop}%
\bibitem [{\citenamefont {Werinos}\ \emph {et~al.}(2016)\citenamefont
  {Werinos}, \citenamefont {Antrekowitsch}, \citenamefont {Ebner},
  \citenamefont {Prillhofer}, \citenamefont {Curtin}, \citenamefont
  {Uggowitzer},\ and\ \citenamefont {Pogatscher}}]{werinos2016design}%
  \BibitemOpen
  \bibfield  {author} {\bibinfo {author} {\bibfnamefont {M.}~\bibnamefont
  {Werinos}}, \bibinfo {author} {\bibfnamefont {H.}~\bibnamefont
  {Antrekowitsch}}, \bibinfo {author} {\bibfnamefont {T.}~\bibnamefont
  {Ebner}}, \bibinfo {author} {\bibfnamefont {R.}~\bibnamefont {Prillhofer}},
  \bibinfo {author} {\bibfnamefont {W.}~\bibnamefont {Curtin}}, \bibinfo
  {author} {\bibfnamefont {P.~J.}\ \bibnamefont {Uggowitzer}},\ and\ \bibinfo
  {author} {\bibfnamefont {S.}~\bibnamefont {Pogatscher}},\ }\bibfield  {title}
  {\bibinfo {title} {Design strategy for controlled natural aging in al--mg--si
  alloys},\ }\href@noop {} {\bibfield  {journal} {\bibinfo  {journal} {Acta
  Materialia}\ }\textbf {\bibinfo {volume} {118}},\ \bibinfo {pages} {296}
  (\bibinfo {year} {2016})}\BibitemShut {NoStop}%
\bibitem [{\citenamefont {Henkelman}\ \emph {et~al.}(2000)\citenamefont
  {Henkelman}, \citenamefont {Uberuaga},\ and\ \citenamefont
  {J{\'o}nsson}}]{henkelman2000climbing}%
  \BibitemOpen
  \bibfield  {author} {\bibinfo {author} {\bibfnamefont {G.}~\bibnamefont
  {Henkelman}}, \bibinfo {author} {\bibfnamefont {B.~P.}\ \bibnamefont
  {Uberuaga}},\ and\ \bibinfo {author} {\bibfnamefont {H.}~\bibnamefont
  {J{\'o}nsson}},\ }\bibfield  {title} {\bibinfo {title} {A climbing image
  nudged elastic band method for finding saddle points and minimum energy
  paths},\ }\href@noop {} {\bibfield  {journal} {\bibinfo  {journal} {The
  Journal of chemical physics}\ }\textbf {\bibinfo {volume} {113}},\ \bibinfo
  {pages} {9901} (\bibinfo {year} {2000})}\BibitemShut {NoStop}%
\bibitem [{\citenamefont {Henkelman}\ and\ \citenamefont
  {J{\'o}nsson}(2000)}]{henkelman2000improved}%
  \BibitemOpen
  \bibfield  {author} {\bibinfo {author} {\bibfnamefont {G.}~\bibnamefont
  {Henkelman}}\ and\ \bibinfo {author} {\bibfnamefont {H.}~\bibnamefont
  {J{\'o}nsson}},\ }\bibfield  {title} {\bibinfo {title} {Improved tangent
  estimate in the nudged elastic band method for finding minimum energy paths
  and saddle points},\ }\href@noop {} {\bibfield  {journal} {\bibinfo
  {journal} {The Journal of chemical physics}\ }\textbf {\bibinfo {volume}
  {113}},\ \bibinfo {pages} {9978} (\bibinfo {year} {2000})}\BibitemShut
  {NoStop}%
\bibitem [{\citenamefont {Kresse}\ and\ \citenamefont
  {Furthm{\"u}ller}(1996{\natexlab{a}})}]{kresse1996efficiency}%
  \BibitemOpen
  \bibfield  {author} {\bibinfo {author} {\bibfnamefont {G.}~\bibnamefont
  {Kresse}}\ and\ \bibinfo {author} {\bibfnamefont {J.}~\bibnamefont
  {Furthm{\"u}ller}},\ }\bibfield  {title} {\bibinfo {title} {Efficiency of
  ab-initio total energy calculations for metals and semiconductors using a
  plane-wave basis set},\ }\href@noop {} {\bibfield  {journal} {\bibinfo
  {journal} {Computational materials science}\ }\textbf {\bibinfo {volume}
  {6}},\ \bibinfo {pages} {15} (\bibinfo {year}
  {1996}{\natexlab{a}})}\BibitemShut {NoStop}%
\bibitem [{\citenamefont {Kresse}\ and\ \citenamefont
  {Furthm{\"u}ller}(1996{\natexlab{b}})}]{kresse1996efficient}%
  \BibitemOpen
  \bibfield  {author} {\bibinfo {author} {\bibfnamefont {G.}~\bibnamefont
  {Kresse}}\ and\ \bibinfo {author} {\bibfnamefont {J.}~\bibnamefont
  {Furthm{\"u}ller}},\ }\bibfield  {title} {\bibinfo {title} {Efficient
  iterative schemes for ab initio total-energy calculations using a plane-wave
  basis set},\ }\href@noop {} {\bibfield  {journal} {\bibinfo  {journal}
  {Physical review B}\ }\textbf {\bibinfo {volume} {54}},\ \bibinfo {pages}
  {11169} (\bibinfo {year} {1996}{\natexlab{b}})}\BibitemShut {NoStop}%
\bibitem [{\citenamefont {BLOCHL}(1994)}]{BLOCHL94PAW}%
  \BibitemOpen
  \bibfield  {author} {\bibinfo {author} {\bibfnamefont {P.}~\bibnamefont
  {BLOCHL}},\ }\bibfield  {title} {\bibinfo {title} {Projector augmented-wave
  method},\ }\href {https://doi.org/10.1103/PhysRevB.50.17953} {\bibfield
  {journal} {\bibinfo  {journal} {PHYSICAL REVIEW B}\ }\textbf {\bibinfo
  {volume} {50}},\ \bibinfo {pages} {17953} (\bibinfo {year}
  {1994})}\BibitemShut {NoStop}%
\bibitem [{\citenamefont {Perdew}\ \emph {et~al.}(1996)\citenamefont {Perdew},
  \citenamefont {Burke},\ and\ \citenamefont {Ernzerhof}}]{Perdew96PBE}%
  \BibitemOpen
  \bibfield  {author} {\bibinfo {author} {\bibfnamefont {J.}~\bibnamefont
  {Perdew}}, \bibinfo {author} {\bibfnamefont {K.}~\bibnamefont {Burke}},\ and\
  \bibinfo {author} {\bibfnamefont {M.}~\bibnamefont {Ernzerhof}},\ }\bibfield
  {title} {\bibinfo {title} {Generalized gradient approximation made simple},\
  }\href {https://doi.org/10.1103/PhysRevLett.77.3865} {\bibfield  {journal}
  {\bibinfo  {journal} {PHYSICAL REVIEW LETTERS}\ }\textbf {\bibinfo {volume}
  {77}},\ \bibinfo {pages} {3865} (\bibinfo {year} {1996})}\BibitemShut
  {NoStop}%
\bibitem [{\citenamefont {Berg}\ \emph {et~al.}(2001)\citenamefont {Berg},
  \citenamefont {Gj{\o}nnes}, \citenamefont {Hansen}, \citenamefont {Li},
  \citenamefont {Knutson-Wedel}, \citenamefont {Schryvers},\ and\ \citenamefont
  {Wallenberg}}]{berg2001gp}%
  \BibitemOpen
  \bibfield  {author} {\bibinfo {author} {\bibfnamefont {L.}~\bibnamefont
  {Berg}}, \bibinfo {author} {\bibfnamefont {J.}~\bibnamefont {Gj{\o}nnes}},
  \bibinfo {author} {\bibfnamefont {V.}~\bibnamefont {Hansen}}, \bibinfo
  {author} {\bibfnamefont {X.}~\bibnamefont {Li}}, \bibinfo {author}
  {\bibfnamefont {M.}~\bibnamefont {Knutson-Wedel}}, \bibinfo {author}
  {\bibfnamefont {D.}~\bibnamefont {Schryvers}},\ and\ \bibinfo {author}
  {\bibfnamefont {L.}~\bibnamefont {Wallenberg}},\ }\bibfield  {title}
  {\bibinfo {title} {Gp-zones in al--zn--mg alloys and their role in artificial
  aging},\ }\href@noop {} {\bibfield  {journal} {\bibinfo  {journal} {Acta
  materialia}\ }\textbf {\bibinfo {volume} {49}},\ \bibinfo {pages} {3443}
  (\bibinfo {year} {2001})}\BibitemShut {NoStop}%
\bibitem [{\citenamefont {Zhuravlev}\ \emph {et~al.}(2017)\citenamefont
  {Zhuravlev}, \citenamefont {Barabash}, \citenamefont {An},\ and\
  \citenamefont {Belashchenko}}]{zhuravlev2017phase}%
  \BibitemOpen
  \bibfield  {author} {\bibinfo {author} {\bibfnamefont {I.}~\bibnamefont
  {Zhuravlev}}, \bibinfo {author} {\bibfnamefont {S.}~\bibnamefont {Barabash}},
  \bibinfo {author} {\bibfnamefont {J.}~\bibnamefont {An}},\ and\ \bibinfo
  {author} {\bibfnamefont {K.}~\bibnamefont {Belashchenko}},\ }\bibfield
  {title} {\bibinfo {title} {Phase stability, ordering tendencies, and
  magnetism in single-phase fcc au-fe nanoalloys},\ }\href@noop {} {\bibfield
  {journal} {\bibinfo  {journal} {Physical Review B}\ }\textbf {\bibinfo
  {volume} {96}},\ \bibinfo {pages} {134109} (\bibinfo {year}
  {2017})}\BibitemShut {NoStop}%
\bibitem [{\citenamefont {Monkhorst}\ and\ \citenamefont
  {Pack}(1976)}]{monkhorst1976special}%
  \BibitemOpen
  \bibfield  {author} {\bibinfo {author} {\bibfnamefont {H.~J.}\ \bibnamefont
  {Monkhorst}}\ and\ \bibinfo {author} {\bibfnamefont {J.~D.}\ \bibnamefont
  {Pack}},\ }\bibfield  {title} {\bibinfo {title} {Special points for
  brillouin-zone integrations},\ }\href@noop {} {\bibfield  {journal} {\bibinfo
   {journal} {Physical review B}\ }\textbf {\bibinfo {volume} {13}},\ \bibinfo
  {pages} {5188} (\bibinfo {year} {1976})}\BibitemShut {NoStop}%
\bibitem [{\citenamefont {Zunger}\ \emph {et~al.}(1990)\citenamefont {Zunger},
  \citenamefont {Wei}, \citenamefont {Ferreira},\ and\ \citenamefont
  {Bernard}}]{zunger1990special}%
  \BibitemOpen
  \bibfield  {author} {\bibinfo {author} {\bibfnamefont {A.}~\bibnamefont
  {Zunger}}, \bibinfo {author} {\bibfnamefont {S.-H.}\ \bibnamefont {Wei}},
  \bibinfo {author} {\bibfnamefont {L.}~\bibnamefont {Ferreira}},\ and\
  \bibinfo {author} {\bibfnamefont {J.~E.}\ \bibnamefont {Bernard}},\
  }\bibfield  {title} {\bibinfo {title} {Special quasirandom structures},\
  }\href@noop {} {\bibfield  {journal} {\bibinfo  {journal} {Physical Review
  Letters}\ }\textbf {\bibinfo {volume} {65}},\ \bibinfo {pages} {353}
  (\bibinfo {year} {1990})}\BibitemShut {NoStop}%
\bibitem [{\citenamefont {Sheppard}\ \emph {et~al.}(2008)\citenamefont
  {Sheppard}, \citenamefont {Terrell},\ and\ \citenamefont
  {Henkelman}}]{sheppard2008optimization}%
  \BibitemOpen
  \bibfield  {author} {\bibinfo {author} {\bibfnamefont {D.}~\bibnamefont
  {Sheppard}}, \bibinfo {author} {\bibfnamefont {R.}~\bibnamefont {Terrell}},\
  and\ \bibinfo {author} {\bibfnamefont {G.}~\bibnamefont {Henkelman}},\
  }\bibfield  {title} {\bibinfo {title} {Optimization methods for finding
  minimum energy paths},\ }\href@noop {} {\bibfield  {journal} {\bibinfo
  {journal} {The Journal of chemical physics}\ }\textbf {\bibinfo {volume}
  {128}},\ \bibinfo {pages} {134106} (\bibinfo {year} {2008})}\BibitemShut
  {NoStop}%
\bibitem [{\citenamefont {Parzen}(1962)}]{parzen1962estimation}%
  \BibitemOpen
  \bibfield  {author} {\bibinfo {author} {\bibfnamefont {E.}~\bibnamefont
  {Parzen}},\ }\bibfield  {title} {\bibinfo {title} {On estimation of a
  probability density function and mode},\ }\href@noop {} {\bibfield  {journal}
  {\bibinfo  {journal} {The annals of mathematical statistics}\ }\textbf
  {\bibinfo {volume} {33}},\ \bibinfo {pages} {1065} (\bibinfo {year}
  {1962})}\BibitemShut {NoStop}%
\bibitem [{\citenamefont {Slater}(1964)}]{Slater64}%
  \BibitemOpen
  \bibfield  {author} {\bibinfo {author} {\bibfnamefont {J.~C.}\ \bibnamefont
  {Slater}},\ }\bibfield  {title} {\bibinfo {title} {Atomic radii in
  crystals},\ }\href {https://doi.org/10.1063/1.1725697} {\bibfield  {journal}
  {\bibinfo  {journal} {The Journal of Chemical Physics}\ }\textbf {\bibinfo
  {volume} {41}},\ \bibinfo {pages} {3199} (\bibinfo {year}
  {1964})}\BibitemShut {NoStop}%
\bibitem [{\citenamefont {Landau}\ and\ \citenamefont
  {Lifshitz}(1980)}]{Landau}%
  \BibitemOpen
  \bibfield  {author} {\bibinfo {author} {\bibfnamefont {L.}~\bibnamefont
  {Landau}}\ and\ \bibinfo {author} {\bibfnamefont {E.}~\bibnamefont
  {Lifshitz}},\ }\href@noop {} {\emph {\bibinfo {title} {Statistical
  Physics}}}\ (\bibinfo  {publisher} {Butterworth-Heinemann},\ \bibinfo {year}
  {1980})\BibitemShut {NoStop}%
\bibitem [{\citenamefont {Lindsey}\ and\ \citenamefont
  {Fultz}(1994)}]{lindsey1994microstructural}%
  \BibitemOpen
  \bibfield  {author} {\bibinfo {author} {\bibfnamefont {T.}~\bibnamefont
  {Lindsey}}\ and\ \bibinfo {author} {\bibfnamefont {B.}~\bibnamefont
  {Fultz}},\ }\bibfield  {title} {\bibinfo {title} {Microstructural dependence
  of vacancy diffusion in ordered alloys},\ }\href@noop {} {\bibfield
  {journal} {\bibinfo  {journal} {Journal of applied physics}\ }\textbf
  {\bibinfo {volume} {75}},\ \bibinfo {pages} {1467} (\bibinfo {year}
  {1994})}\BibitemShut {NoStop}%
\bibitem [{\citenamefont {Rautiainen}(1998)}]{rautiainen1998modelling}%
  \BibitemOpen
  \bibfield  {author} {\bibinfo {author} {\bibfnamefont {T.}~\bibnamefont
  {Rautiainen}},\ }\emph {\bibinfo {title} {Modelling microstructural evolution
  in binary alloys}},\ \href@noop {} {Ph.D. thesis},\ \bibinfo  {school}
  {University of Oxford} (\bibinfo {year} {1998})\BibitemShut {NoStop}%
\bibitem [{\citenamefont {Xie}\ and\ \citenamefont
  {Grossman}(2018)}]{xie2018crystal}%
  \BibitemOpen
  \bibfield  {author} {\bibinfo {author} {\bibfnamefont {T.}~\bibnamefont
  {Xie}}\ and\ \bibinfo {author} {\bibfnamefont {J.~C.}\ \bibnamefont
  {Grossman}},\ }\bibfield  {title} {\bibinfo {title} {Crystal graph
  convolutional neural networks for an accurate and interpretable prediction of
  material properties},\ }\href@noop {} {\bibfield  {journal} {\bibinfo
  {journal} {Physical review letters}\ }\textbf {\bibinfo {volume} {120}},\
  \bibinfo {pages} {145301} (\bibinfo {year} {2018})}\BibitemShut {NoStop}%
\bibitem [{\citenamefont {Chen}\ \emph
  {et~al.}(2021{\natexlab{a}})\citenamefont {Chen}, \citenamefont {Andrejevic},
  \citenamefont {Smidt}, \citenamefont {Ding}, \citenamefont {Xu},
  \citenamefont {Chi}, \citenamefont {Nguyen}, \citenamefont {Alatas},
  \citenamefont {Kong},\ and\ \citenamefont {Li}}]{chen2021direct}%
  \BibitemOpen
  \bibfield  {author} {\bibinfo {author} {\bibfnamefont {Z.}~\bibnamefont
  {Chen}}, \bibinfo {author} {\bibfnamefont {N.}~\bibnamefont {Andrejevic}},
  \bibinfo {author} {\bibfnamefont {T.}~\bibnamefont {Smidt}}, \bibinfo
  {author} {\bibfnamefont {Z.}~\bibnamefont {Ding}}, \bibinfo {author}
  {\bibfnamefont {Q.}~\bibnamefont {Xu}}, \bibinfo {author} {\bibfnamefont
  {Y.-T.}\ \bibnamefont {Chi}}, \bibinfo {author} {\bibfnamefont {Q.~T.}\
  \bibnamefont {Nguyen}}, \bibinfo {author} {\bibfnamefont {A.}~\bibnamefont
  {Alatas}}, \bibinfo {author} {\bibfnamefont {J.}~\bibnamefont {Kong}},\ and\
  \bibinfo {author} {\bibfnamefont {M.}~\bibnamefont {Li}},\ }\bibfield
  {title} {\bibinfo {title} {Direct prediction of phonon density of states with
  euclidean neural networks},\ }\href@noop {} {\bibfield  {journal} {\bibinfo
  {journal} {Advanced Science}\ }\textbf {\bibinfo {volume} {8}},\ \bibinfo
  {pages} {2004214} (\bibinfo {year} {2021}{\natexlab{a}})}\BibitemShut
  {NoStop}%
\bibitem [{\citenamefont {Flynn}(1968)}]{Flynn68}%
  \BibitemOpen
  \bibfield  {author} {\bibinfo {author} {\bibfnamefont {C.~P.}\ \bibnamefont
  {Flynn}},\ }\bibfield  {title} {\bibinfo {title} {Atomic migration in
  monatomic crystals},\ }\href {https://doi.org/10.1103/PhysRev.171.682}
  {\bibfield  {journal} {\bibinfo  {journal} {Phys. Rev.}\ }\textbf {\bibinfo
  {volume} {171}},\ \bibinfo {pages} {682} (\bibinfo {year}
  {1968})}\BibitemShut {NoStop}%
\bibitem [{\citenamefont {Angsten}\ \emph {et~al.}(2014)\citenamefont
  {Angsten}, \citenamefont {Mayeshiba}, \citenamefont {Wu},\ and\ \citenamefont
  {Morgan}}]{Angsten14}%
  \BibitemOpen
  \bibfield  {author} {\bibinfo {author} {\bibfnamefont {T.}~\bibnamefont
  {Angsten}}, \bibinfo {author} {\bibfnamefont {T.}~\bibnamefont {Mayeshiba}},
  \bibinfo {author} {\bibfnamefont {H.}~\bibnamefont {Wu}},\ and\ \bibinfo
  {author} {\bibfnamefont {D.}~\bibnamefont {Morgan}},\ }\bibfield  {title}
  {\bibinfo {title} {Elemental vacancy diffusion database from high- throughput
  first-principles calculations for fcc and hcp structures},\ }\href
  {https://doi.org/10.1088/1367-2630/16/1/015018} {\bibfield  {journal}
  {\bibinfo  {journal} {New Journal of Physics}\ }\textbf {\bibinfo {volume}
  {88}},\ \bibinfo {pages} {015018} (\bibinfo {year} {2014})}\BibitemShut
  {NoStop}%
\bibitem [{\citenamefont {Chen}\ \emph
  {et~al.}(2021{\natexlab{b}})\citenamefont {Chen}, \citenamefont {Xu},
  \citenamefont {Lin}, \citenamefont {Lv}, \citenamefont {Bo},\ and\
  \citenamefont {Zhu}}]{Chen21LiIon}%
  \BibitemOpen
  \bibfield  {author} {\bibinfo {author} {\bibfnamefont {R.}~\bibnamefont
  {Chen}}, \bibinfo {author} {\bibfnamefont {Z.}~\bibnamefont {Xu}}, \bibinfo
  {author} {\bibfnamefont {Y.}~\bibnamefont {Lin}}, \bibinfo {author}
  {\bibfnamefont {B.}~\bibnamefont {Lv}}, \bibinfo {author} {\bibfnamefont
  {S.-H.}\ \bibnamefont {Bo}},\ and\ \bibinfo {author} {\bibfnamefont
  {H.}~\bibnamefont {Zhu}},\ }\bibfield  {title} {\bibinfo {title} {Influence
  of structural distortion and lattice dynamics on li-ion diffusion in
  li3ocl1–xbrx superionic conductors},\ }\href
  {https://doi.org/10.1021/acsaem.0c02519} {\bibfield  {journal} {\bibinfo
  {journal} {ACS Applied Energy Materials}\ }\textbf {\bibinfo {volume} {4}},\
  \bibinfo {pages} {2107} (\bibinfo {year} {2021}{\natexlab{b}})}\BibitemShut
  {NoStop}%
\bibitem [{\citenamefont {HALL}\ and\ \citenamefont {WOLYNES}(1987)}]{Hall87}%
  \BibitemOpen
  \bibfield  {author} {\bibinfo {author} {\bibfnamefont {R.}~\bibnamefont
  {HALL}}\ and\ \bibinfo {author} {\bibfnamefont {P.}~\bibnamefont {WOLYNES}},\
  }\bibfield  {title} {\bibinfo {title} {The aperiodic crystal picture and
  free-energy barriers in glasses},\ }\href {https://doi.org/10.1063/1.452045}
  {\bibfield  {journal} {\bibinfo  {journal} {JOURNAL OF CHEMICAL PHYSICS}\
  }\textbf {\bibinfo {volume} {86}},\ \bibinfo {pages} {2943} (\bibinfo {year}
  {1987})}\BibitemShut {NoStop}%
\bibitem [{\citenamefont {Tsai}\ \emph {et~al.}(2013)\citenamefont {Tsai},
  \citenamefont {Tsai},\ and\ \citenamefont {Yeh}}]{tsai2013sluggish}%
  \BibitemOpen
  \bibfield  {author} {\bibinfo {author} {\bibfnamefont {K.-Y.}\ \bibnamefont
  {Tsai}}, \bibinfo {author} {\bibfnamefont {M.-H.}\ \bibnamefont {Tsai}},\
  and\ \bibinfo {author} {\bibfnamefont {J.-W.}\ \bibnamefont {Yeh}},\
  }\bibfield  {title} {\bibinfo {title} {Sluggish diffusion in co-cr-fe-mn-ni
  high-entropy alloys},\ }\href@noop {} {\bibfield  {journal} {\bibinfo
  {journal} {Acta Materialia}\ }\textbf {\bibinfo {volume} {61}},\ \bibinfo
  {pages} {4887} (\bibinfo {year} {2013})}\BibitemShut {NoStop}%
\bibitem [{\citenamefont {Osetsky}\ \emph {et~al.}(2020)\citenamefont
  {Osetsky}, \citenamefont {Barashev}, \citenamefont {B{\'e}land},
  \citenamefont {Yao}, \citenamefont {Ferasat},\ and\ \citenamefont
  {Zhang}}]{osetsky2020tunable}%
  \BibitemOpen
  \bibfield  {author} {\bibinfo {author} {\bibfnamefont {Y.}~\bibnamefont
  {Osetsky}}, \bibinfo {author} {\bibfnamefont {A.~V.}\ \bibnamefont
  {Barashev}}, \bibinfo {author} {\bibfnamefont {L.~K.}\ \bibnamefont
  {B{\'e}land}}, \bibinfo {author} {\bibfnamefont {Z.}~\bibnamefont {Yao}},
  \bibinfo {author} {\bibfnamefont {K.}~\bibnamefont {Ferasat}},\ and\ \bibinfo
  {author} {\bibfnamefont {Y.}~\bibnamefont {Zhang}},\ }\bibfield  {title}
  {\bibinfo {title} {Tunable chemical complexity to control atomic diffusion in
  alloys},\ }\href@noop {} {\bibfield  {journal} {\bibinfo  {journal} {npj
  Computational Materials}\ }\textbf {\bibinfo {volume} {6}},\ \bibinfo {pages}
  {1} (\bibinfo {year} {2020})}\BibitemShut {NoStop}%
\end{thebibliography}%


\begin{thebibliography}{4}%
\makeatletter
\providecommand \@ifxundefined [1]{%
 \@ifx{#1\undefined}
}%
\providecommand \@ifnum [1]{%
 \ifnum #1\expandafter \@firstoftwo
 \else \expandafter \@secondoftwo
 \fi
}%
\providecommand \@ifx [1]{%
 \ifx #1\expandafter \@firstoftwo
 \else \expandafter \@secondoftwo
 \fi
}%
\providecommand \natexlab [1]{#1}%
\providecommand \enquote  [1]{``#1''}%
\providecommand \bibnamefont  [1]{#1}%
\providecommand \bibfnamefont [1]{#1}%
\providecommand \citenamefont [1]{#1}%
\providecommand \href@noop [0]{\@secondoftwo}%
\providecommand \href [0]{\begingroup \@sanitize@url \@href}%
\providecommand \@href[1]{\@@startlink{#1}\@@href}%
\providecommand \@@href[1]{\endgroup#1\@@endlink}%
\providecommand \@sanitize@url [0]{\catcode `\\12\catcode `\$12\catcode
  `\&12\catcode `\#12\catcode `\^12\catcode `\_12\catcode `\%12\relax}%
\providecommand \@@startlink[1]{}%
\providecommand \@@endlink[0]{}%
\providecommand \url  [0]{\begingroup\@sanitize@url \@url }%
\providecommand \@url [1]{\endgroup\@href {#1}{\urlprefix }}%
\providecommand \urlprefix  [0]{URL }%
\providecommand \Eprint [0]{\href }%
\providecommand \doibase [0]{https://doi.org/}%
\providecommand \selectlanguage [0]{\@gobble}%
\providecommand \bibinfo  [0]{\@secondoftwo}%
\providecommand \bibfield  [0]{\@secondoftwo}%
\providecommand \translation [1]{[#1]}%
\providecommand \BibitemOpen [0]{}%
\providecommand \bibitemStop [0]{}%
\providecommand \bibitemNoStop [0]{.\EOS\space}%
\providecommand \EOS [0]{\spacefactor3000\relax}%
\providecommand \BibitemShut  [1]{\csname bibitem#1\endcsname}%
\let\auto@bib@innerbib\@empty
\bibitem [{\citenamefont {Wigner}(1938)}]{wigner1938transition_si}%
  \BibitemOpen
  \bibfield  {author} {\bibinfo {author} {\bibfnamefont {E.}~\bibnamefont
  {Wigner}},\ }\bibfield  {title} {\bibinfo {title} {The transition state
  method},\ }\href@noop {} {\bibfield  {journal} {\bibinfo  {journal}
  {Transactions of the Faraday Society}\ }\textbf {\bibinfo {volume} {34}},\
  \bibinfo {pages} {29} (\bibinfo {year} {1938})}\BibitemShut {NoStop}%
\bibitem [{\citenamefont {Schmidt}\ and\ \citenamefont
  {Bocquet}(1998)}]{schmidt1998calculation_si}%
  \BibitemOpen
  \bibfield  {author} {\bibinfo {author} {\bibfnamefont {C.}~\bibnamefont
  {Schmidt}}\ and\ \bibinfo {author} {\bibfnamefont {J.}~\bibnamefont
  {Bocquet}},\ }\bibfield  {title} {\bibinfo {title} {Calculation of the
  diffusion parameters in an ordered ni3al-alloy for a relaxed lattice},\
  }\href@noop {} {\bibfield  {journal} {\bibinfo  {journal} {MRS Online
  Proceedings Library (OPL)}\ }\textbf {\bibinfo {volume} {527}} (\bibinfo
  {year} {1998})}\BibitemShut {NoStop}%
\bibitem [{\citenamefont {Henkelman}\ \emph {et~al.}(2000)\citenamefont
  {Henkelman}, \citenamefont {Uberuaga},\ and\ \citenamefont
  {J{\'o}nsson}}]{henkelman2000climbing_si}%
  \BibitemOpen
  \bibfield  {author} {\bibinfo {author} {\bibfnamefont {G.}~\bibnamefont
  {Henkelman}}, \bibinfo {author} {\bibfnamefont {B.~P.}\ \bibnamefont
  {Uberuaga}},\ and\ \bibinfo {author} {\bibfnamefont {H.}~\bibnamefont
  {J{\'o}nsson}},\ }\bibfield  {title} {\bibinfo {title} {A climbing image
  nudged elastic band method for finding saddle points and minimum energy
  paths},\ }\href@noop {} {\bibfield  {journal} {\bibinfo  {journal} {The
  Journal of chemical physics}\ }\textbf {\bibinfo {volume} {113}},\ \bibinfo
  {pages} {9901} (\bibinfo {year} {2000})}\BibitemShut {NoStop}%
\bibitem [{\citenamefont {Henkelman}\ and\ \citenamefont
  {J{\'o}nsson}(2000)}]{henkelman2000improved_si}%
  \BibitemOpen
  \bibfield  {author} {\bibinfo {author} {\bibfnamefont {G.}~\bibnamefont
  {Henkelman}}\ and\ \bibinfo {author} {\bibfnamefont {H.}~\bibnamefont
  {J{\'o}nsson}},\ }\bibfield  {title} {\bibinfo {title} {Improved tangent
  estimate in the nudged elastic band method for finding minimum energy paths
  and saddle points},\ }\href@noop {} {\bibfield  {journal} {\bibinfo
  {journal} {The Journal of chemical physics}\ }\textbf {\bibinfo {volume}
  {113}},\ \bibinfo {pages} {9978} (\bibinfo {year} {2000})}\BibitemShut
  {NoStop}%
\end{thebibliography}%

\end{document}


\preprint{APS/123-QED}

\title{Supplementary Information: Mechanism of Local Lattice Distortion Effects on Vacancy Migration Barriers in FCC Alloys}

\author{Zhucong Xi}
\altaffiliation[]{Department of Materials Science and Engineering, University of Michigan, Ann Arbor, Michigan, 48109, USA}
\author{Mingfei Zhang}
\altaffiliation[]{Department of Materials Science and Engineering, University of Michigan, Ann Arbor, Michigan, 48109, USA}
\author{Louis G. Hector Jr.}
\altaffiliation[]{GM Global Technical Center, General Motors Company, Warren, Michigan 48092, USA}
\author{Amit Misra}
\altaffiliation[]{Department of Materials Science and Engineering, University of Michigan, Ann Arbor, Michigan 48109, USA}
\author{Liang Qi}
\altaffiliation[]{Department of Materials Science and Engineering, University of Michigan, Ann Arbor, Michigan 48109, USA}
\email{qiliang@umich.edu}

\date{\today}

\maketitle

\clearpage

\section{}
\label{sec:kpoints}
\begin{figure}[ht]
    \centering
    \includegraphics[width=1\textwidth]{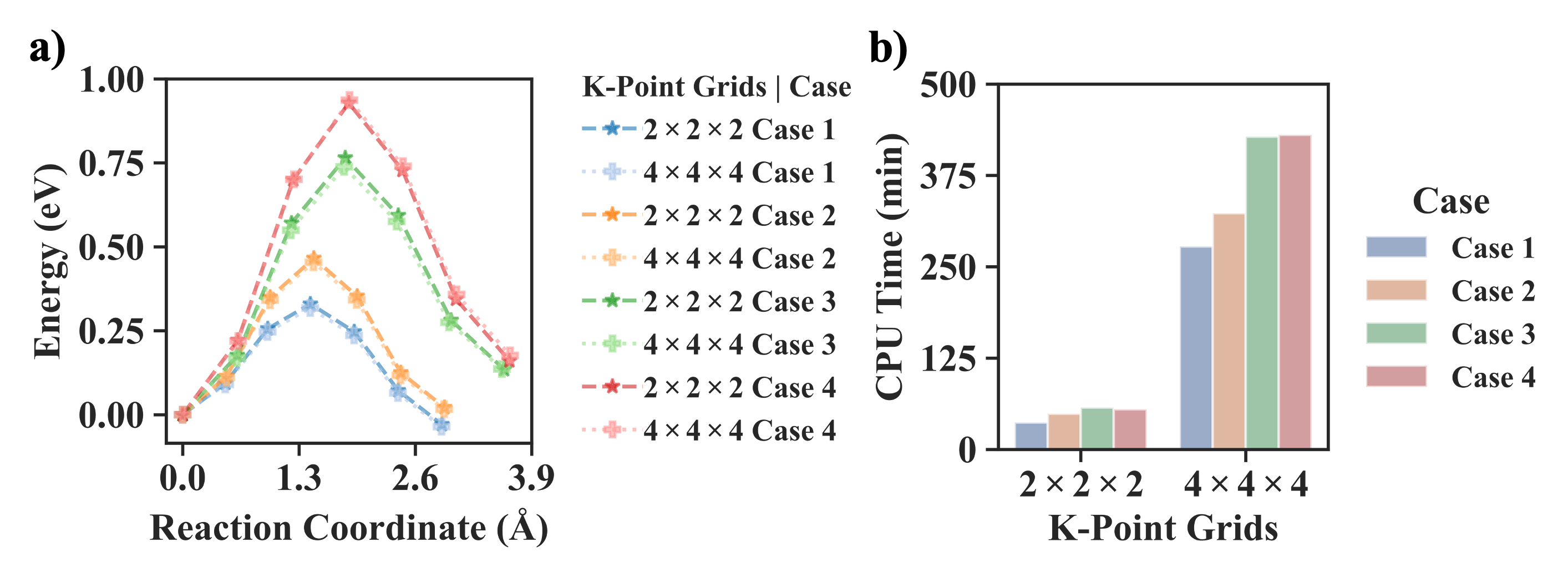}
    \caption{Plots of k-point grid convergence test of vacancy migration barrier calculations of the four selected cases. (a): Minimum energy paths along the reaction coordinates of selected cases using $2\times2\times2$ and $4\times4\times4$ k-point grids. (b): The total CPU time (mins) elapsed using $2\times2\times2$ and $4\times4\times4$ k-point grids of selected cases.}
    \label{fig:kp}
\end{figure}

\textbf{K-Point Convergence}\\
Four cases were randomly selected from the model Al-Mg-Zn alloy systems as shown in the main text Fig. 2 for the k-point convergence tests. Their minimum energy paths were re-calculated using different k-point grids. As shown in \cref{fig:kp} (a), there are no significant differences in the energy landscape (reaction coordinates and migration barriers) computed with the $2\times2\times2$ k-point grids of the k-mesh spacing of $0.194\text{\AA}^{-1}\times0.194\text{\AA}^{-1}\times0.194\text{\AA}^{-1}$ and the $4\times4\times4$ k-point grids of the k-mesh spacing of $0.097\text{\AA}^{-1}\times0.097\text{\AA}^{-1}\times0.097\text{\AA}^{-1}$. However, the CI-NEB calculations were tested on 5 Intel Xeon Phi 7250 Processors. It was found that calculations with $4\times4\times4$ k-point grids required 5-10 times longer simulation time than $2\times2\times2$ grids. Considering this, we chose the $2\times2\times2$ k-point grid with essentially no compromise in accuracy.
\clearpage

\section{}
\label{sec:rs}
\begin{figure}[ht]
    \centering
    \includegraphics[width=1\textwidth]{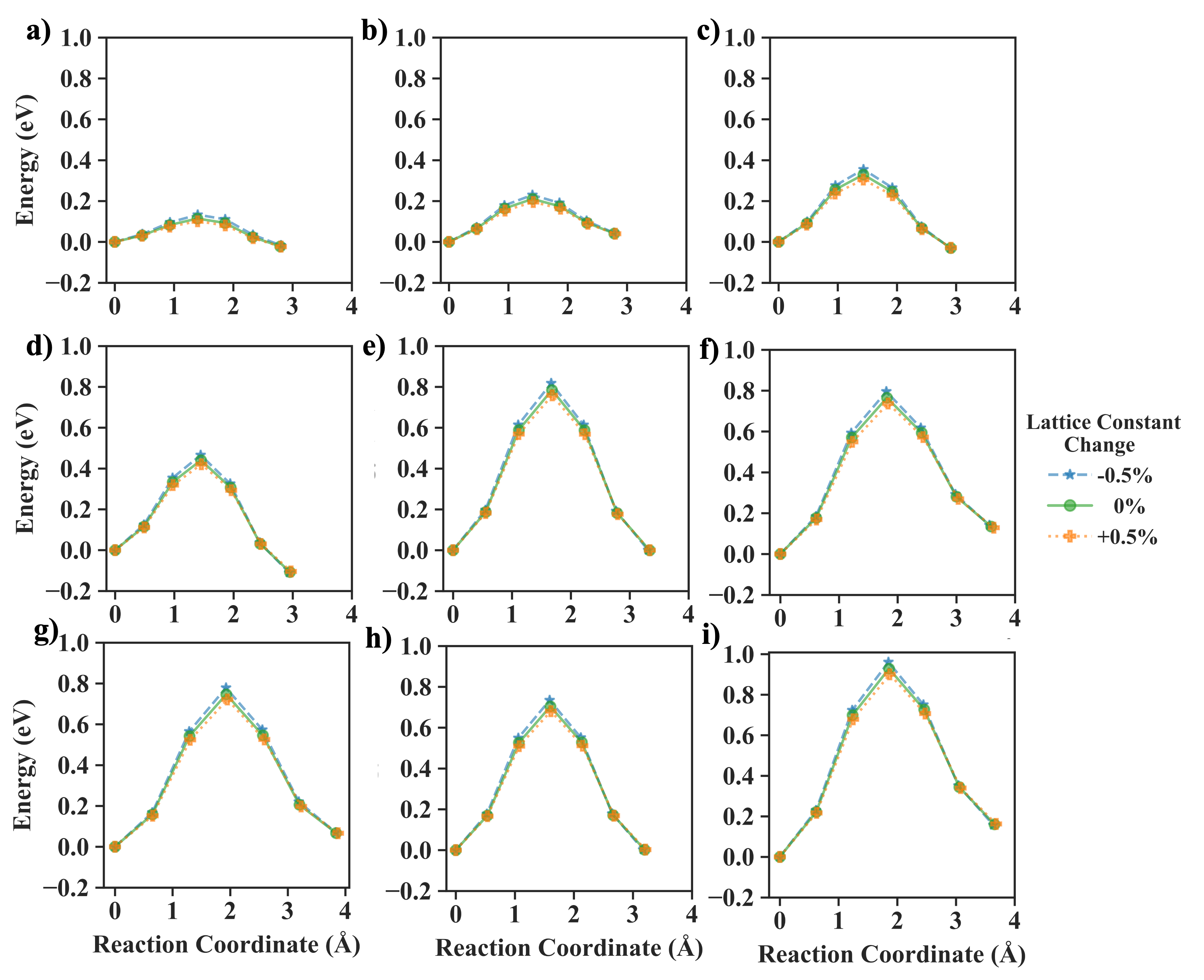}
    \caption{(a)-(i): Plots of energy landscapes of nine randomly selected cases of the original state (with the lattice constant of 4.046 {\r{A}} under no additional strain) and under $\pm0.5\%$ strain. Green lines represent the energy landscape of the original lattice constant cases. Blue dashed lines denote the cases where the lattice constants are shrunk by $0.5\%$, and orange dotted lines represent $0.5\%$ dilating cases. Calculations of the same case with different lattice constants output essentially identical energy landscapes.}
    \label{fig:lattice_constant}
\end{figure}

\textbf{Effects of Lattice Constant Variations}\\
In Al-Mg-Zn alloys, the local lattice constant can change due to local variations of chemical composition or residual stress. To investigate the effect of lattice constant variations on the energy landscapes, we performed DFT+CI-NEB calculations on nine randomly selected cases from the model Al-Mg-Zn alloy systems as shown in Fig. 2 of the main text. The lattice constant in all cases was 4.046 {\r{A}} as discussed in the main text. Lattice constants were then increased and decreased by $0.5\%$ to simulate variations due to the local composition or residual elastic stress. This magnitude of lattice constant variations, $0.5\%$ , is sufficient because it is much larger than the relative lattice constant difference between 7XXX series Al alloys and pure Al obtained from our DFT calculations. The calculated energy landscapes are shown in \cref{fig:lattice_constant}. There is no significant difference in energy changes between $\Delta E$, $\Delta E_{\text{a}}$, or $D_{\text{MEP}}$ 
between the three calculations with different strain in all selected cases. The maximum differences in $\Delta E$, $\Delta E_{\text{a}}$, or $D_{\text{MEP}}$ between the $\pm0.5\%$ strain calculations are 0.013 eV, 0.061 eV, and 0.075 {\r{A}}, respectively. Thus, the contributions of lattice constants (within $\pm0.5\%$) to the energy landscape of vacancy migration were negligible relative to other factors from variations in local chemical compositions.
\clearpage

\section{}
\label{sec:phonon}
\textbf{Verification of Transition State with Phonon Calculations}\\
On the energy landscape, the first derivative of the energy of transition states with respect to the coordinates is zero. In addition, the transition state should meet the requirements for first-order saddle points. That is, the curvature (second derivative of the energy with respect to the coordinates) of the transition state is negative only in the direction of the reaction coordinate and positive along any other directions. Hence, the eigenvalues of the energy second derivative matrix (Hessian matrix) of the transition state structure have a single negative value, which gives an eigenfrequency of the transition state \cite{wigner1938transition_si,schmidt1998calculation_si}. 

\begin{figure}[ht]
    \centering
    \includegraphics[width=1\textwidth]{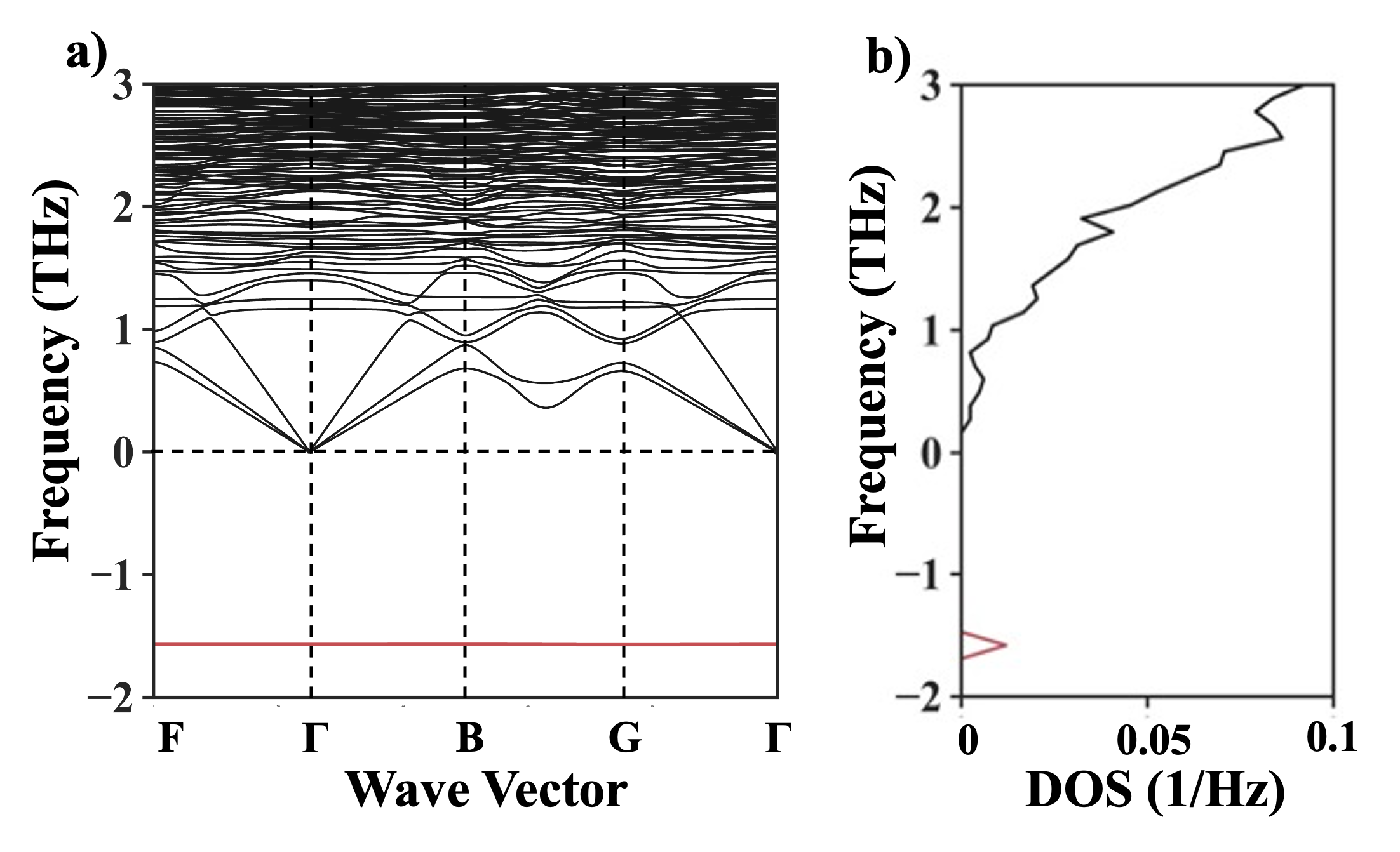}
    \caption{Phonon dispersion curves (a) and phonon DOS (b) of one randomly selected configuration of Al-Mg-Zn ternary alloy system (truncated at 3 THz). Negative frequencies represent imaginary frequencies. A single (red) branch is shown in the negative range of both plots that spans the irreducible wedge of the Brillouin zone.}
    \label{fig:phonon}
\end{figure}

To verify that the structures from the CI-NEB calculations were true saddle point configurations, we performed phonon calculations on four randomly selected transition structures from the model Al alloys systems as shown in Fig. 2 of the main text. There were 255 atoms in each supercell, and each symmetry-unique atom was given $\pm0.02$\r{A} displacements. Only one single imaginary frequency was found in their vibrational spectra. \cref{fig:phonon} shows phonon dispersion curves (a) and phonon density of states (DOS) (b) of one example of Al-Mg-Zn ternary alloys. The frequencies in the plot are truncated at 3 THz. The single negative frequency representing the imaginary frequency is represented by the horizontal red branch in \cref{fig:phonon} (a) and its corresponding DOS is shown in \cref{fig:phonon} (b). This supports our contention that the structures from CI-NEB calculated in this study were indeed saddle points on the respective potential landscapes.
\clearpage

\section{}
\label{sec:distance}
\textbf{Vacancy Migration Distances}\\
In the CI-NEB method, the minimum energy path (MEP) connecting the initial and final states is determined by searching for the path where the spring force between adjacent image configurations only points along the tangent direction of this path. The maxima on the MEP are saddle points on the energy landscape with zero first derivative of the reaction coordinate, and the energy of the highest saddle point minus the energy of the initial state gives $\Delta E_{\text{a}}$ \cite{henkelman2000climbing_si,henkelman2000improved_si}. For all CI-NEB calculations that we performed with different model Al alloys, the MEPs are simple: each MEP only has one maximum at the transition state with two local minima corresponding to the initial and final states, respectively. The relative distance between the initial and final states along the MEP, $D_{\text{MEP}}$, is a natural choice for the reaction coordinate. The sign of $D_{\text{MEP}}$ is positive if the states migrate from the initial to the final and negative if the states migrate from the final to the initial states. The magnitude of $D_{\text{MEP}}$ is defined as the length of the high-dimensional curvature of the MEP as follows:
\begin{equation}
    \label{eq:D_MEP_SI}
    D_{\text{MEP}} = \int\limits_{\text{MEP}} {ds} = \lim\limits_{N\to \infty} \sum\limits_{j=0}^{N}D_{\text{RHD}}(I_{j}, I_{j+1})
\end{equation}
Here, $N$ is the number of intermediate images inserted between the initial and final states, and $I_j$ represents the configuration of the $j^{\text{th}}$ intermediate image. Specifically, $I_0=I_{\text{i}}$ and $I_{N+1}=I_{\text{f}}$ denote the initial and final configurations, respectively. $D_{\text{RHD}}(I_{a}, I_{b})$ is a function that returns the magnitude of the relative high-dimensional distance between $I_{a}$ and $I_{b}$:
\begin{equation}
    D_{\text{RHD}}(I_{a}, I_{b}) = \sqrt{\sum\limits_{k=1}^{N_{\text{atom}}} \left(\left(\pmb x_{b, k} - \pmb x_{a, k}\right)^\text{T} \left(\pmb x_{b, k} - \pmb x_{a, k}\right) \right)}
\end{equation}
Here, $\pmb x_{j, k}$ is a three-dimensional vector representing the Cartesian positions of the $k^{\text{th}}$ atom in the $j^{\text{th}}$ image, and $N_{\text{atom}}$ is the total number of atoms in each configuration. Since only 5 intermediate images were chosen between the relaxed initial and final images for all CI-NEB calculations, \cref{eq:D_MEP_SI} reduces to:
\begin{equation}
    D_{\text{MEP}} = \sum\limits_{i=0}^{5}D_{\text{RHD}}(I_{j}, I_{j+1})
\end{equation}

Besides $D_{\text{MEP}}$, the relative high-dimensional distance $D_{\text{total}}$ between the initial and final states, and the distance $D$ of the migrating atom between two adjacent equilibrium positions (its Cartesian positions in initial and final states) can also be utilized to quantify the local lattice distortion effects on the MEP due to changes of local chemical compositions:
\begin{equation}
    D_{\text{total}} = D_{\text{RHD}}(I_{\text{i}}, I_{\text{f}})
\end{equation}
\begin{equation}
    D = \sqrt{ \left(\pmb x_{\text{f}} - \pmb x_{\text{i}}\right)^\text{T} \left(\pmb x_{\text{f}} - \pmb x_{\text{i}}\right)}
\end{equation}
Here, $\pmb x_{\text{f}}$ and $\pmb x_{\text{i}}$ denote the Cartesian positions of the migrating atom in the supercell of the relaxed initial and final states, respectively.

\begin{figure}[ht]
    \centering
    \includegraphics[width=1\textwidth]{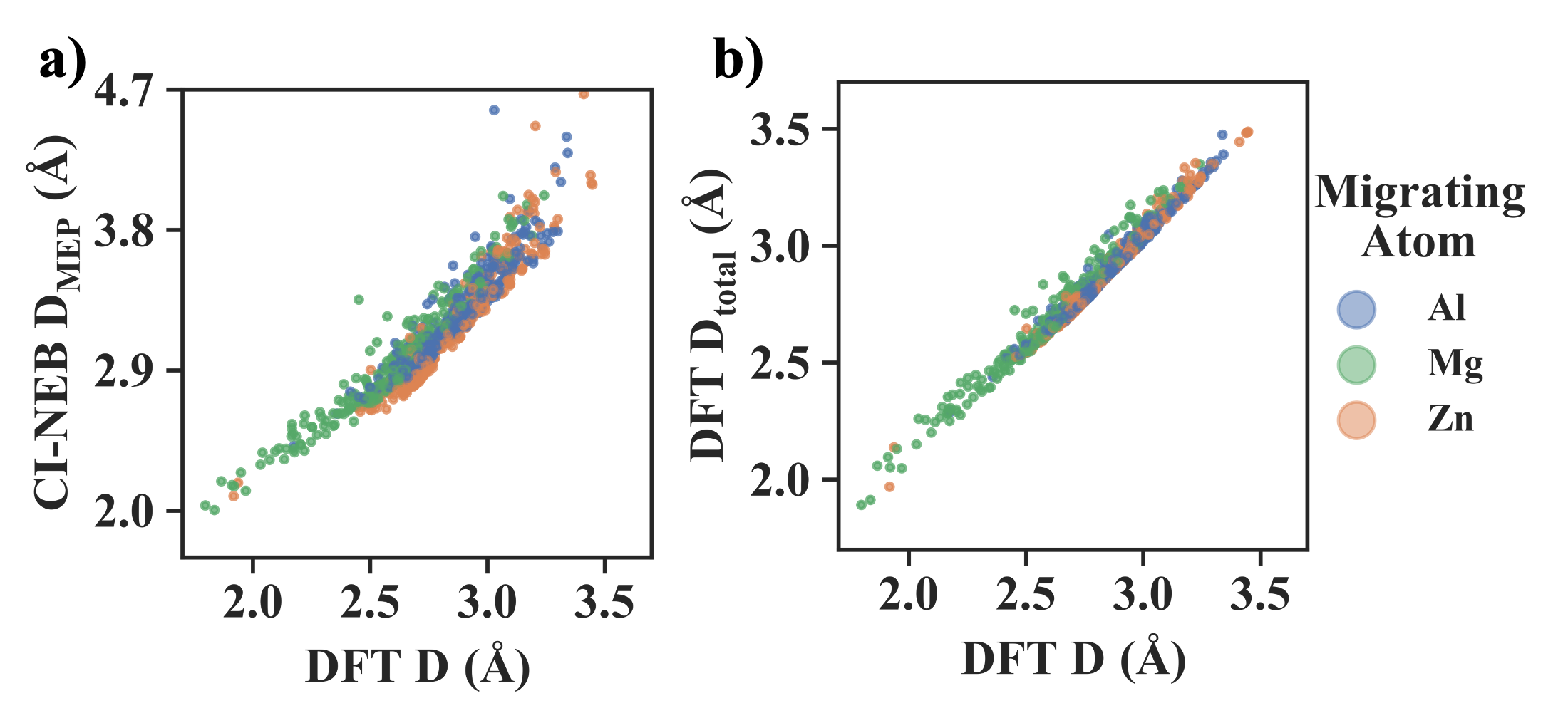}
    \caption{(a) Correlations between DFT+CI-NEB calculated $D_{\text{MEP}}$ and DFT calculated $D$. (b) Correlations between DFT calculated $D_{\text{total}}$ and DFT calculated $D$. Strong positive correlations are shown in both subfigures suggesting $D$ can also be utilized to quantify the local lattice distortion effects.}
    \label{fig:distance}
\end{figure}

The correlations between $D_{\text{MEP}}$ and $D$ are plotted in  \cref{fig:distance} (a), and the correlations between $D_{\text{total}}$ and $D$ are plotted in \cref{fig:distance} (b). Note the strong positive (almost linear) correlations between these three distance variables. Nevertheless, to obtain $D_{\text{MEP}}$, CI-NEB calculations must still be conducted, while $D_{\text{total}}$ and $D$ can be directly acquired from geometry-optimized initial and final state configurations. Thus, $D_{\text{total}}$ and $D$ can be approximated as the reaction coordinate as well in our Al alloy systems without performing CI-NEB calculations.

\clearpage

\section{}
\label{sec:kf}
\textbf{Vibration Spring Constants}\\
The Hessian matrix, $\pmb H$ (the matrix of the second derivatives of the energy relative to the atomic positions) can be calculated using phonon calculations implemented by VASP. These calculations also give us the vibrational frequencies of a supercell, i.e., vibration spring constant, $k_f$, at the initial and final states on the energy landscapes. $\pmb H$ would be a $3N_{\text{atom}}$ dimensional matrix, where $N_{\text{atom}}$ is the total number of atoms in the supercell. The eigenvectors and eigenvalues of $\pmb H$ give the vibration directions and corresponding second derivatives along the vibration directions. For an atom that undergoes harmonic motion, its potential energy can be expressed by a parabolic equation with respect to the displacement $x$:
\begin{equation}
    V =\frac{1}{2}k_f x^2
\end{equation}
With vectorization, we can extend the equation to many atoms:
\begin{equation}
    V =\frac{1}{2} \pmb x^\text{T} \pmb H \pmb x
\end{equation}
Here, the vibration spring constant of a migrating atom can be acquired by finding the eigenvalue of $\pmb H$, of which the corresponding eigenvector is along the direction of migration energy minimum path (MEP).

However, it is expensive to calculate $\pmb H$ for all 255 atoms in a supercell for all the investigated cases. In fact, the ratio of $D$ to $D_{\text{total}}$ is always above $90 \%$ for all CI-NEB cases. This shows that the migrating atom moves a distance much larger than that of other atoms in the supercell during a migration event. Thus, we can fix the positions of atoms far away from the migrating atom and the vacancy since the atomic positions of the far away atoms are nearly stationary during the vacancy migration process: this will speed up calculations of $\pmb H$. The calculated vibration spring constant $k_f$ under the fixed-atom condition is assumed to be correlated with the vibration spring constant $k_f$ when all atoms in the supercells were displaced (defined as $k_{f*}$).

To find the correlation between $k_f$ under the fixed-atom conditions and $k_{f*}$, we randomly selected 6 supercells from our vacancy migration event database and calculated their vibration spring constants for initial and final states using three different types of strategies: 
\begin{itemize}
\item{Type 1: Only the migrating atom is displaced during calculations of $\pmb H$, but all other atoms are fixed. Results are denoted as $k_{f0}$;}
\item{Type 2: Only the migrating atom and the 1$^{\text{st}}$ nearest neighbors of the migration-atom-vacancy pair are displaced during calculations of $\pmb H$, but all other atoms are fixed. Results are denoted as $k_{f1}$;}
\item{Type 3: The migrating atom plus both the 1$^{\text{st}}$ and 2$^{\text{nd}}$ nearest neighbors of the migration-atom-vacancy pair are displaced, but all other atoms are fixed. Results are denoted as $k_{f2}$.}
\end{itemize}

\begin{figure}[ht]
    \centering
    \includegraphics[width=1\textwidth]{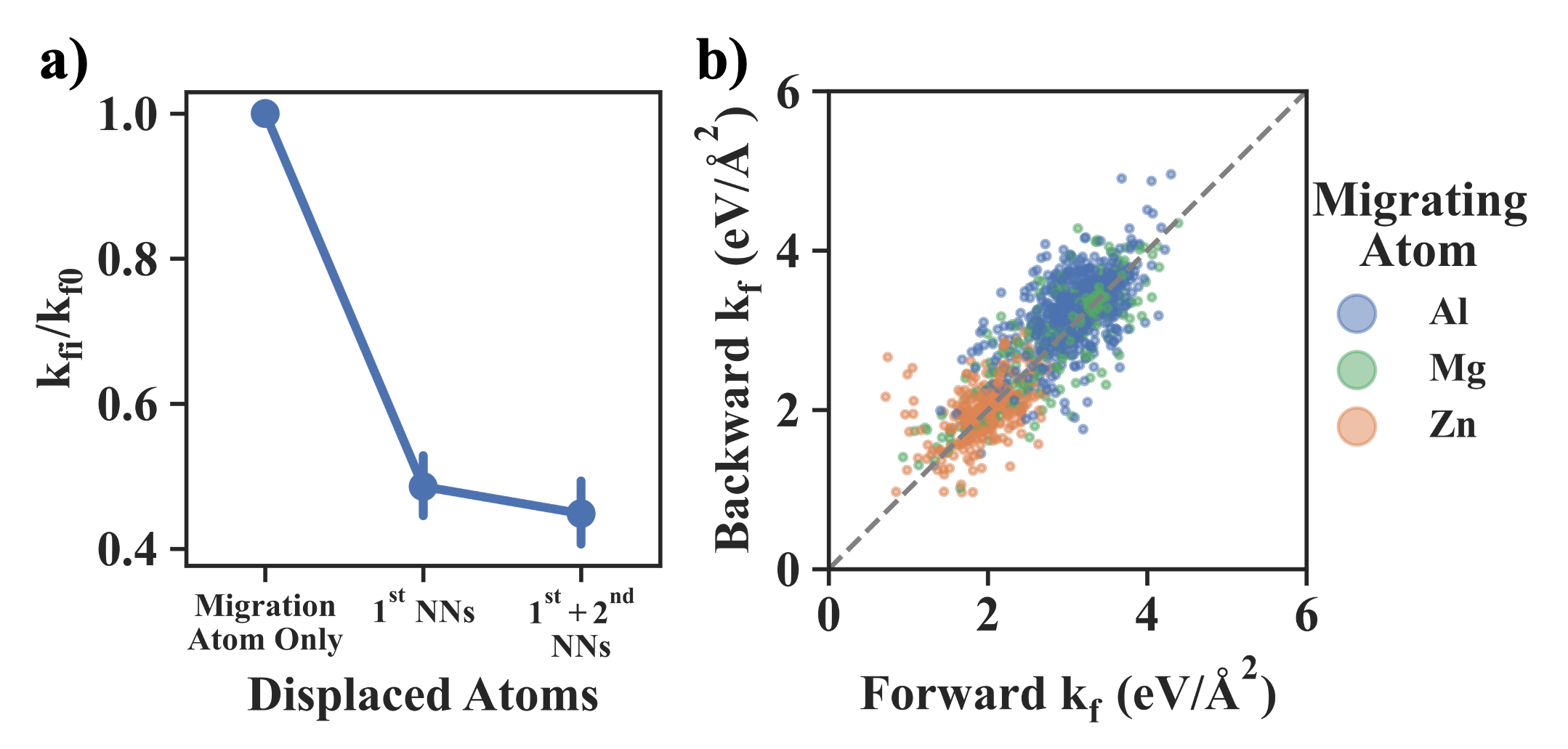}
    \caption{(a) Examples of the ratio of $k_{f}$ with displaced neighbors ($k_{fi}$ from Type 2 and Type 3 calculations) to $k_{f}$ with migrating atom displaced only ($k_{f0}$ from Type 1 calculations). Here all $k_f$ values are the average results of the 6 investigated supercells. (b) Correlations between the calculated forward $k_{f}$ based on the initial state and backward $k_{f}$ based on the final state. Forward $k_f$ and the backward $k_f$ are roughly linearly correlated.}
    \label{fig:kf}
\end{figure}

In the 1$^{\text{st}}$ nearest neighbor relaxation cases (Type 2), there are 19 atoms displaced, while in the 1$^{\text{st}}$ and 2$^{nd}$ nearest neighbor relaxation cases (Type 3), there are overall 27 atoms displaced. The averaged relative intensities ($k_{fi}/k_{f0}$, where $i$ can be $0$, $1$, or $2$) of these 6 supercells are shown in \cref{fig:kf} (a). With more atoms displaced, $k_{fi}$ drops and converges to around one-half of $k_{f0}$, obtained from the only migrating atom displaced cases (Type 1), indicating there is an approximately linear correlation between $k_f$ under the fixed-atom conditions (Type 1) and $k_{f*}$.

Based on this linear correlation, we obtained the second derivative of the energy landscape along the MEP of each CI-NEB calculation in an efficient way. We fixed other atoms and only allowed the migrating atom to have a displacement in geometry-optimized initial and final configurations (Type 1 strategy mentioned above) during $\pmb H$ calculations. Hence, $\pmb H$ with $3\times3$ dimensions were obtained. For each migration event, the migrating atom was located at the initial and final states, generating two vibration spring constants, the forward $k_f$ denoting the spring force for the migrating atom jumping from initial state to the final state, and the backward $k_f$ denoting the spring force for the migrating atom jumping from final state to the initial state. The correlations between these two types of spring constants from all migration cases are shown in  \cref{fig:kf} (b). The forward $k_f$ and the backward $k_f$ are approximately linearly related for most of the migration events, even though for individual cases the forward $k_f$ and backward $k_f$ can be different. We can use the average value of these two types of $k_f$ to describe the changes of the second derivative of the energy landscape along the MEP at local-minimum states. These average $k_f$ of both the initial and final states are used in Eq. (9) of the main text.

\clearpage

\section{}
\label{sec:dimensions}
\textbf{Dimensionalities of Combined Feature Vectors}\\
In main text, we discussed sets of single atoms that shared the same symmetry among the $1^{\text{st}}$ nearest neighboring sites of the vacancy and the migrating atom of the $mmm$ point group as shown in the main text Fig. 7 (b). Here we list sets of single atoms and 2-atom clusters:
\begin{itemize}
\item{Single atom set \#1: 1, 18;}
\item{Single atom set \#2: 2, 3, 4, 5, 14, 15, 16, 17;}
\item{Single atom set \#3: 6, 7, 12, 13;}
\item{Single atom set \#4: 8, 9, 10, 22;}
\item{2-Atom cluster (the same symmetry) set \#1: [8,9], [10, 11];}
\item{2-Atom cluster (the same symmetry) set \#2: [6, 12], [7, 13];}
\item{2-Atom cluster (the same symmetry) set \#3: [2, 3], [4, 5], [14, 15], [16, 17];}
\item{2-Atom cluster (different symmetry) set \#1: [8, 6], [8, 12], [9, 7], [9, 13], [10, 7], [10, 13], [11, 6], [11, 12];}
\item{2-Atom cluster (different symmetry) set \#2: [8, 2], [8, 14], [9, 3], [9, 15], [10, 4], [10, 16], [11, 5], [11, 17];}
\item{2-Atom cluster (different symmetry) set \#3: [6, 2], [6, 5], [7, 3], [7, 4], [12, 14], [12, 17], [13, 15], [13, 16];}
\item{2-Atom cluster (different symmetry) set \#4: [2, 1], [3, 1], [4, 1], [5, 1], [14, 18], [15, 18], [16, 18], [17, 18].}
\end{itemize}
Here, single numbers of each set represent the atom indexes as labelled in Fig. 7 (b) of the main text. Atoms listed in square brackets of each set denote the atom indexes of 2-atom clusters also as labelled in Fig. 7 (b) of the main text. It is notable that for 2-atom clusters where two atoms are from different symmetry sites (different color of spheres as shown in Fig. 7 (b) of the main text), the order of indexes in each pair is important because these clusters are orientation-sensitive. After the average operation is performed to each set listed above based on the $mmm$ point group symmetry as illustrated by Eq. (13) of the main text, there are four single atom sets (each set is represented by a feature vector $\in \mathbb{R}^3$ as Eq. (10) in the main text), three symmetric 2-atom cluster set (2 atoms at symmetrically equivalent sites and each set is represented by a feature vector $\in \mathbb{R}^6$ as Eq. (11) in the main text), and four unsymmetrical 2-atom cluster sets (2 atoms at symmetrically different sites and each set is represented by a feature vector $\in \mathbb{R}^9$ as Eq. (12) in the main text). So, the total dimensionality of this combined feature vector is $4\times3 + 3\times6 + 4\times9 = 66$.

The numbers of different sets with different nearest neighbor lattice sites are summarized in \cref{tbl:dimensionality} when the cutoff is increased to $3^{\text{rd}}$ nearest neighbor distance,.
\begin{table}[ht]
    \centering 
    \caption{The number of sets of single atom and 2-atom clusters that share the same symmetry.}
    \label{tbl:dimensionality}
    \vspace{5mm} 
    \begin{tabular}{|c|c|c|c|c|c|c|}
\hline
\textbf{Sites} & \multicolumn{2}{|c|}{$\pmb{1^{\text{st}}}$ \textbf{NN}} & \multicolumn{2}{|c|}{$\pmb{1^{\text{st}}+2^{\text{nd}}}$ \textbf{NN}} & \multicolumn{2}{|c|}{$\pmb{1^{\text{st}}+2^{\text{nd}}+3^{\text{rd}}}$ \textbf{NN}} \\
\hline
\textbf{Symmetry} & $\pmb{mmm}$ & $\pmb{mm2}$ & $\pmb{mmm}$ & $\pmb{mm2}$ & $\pmb{mmm}$ & $\pmb{mm2}$ \\
\hline
Single atom sets & 4 & 7 & 6 & 11 & 11 & 20\\
 \hline
First nearest 2-Atom cluster (the same symmetry) sets & 3 & 3 & 4 & 3 & 6 & 5\\
 \hline
Second nearest 2-Atom cluster (the same symmetry) sets & 2 & 3 & 2 & 3 & 5 & 8\\
 \hline
Third nearest 2-Atom cluster (the same symmetry) sets & 2 & 3 & 2 & 3 & 3 & 5 \\
 \hline
First nearest 2-Atom cluster (different symmetry) sets & 4 & 9 & 9 & 20 & 26 & 54 \\
 \hline
Second nearest 2-Atom cluster (different symmetry) sets & 2 & 4 & 2 & 4 & 10 & 20 \\
 \hline
Third nearest 2-Atom cluster (different symmetry) sets & 4 & 10 & 9 & 20 & 30 & 63 \\
\hline
\textbf{Dimensionality of Combined Feature Vector} & \textbf{144} & \textbf{282} & \textbf{246} & \textbf{483} & \textbf{711} & \textbf{1401} \\
\hline
    \end{tabular}
\end{table}

\clearpage

\section{}
\label{sec:pca}
\textbf{Principal Component Analysis for Dimensionality Reduction}\\
Principal component analysis (PCA) is applied to reduce the dimensionality of the feature vectors. PCA is a process of computing the principal components of a set of data and using them to change the data basis. It ignores the less important components. 

\begin{figure}[ht]
     \centering
     \includegraphics[width=1\textwidth]{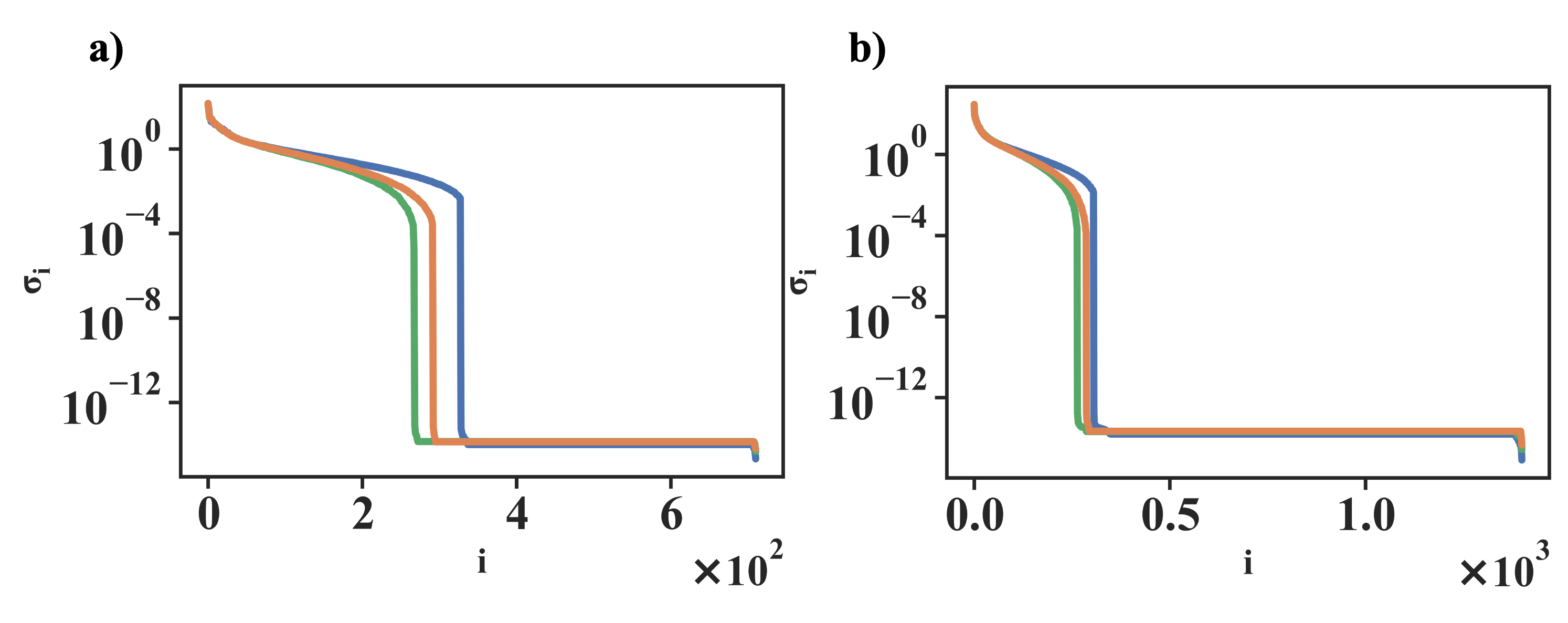}
     \caption{Distributions of the singular values of the covariance matrix. X axis represents the $i^{\text{th}}$ element in the $\pmb \sigma$ list, and the Y axis represents the magnitude of the $\sigma_i$ of the data of Al (blue), Zn (orange), and Mg (green) migrating atoms. (a) $\sigma_i$ of $\pmb C$ obtained from the training data with the $mmm$ symmetry. The size of the $\pmb \sigma$ list is 1401. (b) $\sigma_i$ of $\pmb C$ obtained from the training data with the $mm2$ symmetry. The size of the $\pmb \sigma$ list is 711. In both cases, the magnitudes of singular values with larger indices are notably greater than those of singular values with smaller indices, indicating the dimensionality of the data can be significantly reduced without damaging the integrity. }
     \label{fig:svd}
\end{figure}

We divided the samples into three different sets based on the chemical type of migrating atom (Al, Mg, and Zn). For each set of data, we built a zero centered and unit-variance scaled feature space $\pmb X_{m\times n}$ that contains all the feature vectors of the training data in this set, where $m$ is the size of the dataset and $n$ is the dimensionality of the feature vector. Then we applied the singular value decomposition to its covariance matrix, $\pmb C=\text{cov}(\pmb X, \pmb X)$, which is:
\begin{equation}
    \pmb U \pmb \Sigma \pmb U^\text{T} = \frac{1}{n} \sum\limits_{i=1}^{n} (\pmb x_i - \bar{\pmb x})(\pmb x_i - \bar{\pmb x})^\text{T}
\end{equation}
Here, $\pmb x_i$ is the $i^{\text{th}}$ row of $\pmb X$, and $\bar{\pmb x}$ is the averaged row of $\pmb X$. $\pmb {\Sigma }$ is a square diagonal of the size $r \times r$, where $r$ is the rank of $\pmb C$, and $\pmb U$ is an $n \times r$ semi-unitary matrix. The diagonal entries of $\pmb {\Sigma}$ are equal to the singular values of $\pmb C$, which are non-negative real numbers sorted in decreasing order representing the information abundance of each dimensionality after change of basis. \cref{fig:svd} (a) shows the distribution of all of the singular values, $\sigma_i$, of $\pmb C$ obtained from the training data with the $mmm$ symmetry for each type of the migrating atoms (Al, Zn and Mg). \cref{fig:svd} (b) shows singular values of $\pmb C$ obtained from the training data with the $mm2$ symmetry.

The dimensionality reduced sample space can be represented as
\begin{equation}
    \hat {\pmb{X}} = \pmb{X} \pmb U[:,1:z]
\end{equation}
Here $\pmb U[:,1:z]$ represents the first $z^{\text{th}}$ columns of the semi-unitary matrix $\pmb U$, and $z$ is the number of dimensionalities we want to keep. Since singular values can be used to describe the information abundance of dimensionalites, the ratio of the sum of first $z$ singular values, $\sum\limits_{i=1}^{z} \sigma_i$, to the sum of all singular values$\sum\limits_{i=1}^{r} \sigma_i$, ranging from $0$ to $100 \%$, denotes the degree of representativeness of the dimensionality-reduced data compared to the original data. The larger this ratio, the more representativeness the processed dataset has. To make sure the integrity of the data is not weakened, we chose a parameter $z$ that makes the ratio, $\frac{\sum\limits_{i=1}^{z} \sigma_i}{\sum\limits_{i=1}^{r} \sigma_i}$,  greater than $99.9 \%$. \cref{tbl:pca} shows the parameter $z$ chosen (reduced dimension) for the different types of migrating atoms (Al, Mg, or Zn) and different symmetries ($mmm$ and $mm2$).

\begin{table}[ht]
    \centering 
    \caption{Dimensionality reduction for the training dataset. Note that the chosen dimension $z$ was selected based on the properties of training dataset; hence it may change since training data are randomly chosen from the whole dataset.}
    \label{tbl:pca}
    \vspace{5mm} 
    \begin{tabular}{|c|c|c|c|c|}
\hline
Symmetry & Migrating atom & Original Dimension & Reduced Dimension $z$ & Reduction in Dimension \\
\hline
$mmm$ & Al & 711 & 286 & $59.77 \%$ \\
 \hline
$mmm$ & Mg & 711 & 210 & $70.46 \%$ \\
 \hline
$mmm$ & Zn & 711 & 231 & $67.51 \%$ \\
 \hline
$mm2$ & Al & 1401 & 273 & $80.51 \%$ \\
 \hline
$mm2$ & Mg & 1401 & 207 & $85.22 \%$ \\
 \hline
$mm2$ & Zn & 1401 & 224 & $84.01 \%$ \\
\hline
    \end{tabular}
\end{table}

\clearpage

\section{}
\label{sec:results}
\begin{figure}[ht]
     \centering
     \includegraphics[width=1\textwidth]{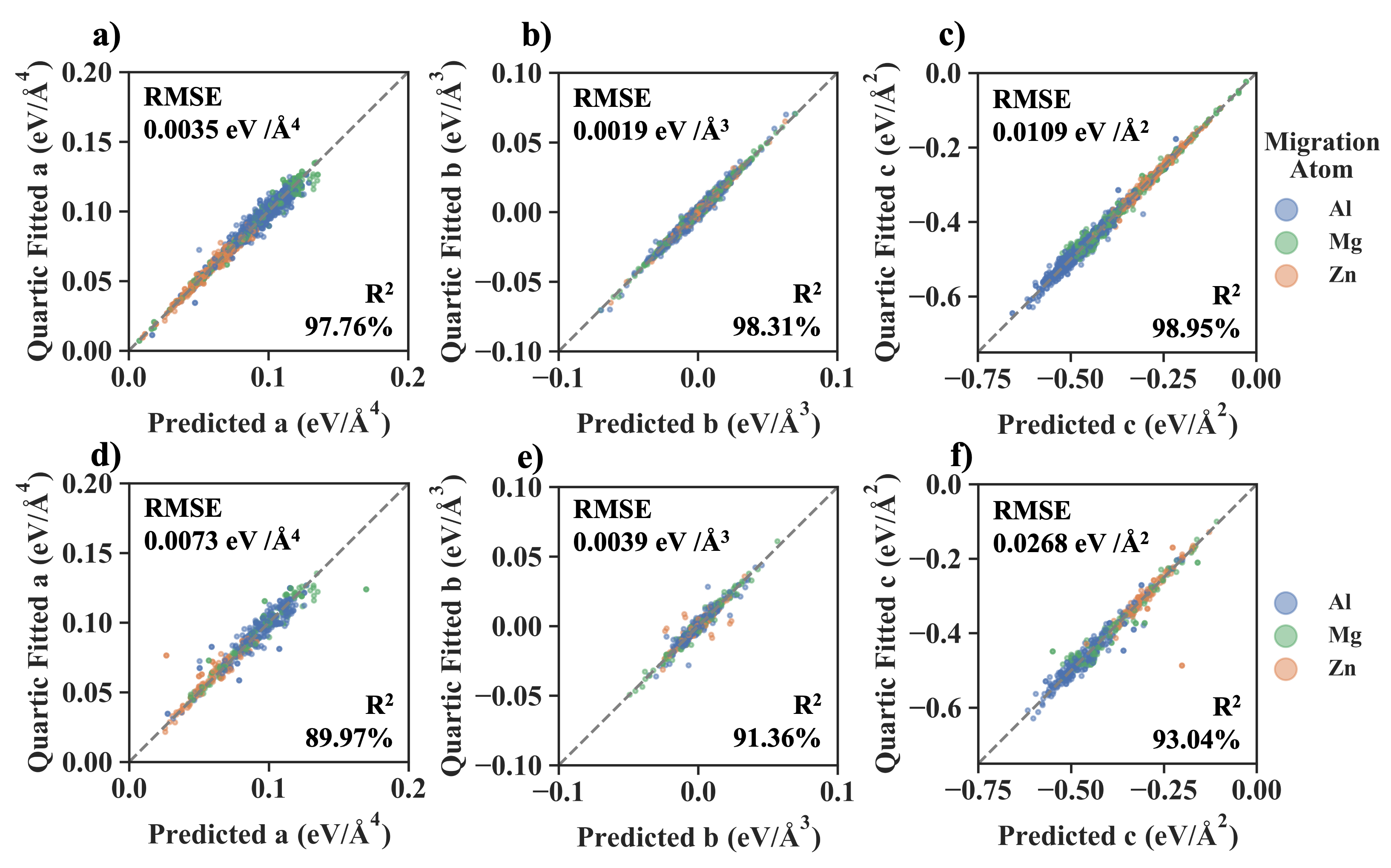}
     \caption{Results from surrogate models of the quartic function coefficients $a$, $b$ and $c$ for the training data (a)-(c) and testing data (d)-(f) compared with quartic fitted results. Different colors represent different types of migrating atoms: Al (blue), Zn (orange), Mg (green). In each sub-figure, the number in the upper-left corner shows the RMSE values, and the number in the bottom-right corner shows the R$^2$ scores. Low values of the RMSE (close to 0) and high values of the $R^2$ (close to 100$\%$) prove the accuracy of our surrogate models.}
     \label{fig:prediction_quartic}
\end{figure}

\begin{figure}[ht]
     \centering
     \includegraphics[width=1\textwidth]{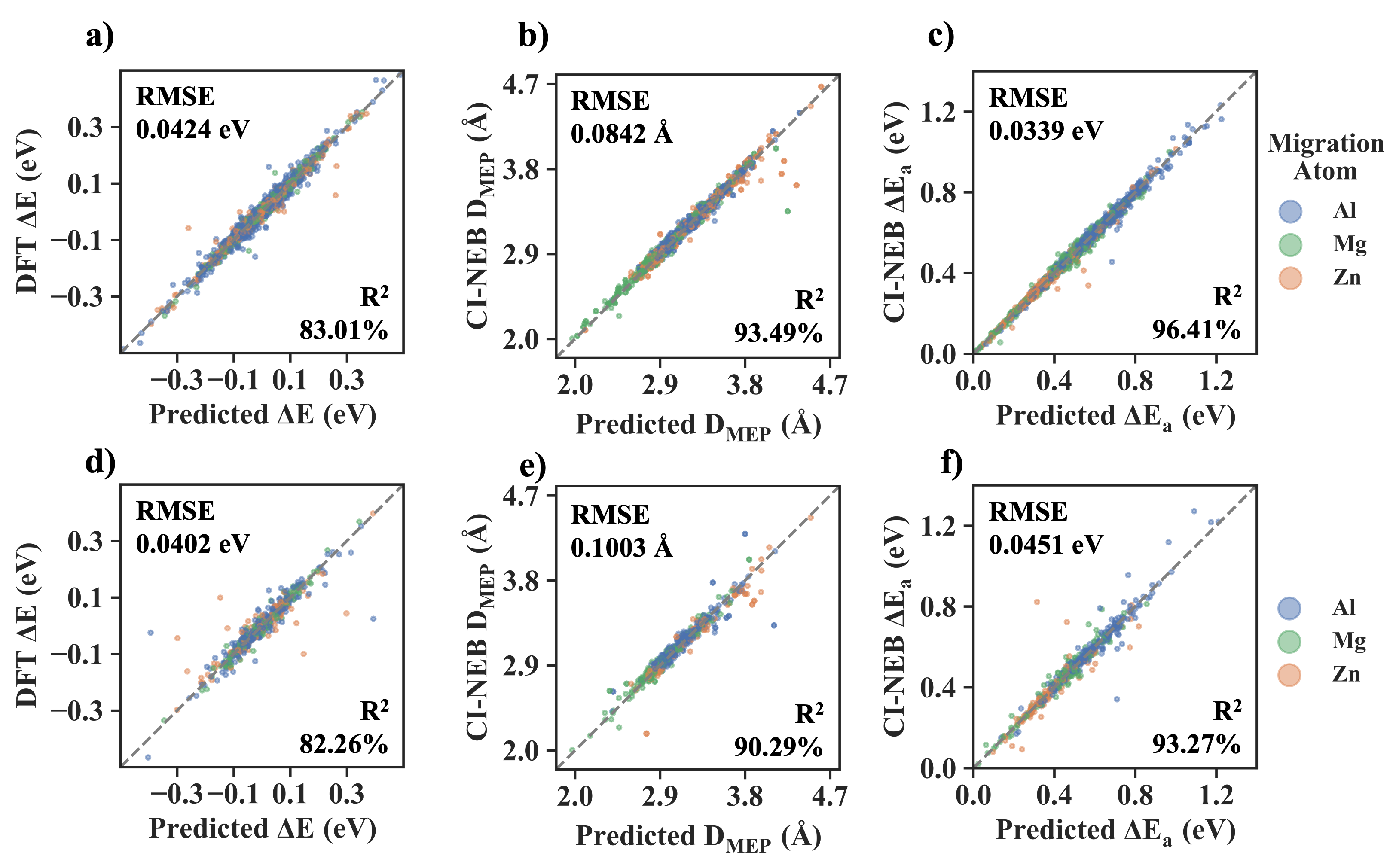}
     \caption{Results from surrogate models of $\Delta E$, $D_{\text{MEP}}$ and $\Delta E_{\text{a}}$ for the training data (a)-(c) and testing data (d)-(f) compared with DFT+CI-NEB calculated results. Different colors represent different types of migrating atoms: Al (blue), Zn (orange), Mg (green). In each sub-figure, the number in the upper-left corner shows the RMSE values, and the number in the bottom-right corner shows the R$^{2}$ scores. Low values of the RMSE (close to 0) and high values of the $R^2$ (close to 100$\%$) prove the accuracy of our surrogate models.}
     \label{fig:prediction_MEP}
\end{figure}

\textbf{Performance of Surrogate Models}\\
The training and testing results of the performance for the surrogate models to predict coefficients of the quartic function of the MEP (Eq. (5) in the main text) are shown in \cref{fig:prediction_quartic}. The training and testing results of the performance for the surrogate models to predict properties of the MEP ($\Delta E_{\text{a}}$, $\Delta E$, and $D_{\text{MEP}}$) are shown in \cref{fig:prediction_MEP}. \cref{fig:prediction_quartic} (a), (b), (c) show the performance of the surrogate quartic parameter $a$, $b$, and $c$ for the training data, respectively. \cref{fig:prediction_quartic} (d), (e), and (f) show that for the testing data. \cref{fig:prediction_MEP} (a), (b), and (c) show the comparison of predictions of the MEP ($\Delta E_{\text{a}}$, $\Delta E$, and $D_{\text{MEP}}$) from our surrogate models (X-axis) with DFT+CI-NEB results (Y-axis) of 2000 training data. \cref{fig:prediction_MEP} (d), (e), and (f) show those of the testing data with low values of the root-mean-square error (RMSE) (close to 0) and high values of the coefficient of determination $R^2$ (close to 100$\%$)

\clearpage

\bibliography{supplementary}